%% file: master.tex
\documentclass[a4paper,fleqn,10pt]{article}
\pdfoutput=1
\input{text/global}
\hypersetup{
  pdfauthor={Stefan Hoeche, Frank Krauss, Marek Schoenherr, Frank Siegert},
  pdftitle={A critical appraisal of NLO+PS matching methods}
}
\preprint{SLAC-PUB 14661\\IPPP/11/67\\DCPT/11/134\\LPN11-58\\
  FR-PHENO-2011-019\\NSF-KITP-11-235\\MCNET-11-24}
\author{Stefan H{\"o}che$^1$, Frank Krauss$^2$,
  Marek Sch{\"o}nherr$^2$, Frank Siegert$^3$}
\title{A critical appraisal of NLO+PS matching methods}
\institute{$^1$ SLAC National Accelerator Laboratory, 
  Menlo Park, CA 94025, USA\\
  $^2$ Institute for Particle Physics Phenomenology,
  Durham University, Durham DH1 3LE, UK\\
  $^3$ Physikalisches Institut,
  Albert-Ludwigs-Universit{\"a}t Freiburg, D-79104 Freiburg, Germany\\}
\begin{document}
\maketitle
\begin{abstract}
  \input{text/abstract}

\end{abstract}
\input{text/intro}
\input{text/catchy}
\input{text/mcatnlo}
\input{text/results}

\input{text/conclusions}
\input{text/acknowledgements}
\appendix
\input{text/appendix}

\bibliographystyle{bib/amsunsrt_modp}
\bibliography{bib/journal}
\end{document}

%% file: text/global.tex
\usepackage{amsmath}
\usepackage{amssymb}
\usepackage{array}
\usepackage{calc}
\usepackage{longtable}
\usepackage{multirow}
\usepackage{pstricks}
\usepackage{graphicx}
\usepackage{xspace}
\usepackage{mciteplus}
\usepackage[a4paper,pdfborder={0 0 0}]{hyperref}
\usepackage[format=hang,labelfont=bf,hypcap=true]{caption}
\usepackage{subfig}
\usepackage{sectsty}
\allsectionsfont{\sffamily}
\subsubsectionfont{\mdseries\itshape\large}
\makeatletter
\DeclareRobustCommand*{\bfseries}{%
  \not@math@alphabet\bfseries\mathbf
  \fontseries\bfdefault\selectfont
  \boldmath
}
\makeatother
\let\spreprint\empty
\newcommand{\preprint}[1]{\def\spreprint{\protect#1}}
\let\sinstitute\empty
\newcommand{\institute}[1]{\def\sinstitute{\protect#1}}
\makeatletter
\renewcommand{\maketitle}{\begingroup
  \null\thispagestyle{empty}%
    \ifx\spreprint\empty
      \vskip 5ex
    \else
      \flushright\large\spreprint\vskip 2ex
    \fi
    \vskip 5ex
    \flushleft
      {\sffamily\bfseries\scalebox{0.95}{\huge\@title}}\vskip 2ex
      \@author\vskip 2ex
      \ifx\sinstitute\empty
      \else
        {\small\sinstitute}
      \fi
    \vskip 5ex
  \endgroup
}
\makeatother
\renewenvironment{abstract}{\begin{center}
  {\large\sffamily\bfseries Abstract: }
  \begin{minipage}[t]{0.75\textwidth}
}{\end{minipage}\end{center}\vskip 10ex}

\numberwithin{equation}{section}
\newcommand{\MCatNLO}{M\protect\scalebox{0.8}{C}@N\protect\scalebox{0.8}{LO}\xspace}
\newcommand{\aMCatNLO}{aM\protect\scalebox{0.8}{C}@N\protect\scalebox{0.8}{LO}\xspace}

\newcommand{\POWHEG}{P\protect\scalebox{0.8}{OWHEG}\xspace}
\newcommand{\PowhegBox}{P\protect\scalebox{0.8}{OWHEG}B\protect\scalebox{0.8}{OX}\xspace}
\newcommand{\LOPS}{LO$\otimes$PS\xspace}

\newcommand{\Herwig}{H\protect\scalebox{0.8}{ERWIG}\xspace}

\newcommand{\Pythia}{P\protect\scalebox{0.8}{YTHIA}\xspace}

\newcommand{\MCFM}{M\protect\scalebox{0.8}{CFM}\xspace}
\newcommand{\Sherpa}{S\protect\scalebox{0.8}{HERPA}\xspace}

\newcommand{\Amegic}{A\protect\scalebox{0.8}{MEGIC++}\xspace}

\newcommand{\DO}{D\O\ }

\newcommand{\ATLAS}{ATLAS\xspace}

\long\def\symbolfootnote[#1]#2{\begingroup%
\def\thefootnote{\fnsymbol{footnote}}\footnote[#1]{#2}\endgroup}

\newcommand{\abs}[1]{\left| #1\right|}
\newcommand{\rbr}[1]{\left( #1\right)}
\newcommand{\abr}[1]{\langle #1\rangle}
\newcommand{\cbr}[1]{\left\{ #1\right\}}
\newcommand{\sbr}[1]{\left[ #1\right]}
\newcommand{\im}{\imath}
\newcommand{\jm}{\jmath}
\newcommand{\args}[1]{\{\vec{#1}\}}
\newcommand{\mathsc}[1]{\MakeUppercase{\scriptstyle #1}}
\newcommand{\rargs}[1]{\{\vec{\mathsc{#1}}\}}

\newcommand{\done}{{\rm d}}
\newcommand{\order}{\mathcal{O}}
\newcommand{\mc}[1]{\mathcal{#1}}
\newcommand{\mr}[1]{\mathrm{#1}}
\newcommand{\mb}[1]{\mathbb{#1}}

\newcommand{\bmap}[3]{b_{#1,#2}(#3)}
\newcommand{\rmap}[3]{r_{\widetilde{#1},\tilde{#2}}(#3)}

\newcommand{\nnb}{\nonumber}
\newcommand{\bea}{\begin{eqnarray}}
\newcommand{\eea}{\end{eqnarray}}
\newcommand{\bi}{\begin{itemize}}
\newcommand{\ei}{\end{itemize}}

\newcommand{\changed}[1]{#1}

%% file: text/abstract.tex
In this publication, uncertainties in and differences between the \MCatNLO and \POWHEG 
methods for matching next-to-leading order QCD calculations with parton showers
are discussed.  Implementations of both algorithms within the event generator
\Sherpa \changed{and based on Catani-Seymour subtraction} are employed to assess the impact 
on a representative selection of observables. \changed{In the case of \MCatNLO a substantial 
simplification is achieved by using dipole subtraction terms to generate the first emission.}
A phase space restriction is \changed{employed}, which allows to vary in a transparent way 
the amount of non-singular radiative corrections that are exponentiated.  
Effects on various observables are investigated, using the production of a Higgs boson 
in gluon fusion, with or without an associated jet, as a benchmark process.  
The case of $H$+jet production is presented for the first time
in an NLO+PS matched simulation.
Uncertainties due to scale choices and non-perturbative effects
are explored in the production of $W^{\pm}$ and $Z$ bosons
in association with a jet.  Corresponding 
results are compared to data from the Tevatron and LHC experiments.

%% file: text/intro.tex
\section{Introduction}
\label{sec:intro}

The central topic in the development of Monte-Carlo event generators
during the past decade was
the systematic inclusion of higher order QCD effects.
One of the first approaches, the merging of leading-order (LO) 
multi-jet matrix elements of varying multiplicity with parton showers,
has been pioneered and matured in a series of papers 
\cite{Catani:2001cc,*Mangano:2001xp,*Krauss:2002up,*Lonnblad:2001iq,
  *Lavesson:2007uu,*Alwall:2007fs,*Hoeche:2009rj,
  *Hamilton:2009ne,*Carli:2010cg,*Lonnblad:2011xx},
such that by now it has become the accepted standard for simulating final 
states which include multi-jet topologies.  
It typically yields a very satisfactory description of experimental data,
but due to the lack of virtual corrections it can never achieve the formal 
accuracy of a full next-to-leading order (NLO) calculation.  This shortcoming 
is especially worrisome in the case of Standard Model Higgs-boson production 
through gluon fusion, where a large $K$-factor indicates 
substantial higher-order corrections,
such that theoretical predictions should be made at next-to-leading order 
accuracy or better.

With the development of the \MCatNLO~\cite{Frixione:2002ik} and 
\POWHEG~\cite{Nason:2004rx,*Frixione:2007vw} techniques it became feasible 
to combine NLO accurate matrix-element calculations with parton showers.
This technology is called NLO matching, in contrast to the LO merging described above.
Ultimately, NLO matching techniques allow to carry out a full simulation of events,
including hadronisation, hadron decays and the 
underlying event. While \MCatNLO relies on a subtraction algorithm based on 
the parton shower approximation to collinear divergences in real-emission 
matrix elements, the \POWHEG method effectively constructs a one-emission generator,
similar to a one-step parton shower, with evolution kernels determined by ratios 
of real-emission and Born matrix elements. In this respect the \POWHEG method 
is very similar to traditional matrix-element correction techniques~\cite{Seymour:1994df,
  *Seymour:1994we,*Andre:1997vh,*Norrbin:2000uu}.

Despite having been proposed several years ago, the \MCatNLO and \POWHEG methods
were not applied to processes which can become singular at Born level, such as
di-jet production, until recently. This delay signals several complications which arise 
in reactions with additional light partons. Di-jet production~\cite{Alioli:2010xa}
and $Z+j$ production~\cite{Alioli:2010qp} have now become available through the 
\PowhegBox~\cite{Alioli:2010xd} \changed{and, more} recently, $W^\pm$ plus two jets 
production was implemented~\cite{Frederix:2011ig} using \MCatNLO methods. 

In this publication the \MCatNLO and \POWHEG methods will be compared in some detail,
and open issues related to their implementation and validation will be discussed.
As benchmark processes, the production of Higgs-, $W^\pm$- and $Z$-bosons, 
alone or in association with one hard jet at Born level have been chosen.   
These processes are either signals central to the experimental program at the 
LHC, or they contribute as important backgrounds to many new physics 
searches~\cite{daCosta:2011qk,*Aad:2011hh,*Chatrchyan:2011nd,
  *Chatrchyan:2011ida,*Chatrchyan:2011qs}.
Apart from that, QCD-associated $W^\pm$- and $Z$-production are of great 
interest, to study jet production and evolution in a 
hadron-collider environment at the energy frontier~\cite{Aad:2010pg,Aad:2011xn,
  *Ask:2011cg,*Chatrchyan:2011ig,*Rogan:2009zz}, 
to improve the jet energy scale determination by the experiments, or to study 
multiple parton scattering processes.  At the LHC as well as at the 
Tevatron, typically, good agreement is found 
when comparing respective data with merged leading-order predictions
or with next-to-leading order perturbative QCD predictions, for instance 
from~\cite{Campbell:2002tg,*KeithEllis:2009bu,*Ellis:2009zw,
  Berger:2008sj,*Berger:2009zg,*Berger:2010vm,*Berger:2010zx,*Ita:2011wn}.  
This level of theoretical control and experimental precision suggests that the 
processes chosen here are indeed well-suited to serve as testbed for NLO
matching methods.

The present paper is organised as follows: In Sec.~\ref{sec:catchy} 
basic ideas underlying the \MCatNLO and \POWHEG strategies are reviewed.
Similarities and differences between the two methods are discussed and
potential pitfalls in their implementation are indicated.  
The solution to some of these problems is detailed in Sec.~\ref{sec:mcatnlo},
where the \MCatNLO algorithm is worked out in a formalism similar 
to~\cite{Hoeche:2010pf}. This allows to elaborate on various aspects of the
method regarding non-trivial colour and flavour configurations.
Readers not interested in the technical details may skip this section. 
In Sec.~\ref{sec:perturbative} perturbative uncertainties related
to scale variations and to the different exponentiation in the \MCatNLO and
\POWHEG methods are highlighted.
The impact of non-perturbative 
effects is investigated in Sec.~\ref{sec:nonperturbative} and results for $W^\pm$- 
and $Z$-production are compared to experimental data.
Finally, Sec.~\ref{sec:conclusions} summarises the findings 
of this publication and outlines some possible extensions of the methods
presented here.  

%% file: text/catchy.tex
\section{Event generation with \texorpdfstring{\POWHEG}{POWHEG} and \texorpdfstring{\MCatNLO}{MC@NLO}
  in a nutshell}
\label{sec:catchy}

This section is meant to highlight in a simple language the ideas of how NLO QCD calculations
can be combined with subsequent parton showers. A common notation is
established for \POWHEG and \MCatNLO and critical aspects related to the
practical implementation for processes with non-trivial colour structure
are discussed.

\subsection{Anatomy of NLO calculations}
In order to see how the existing matching algorithms work, consider first the 
structure of an NLO calculation.  The total cross section can be written as
\begin{equation}
  \label{eq:total_xsec_trivial}
  \sigma^{\rm (NLO)} \,=\; \int\done\Phi_B\sbr{\,\mr{B}(\Phi_B)+\tilde{\mr{V}}(\Phi_B)+\mr{I}^{\rm(S)}(\Phi_B)\,} +
    \int\done\Phi_R\sbr{\,\vphantom{\tilde{V}}\mr{R}(\Phi_R)-\mr{D}^{\rm(S)}(\Phi_R)\,}\;,
\end{equation}
where $\done\Phi_B$ and $\done\Phi_R$ denote Born and real-emission phase space,
respectively and $\mr{B}$, $\tilde{\mr{V}}$, $\mr{I}^{\rm(S)}$, $\mr{R}$ and $\mr{D}^{\rm(S)}$, 
are the matrix elements for the Born, virtual, integrated subtraction, 
real emission and real subtraction contribution.
Note that for hadronic initial states $\tilde{\mr{V}}$ is defined such as to include
the collinear mass-factorisation counterterms.
All integrands include parton luminosity and flux factors.
For simplicity it is assumed that the processes under consideration here 
have no identical QCD partons in the final state, a detailed discussion of 
symmetry factors is postponed to Sec.~\ref{sec:mcatnlo}.
In the subtraction formalisms most commonly employed to date~\cite{Catani:1996vz,
  *Catani:2002hc,Frixione:1995ms,*Frixione:1997np}, subtraction terms can
be written as the convolution of Born matrix elements with suitably defined
operators $\tilde{\mr{K}}$, such that schematically
\begin{equation}
\done\Phi_R\,\mr{D}^{\rm(S)}(\Phi_R) \,=\; \done\Phi_B\,\done\Phi_1\,
  \sbr{\mr{B}(\Phi_B)\otimes\tilde{\mr{K}}(\Phi_1)}
\quad\text{and}\quad
\mr{I}^{\rm(S)}(\Phi_B) \,=\; \int\done\Phi_1\sbr{\vphantom{\tilde{\mr{B}}}
  \mr{B}(\Phi_B)\otimes\tilde{\mr{K}}(\Phi_1)}\;,
\end{equation}
indicating that the two related contributions cancel locally in the Born phase space.
This also suggests that the definition of the subtraction kernels $\tilde{\mr{K}}$ is 
somewhat arbitrary, as long as they exhibit the same singularity structure as the full 
real-emission  matrix elements in the soft and collinear limits.

To combine an NLO calculation with a parton shower, it is essential to ensure 
that the result inherits the total cross section from the fixed-order calculation.
This could trivially be achieved by multiplying a leading order plus parton shower 
simulation with a local $K$-factor - the ratio of NLO and LO cross section.  
The problem is that parton showers generate emissions which 
do not necessarily follow the pattern obtained from the real emission matrix
elements, leading to a mismatch of radiation patterns at first order
in the strong coupling constant $\alpha_s$.  In principle, this problem
has been solved by the traditional matrix-element correction procedure 
outlined in \cite{Seymour:1994df,*Seymour:1994we,*Andre:1997vh,*Norrbin:2000uu},
which is implemented for a number of processes in both \Herwig and \Pythia.  The 
essence of this method lies in the fact that, for the processes it is applied 
to, the product of Born-level contribution and parton shower evolution kernel 
$\mr{K}$ is larger than the corresponding real emission term 
$\mr{R}$ in the complete phase space of the extra emission.  This
allows to correct the first (hardest) emission with a factor 
$\mr{R}/(\mr{B}\,\mr{K})$, leading to the desired 
distribution in phase space.  From this it can be seen that such an algorithm 
will fail for cases, where $\mr{R}$ is not always smaller than the 
Born-times-parton-shower result, or where the parton shower is not capable 
of filling the full phase space.  In such cases a more
detailed analysis is necessary.

\subsection{Sudakov form factors}
In this section the construction of the parton shower using 
Sudakov form factors will be briefly recalled.  They are defined as 
\begin{equation}
\label{eq:def_Sudakov_trivial}
\Delta(t,t') \,=\; 
\exp\cbr{-\int_t^{t'}\frac{\done \bar{t}}{\bar{t}}
  \int\done z\int\frac{\done\phi}{2\pi}\,J(\bar t,\,z,\,\phi)\,
  \frac{\alpha_s}{2\pi}\,\mr{K}(\bar t,\,z,\,\phi)} =
\exp\cbr{-\int_t^{t'}\done\Phi_1\;
  \frac{\alpha_s}{2\pi}\;\mr{K}(\Phi_1)}\,,
\end{equation}
where the sequence of emissions is ordered by a parameter $t$ -- typically
related to the virtuality, transverse momentum or opening angle of the emission.  
Furthermore, $z$ is the variable defining the splitting of energy or 
light-cone momentum of the two daughters emerging in the decay of the parton, 
and $\phi$ denotes the azimuthal angle.  Collectively they define the 
one-emission phase space $\Phi_1$ and its Jacobian $J$, while $\mr{K}$ is the splitting kernel 
encoding the detailed kinematical distribution due to soft-collinear
enhancement and spin effects. For simplicity, parton luminosity
factors are included into this kernel, and a detailed discussion is postponed
to Sec.~\ref{sec:mcatnlo}. Note that in all practical 
implementations of Eq.~\eqref{eq:def_Sudakov_trivial} the argument of the 
strong coupling is related to the transverse momentum of the splitting, given
by the decay kinematics, i.e.\ by $\Phi_1$. For the sake of clarity this
argument will be suppressed in this section.

This Sudakov form factor can be interpreted as a no-branching probability 
between the two scales $t$ and $t'$.  Obviously, $1-\Delta(t,t')$ then yields the 
probability that a splitting has taken place in the interval between the two 
scales.  Therefore, to first order in $\alpha_s$, the cross section in the 
parton-shower approximation reads
\begin{equation}
\label{eq:emit_by_shower}
  \sigma_{\rm PS}^{\rm(LO)} \,=\;
  \int\done\Phi_B\,\mr{B}(\Phi_B)\sbr{\,\Delta(t_0,\mu_F^2) +
  \int_{t_0}^{\mu_F^2}\done\Phi_1\frac{\alpha_s}{2\pi}\;
    \mr{K}(\Phi_1)\,\Delta(t,\mu_F^2)\,}\;,
\end{equation} 
where $t$ is determined by the kinematics of the first emission, 
$t = t(\Phi_1)$, such that it is smaller than the suitably chosen, 
process-dependent factorisation scale $\mu_F^2$, and where the arguments of the splitting 
kernel depend on this extra emission kinematics.  The first term in the square bracket
represents the probability of no extra emission to happen above the infrared cut-off $t_0$ 
of the parton shower, while the second term generates one emission above this 
scale.  It can also be seen that the square bracket as a whole integrates to 
one, exposing the probabilistic nature of the parton shower, which 
leaves the total cross section of an event sample unchanged.

\subsection{Matrix element corrections and \texorpdfstring{\POWHEG}{POWHEG}}
Applying matrix element corrections to the parton shower
transforms the equation above, Eq.~\eqref{eq:emit_by_shower}, into
\begin{equation}
\label{eq:emit_by_MEC}
\sigma_{\rm MEC}^{\rm(LO)} \,=\; 
\int\done\Phi_B\,\mr{B}(\Phi_B)\sbr{\,\bar{\Delta}(t_0,\mu_F^2) +
  \int_{t_0}^{\mu_F^2}\done\Phi_1
  \frac{\mr{R}(\Phi_B,\Phi_1)}{\mr{B}(\Phi_B)}\,
  \bar{\Delta}(t,\mu_F^2)\,}\;,
\end{equation} 
where the modified Sudakov form factor $\bar{\Delta}$ reads
\begin{equation}
  \label{eq:def_mod_Sudakov_trivial1}
  \bar{\Delta}(t,t') \,=\; 
  \exp\cbr{-\int_t^{t'}\done\Phi_1\,
  \frac{\mr{R}(\Phi_B,\Phi_1)}{\mr{B}(\Phi_B)}}\,.
\end{equation} 
Again, the square bracket in Eq.~\eqref{eq:emit_by_MEC} integrates to one, 
and the cross section of an event sample is identical to the Born cross section.
However, the distribution of the first (hardest) emission, to first order
in $\alpha_s$ will follow the pattern given by the full real emission
matrix element. \changed{It is thus correct to $\mr{O}(\alpha_s)$ 
if the upper integration limit in the real-emission term is extended 
to the hadronic centre-of-mass energy. This corresponds to a power-shower
approach~\cite{Plehn:2005cq,*Corke:2010zj} and ensures coverage of the 
full phase space. However, it also implies extending the resummation built
into the parton shower beyond the region of its validity and is therefore
unjustified. A more detailed discussion of this point can be found in 
Appendix~\ref{sec:powershower}. The problem can be solved for example
in the \MCatNLO method, see Sec.~\ref{sec:catchy_split}.
}

In order to achieve full $\mr{O}(\alpha_s)$-accuracy in both the cross 
section of the produced event sample and the pattern of the first emission
it is mandatory to replace the differential Born cross section 
$\done\Phi_B\mr{B}$ with an expression that integrates to the full 
NLO cross section of Eq.~\eqref{eq:total_xsec_trivial}.  
This is achieved by the substitution
\begin{equation}\label{eq:def_bbar_trivial}
  B(\Phi_B) \quad \longrightarrow\quad
  \bar{\mr{B}}(\Phi_B) \,=\;
    \mr{B}(\Phi_B)+\tilde{\mr{V}}(\Phi_B)+\mr{I}^{\rm(S)}(\Phi_B)+
    \int\done\Phi_1\left[\vphantom{\tilde{V}}
    \mr{R}(\Phi_B,\Phi_1)-\mr{D}^{\rm(S)}(\Phi_B,\Phi_1)\right]\;.
\end{equation} 
It is straightforward to interpret this term as an NLO-weighted Born-level
cross section, or, put in a slightly different way, a Born-level cross section 
that is augmented with a {\em local} $K$-factor.

Equipped with this definition the expression for the cross section in the
\POWHEG-formalism reads
\begin{equation}
  \label{eq:Powheg_trivial}
  \sigma^{\rm(NLO)}_{\mr{POWHEG}} \,=\; 
    \int\done\Phi_B\,\bar{\mr{B}}(\Phi_B)\,\sbr{\,\bar{\Delta}(t_0) +
    \int_{t_0}\done\Phi_1
    \frac{\mr{R}(\Phi_B,\Phi_1)}{\mr{B}(\Phi_B)}\,
    \bar{\Delta}(t)\,}\;.
\end{equation} 
This expression is accurate up to the first order in the strong coupling 
constant for {\em both} the total cross section of the event sample and the differential
cross section with respect to the kinematics of the first emission.

\subsection{Modified subtraction and \texorpdfstring{\MCatNLO}{MC@NLO}}
\label{sec:catchy_split}
\changed{As indicated above, hard real-emission configurations should not be
included in the exponentiation in Eq.~\eqref{eq:Powheg_trivial}.
There are also contributions which can generically not be obtained from a Born-level 
configuration by simply emitting an extra parton.}  Examples for this include flavour 
processes such as $u\bar u\to c\bar s e^-\bar{\nu}_e$ or kinematical configurations
such as radiation zeroes~\cite{Mikaelian:1979nr,*Goebel:1980es,*Bern:2008qj}.
Therefore, it is well motivated to split the real-emission matrix elements $\mr{R}$
into an infrared-singular (soft) and an infrared-regular (hard) part, $\mr{D}^{\rm(A)}$ and 
$\mr{H}^{\rm(A)}$, respectively: $\mr{R} = \mr{D}^{\rm(A)} + \mr{H}^{\rm(A)}$ \cite{Alioli:2008gx}.
Equation~\eqref{eq:total_xsec_trivial} then transforms into
\begin{equation}
\label{eq:total_xsec_split_trivial}
  \begin{split}
  \sigma^{\rm (NLO)} \,=&\;
  \int\done\Phi_B\left[\,\mr{B}(\Phi_B)+\tilde{\mr{V}}(\Phi_B)+\mr{I}^{\rm(S)}(\Phi_B)\,\right]\\
  &\quad +
  \int\done\Phi_R\left[\,\mr{D}^{\rm(A)}(\Phi_R)-\mr{D}^{\rm(S)}(\Phi_R)\,\right] +
  \int\done\Phi_R\,\mr{H}^{\rm(A)}(\Phi_R)\;,
  \end{split}
\end{equation}
and the cross section, Eq.~\eqref{eq:Powheg_trivial}, can be rewritten as
\begin{equation}
\label{eq:Powheg_split_trivial}
  \sigma^{\rm(NLO)}_{\mr{MC@NLO}} \,= \;
  \int\done\Phi_B\,\bar{\mr{B}}^{\rm(A)}(\Phi_B)\,
  \sbr{\,\bar{\Delta}^{\rm(A)}(t_0) +
  \int_{t_0}\done\Phi_1
  \frac{\mr{D}^{\rm(A)}(\Phi_B,\Phi_1)}{\mr{B}(\Phi_B)}\,
  \bar{\Delta}^{\rm(A)}(t)\,} + 
  \int\done\Phi_R\,\mr{H}^{\rm(A)}(\Phi_R)\;,
\end{equation} 
where
\begin{equation}\label{eq:split_bbar}
  \bar{\mr{B}}^{\rm(A)}(\Phi_B) \,=\;
    \mr{B}(\Phi_B)+\tilde{\mr{V}}(\Phi_B)+\mr{I}^{\rm(S)}(\Phi_B)+
    \int\done\Phi_1\left[\vphantom{\int}\,
    \mr{D}^{\rm(A)}(\Phi_B,\Phi_1)-\mr{D}^{\rm(S)}(\Phi_B,\Phi_1)\,\right]
\end{equation} 
and, with obvious notation,
\begin{equation}
  \label{eq:def_mod_Sudakov_trivial}
  \bar{\Delta}^{\rm(A)}(t,t') \,=\; 
  \exp\cbr{-\int_t^{t'}\done\Phi_1\,
  \frac{\mr{D}^{\rm(A)}(\Phi_B,\Phi_1)}{\mr{B}(\Phi_B)}}\,.
\end{equation} 
In this manner, the non-singular contributions at real-emission level 
are correctly captured \changed{and their exponentiation can be avoided.
This is how the \MCatNLO formalism works~\cite{Frixione:2002ik}.}

\subsection{Subtleties in practical implementations of \texorpdfstring{\POWHEG}{POWHEG} and \texorpdfstring{\MCatNLO}{MC@NLO}}
\label{sec:subtleties}
At the level of detail of the discussion up to now, an implementation of both
methods, \POWHEG and \MCatNLO, seems quite straightforward. However, there are 
a number of subtleties, which have not been presented yet in a coherent fashion 
in a journal publication.
\begin{enumerate}
\item Subtraction of infrared divergent sub-leading colour terms in \MCatNLO\\
  \changed{The parton shower is based on a leading logarithmic and leading colour approximation.
  Using it, without modification, to define modified subtraction terms $\mr{D}^{\rm(A)}$
  will therefore necessarily miss divergences in sub-leading colour configurations
  and will result in infinite integration results in Eq.~\eqref{eq:split_bbar}.}
  This presents a clear obstacle to apply the method to processes where infrared divergences 
  in such sub-leading colour configurations emerge.  In \cite{Frixione:2003ei}, and referring 
  to ideas outlined in the original \MCatNLO publication \cite{Frixione:2002ik}, this 
  problem was overcome for the case of heavy flavour hadro-production 
  processes of the type $gg\to Q\bar Q$ and $q\bar q\to Q\bar Q$ at Born-level.
  \changed{An additional factor was introduced, which multiplies the complete integrand 
  in Eq.~\eqref{eq:split_bbar} and tends to zero as emissions become soft. 
  This same method is applied in the case of the recently presented 
  \aMCatNLO~\cite{Frederix:2011qg,Frederix:private}.}

  \changed{We propose a different, exact and process-independent solution here, which relies 
  on choosing $\mr{D}^{\rm(A)}=\mr{D}^{\rm(S)}$, leading to a tremendous simplification 
  of Eq.~\eqref{eq:split_bbar}. 
  In the remainder of this publication we will refer to this scheme as the \MCatNLO method.
  Technical details of how $\mr{D}^{\rm(S)}$ is exponentiated into a Sudakov form factor
  will be given in Sec.~\ref{sec:mcatnlo}.}
\item Choice of scales in \POWHEG and \MCatNLO\\
  It was argued in~\cite{Nason:2004rx,*Frixione:2007vw}, 
  based on~\cite{Dokshitzer:1978hw}, that the proper choice
  of scale for the strong coupling in the kernel $\mr{R}/\mr{B}$ of the 
  \POWHEG Sudakov form factor is given by the transverse momentum 
  of the emission generated in the branching process. Similarly,
  it was pointed out in~\cite{Frixione:2002ik} that the scale in 
  the evolution kernel of an \MCatNLO should be the transverse
  momentum. This particular choice of scale effectively resums a 
  certain class of universal higher-order corrections. It typically
  leads to a very good agreement between the results from existing
  parton shower algorithms and experimental data and is therefore
  employed in all standard Monte-Carlo event generators~\cite{Buckley:2011ms}.

  Such a scale choice ultimately implies higher-order corrections 
  to the event kinematics. The key point is that the differential NLO cross section
  defined by Eq.~\eqref{eq:def_bbar_trivial} can be evaluated with arbitrary scales.
  However, the functional form of the scale must be infrared and collinear safe.
  This is not the case for the transverse momentum in the branching process, 
  because, in contrast to the Sudakov branching algorithm, there is no infrared cutoff 
  in the calculation of $\bar{\mr{B}}$. Thus, the first-order expansion of the Sudakov 
  form factor does not reproduce the exact same real-correction / subtraction terms 
  as the NLO calculation. \changed{The differences are numerically large in the logarithmically 
  enhanced regions of phase space, where it is not expected that \POWHEG or \MCatNLO
  reproduce the exact leading-order result. But even the region of hard radiation
  can be affected, depending on how much phase space is covered by $\mr{D}^{\rm(A)}$.}
\item Exponentiation of non-leading logarithms in \POWHEG\\
  Ultimately, there is a last subtlety, which has also been discussed
  in \cite{Alioli:2010xd}, summarising some previous work by the
  same authors. The evolution kernel $\mr{R}/\mr{B}$ in the \POWHEG
  method \changed{generates subleading} logarithms and it is somewhat
  questionable in how far they should be exponentiated.  The difference compared to
  evolution kernels in the parton shower may lead to sizable effects, up to orders 
  of magnitude, when results for the hard, non-logarithmic tails of distributions in 
  the \POWHEG approach are compared with those of pure NLO calculations, 
  see for instance \cite{Alioli:2008tz}.

  \changed{In this context, it is important to distinguish two effects:
  One is exponentiation of $\mr{R}/\mr{B}$ beyond the factorization scale 
  $\mu_F$ (cf.\ Eq.~\eqref{eq:Powheg_trivial}),
  the other is the difference in the functional form of the evolution kernels itself.
  The former effect can be emulated in our implementation of the \MCatNLO formalism
  and we will examine it more closely in Secs.~\ref{sec:ggh} and~\ref{sec:gghj}.
  We use CS subtraction and a parton shower built on CS subtraction kernels~\cite{Schumann:2007mg}.  
  Initially the starting scales of the parton shower will be maximised for the first emission, 
  in order to fill the full phase space, similar to the \POWHEG method and a power-shower
  approach~\cite{Plehn:2005cq,*Corke:2010zj}. To show that the main difference between
  \POWHEG and \MCatNLO lies in the enlarged phase space in \POWHEG,
  a simple phase-space constraint is added to the subtraction~\cite{Nagy:2003tz},
  limiting the logarithmically enhanced region of the real-emission
  contribution and adding the hard regions to the regular contribution.  
  The impact of the choice of this constraint is analysed quantitatively 
  in Sec.~\ref{sec:perturbative}. The respective variations in the result are 
  referred to as ``exponentiation uncertainties''. It is important to note that
  this uncertainty is minimised in the \MCatNLO method, as the phase space constraint 
  is determined by the resummation scale. In this publication we simply use the
  \MCatNLO framework to make the problem in \POWHEG explicit.}
\end{enumerate}

%% file: text/mcatnlo.tex
\section{\texorpdfstring{\POWHEG}{POWHEG} and \texorpdfstring{\MCatNLO}{MC@NLO} in detail}
\label{sec:mcatnlo}

When combining next-to-leading order matrix elements with parton showers, 
the logarithmic corrections encoded in the Sudakov form factor must be matched 
to the full next-to-leading order prediction of the parton-level calculation. 
This is usually achieved either by subtracting the first-order expansion of the 
resummed result from the NLO calculation (\MCatNLO with $\mr{D}^{\rm(A)}=\mr{D}^{\rm(S)}$),
or by exponentiating the full real-emission correction in the resummation (\POWHEG).
One can also construct mixed schemes, as argued in Sec.~\ref{sec:catchy_split}.
The difference which is induced by the exponentiation of subleading corrections vanishes 
for soft or collinear parton emission. However, it might play an important role in other
regions of the phase space. In the following, a formalism that allows
to discuss the difference between the \MCatNLO and \POWHEG methods in more detail is reviewed.

\subsection{Notation and definitions}
Denoting sets of $n$ particles in a $2\to (n-2)$ process by 
$\args{a}=\{a_1,\ldots,a_n\}$, while their respective flavours and momenta
are specified separately through $\args{f\,}=\{f_1,\ldots,f_n\}$ and
$\args{p}=\{p_1,\ldots,p_n\}$, the generic expression for a fully differential 
Born-level cross section can be written as a sum over all contributing 
flavour configurations:
\begin{equation}
  \done\sigma_B(\args{p})\,=\;
  \sum_{\args{f\,}}\done\sigma_B(\args{a})\;,
  \qquad\text{where}\qquad
  \done\sigma_B(\args{a})\,=\;\done\Phi_B(\args{p})\,\mr{B}(\args{a})\;,
\end{equation}
The individual terms in the sum are given by
\begin{equation}
  \begin{split}
  \mr{B}(\args{a})\,=&\;\mc{L}(\args{a})\,\mc{B}(\args{a})\;,
  &\mc{B}(\args{a})\,=&\;
  \frac{1}{F(\args{p})}\,\frac{1}{S(\args{f\,})}\,
  \abs{\mc{M}_B}^2(\args{a})\;,\\
  \done\Phi_B(\args{p})\,=&\;\frac{\done x_1}{x_1}\frac{\done x_2}{x_2}\,
  \done{\it\Phi}_B(\args{p})\;,
  &\mc{L}(\args{a};\mu^2)\,=&\;x_1f_{f_1}(x_1,\mu^2)\;x_2f_{f_2}(x_2,\mu^2)\;,
  \end{split}
\end{equation}
where $\abs{\mc{M}_B}^2(\args{a})$ denotes the partonic matrix element 
squared, $\done{\it\Phi}_B(\args{p})$ is the corresponding differential 
$n$-particle partonic phase-space element, $S(\args{f\,})$ is the symmetry 
factor, $F(\args{p})$ is the flux factor, and $\mc{L}$ is the parton 
luminosity~\footnote{ In the case of leptonic initial states, and ignoring QED 
initial-state radiation, the parton distribution functions $f(x,\mu^2)$ are 
replaced by $\delta(1-x)$.}.

In a similar fashion, the real-emission part of the QCD next-to-leading order
cross section can be written as a sum, depending on parton configurations
$\{\mathsc{a}_1,\ldots,\mathsc{a}_{n+1}\}$, by replacing the Born-level matrix elements $\mc{B}$ 
with the real-emission matrix elements $\mc{R}$ and the Born-level phase space
$\done\Phi_B$ with the real-emission phase-space $\done\Phi_R$.

It is then useful to introduce a notation for mappings from real-emission 
parton configurations to Born-level parton configurations and vice versa.
They are given by (cf.~\cite{Hoeche:2010pf})
\begin{align}\label{eq:parton_maps}
  \bmap{ij}{k}{\rargs{a}}\,=&\;\left\{\begin{array}{c}
    \rargs{f\,}\setminus\{\mathsc{f}_i,\mathsc{f}_j\}\cup\{f_{\widetilde{\im\jm}}\}\\
    \rargs{p\,}\to\args{p\;}
    \end{array}\right.
  &&\text{and}
  &\rmap{\im\jm}{k}{\mathsc{f}_i,\Phi_{R|B}^{ij,k}\,;\args{a}}\,=&\;
  \left\{\begin{array}{c}
    \args{f\,}\setminus\{f_{\widetilde{\im\jm}}\}\cup\{\mathsc{f}_i,\mathsc{f}_j\}\\
    \args{p\;}\to\args{\mathsc{p}\,}
    \end{array}\right.\;.
\end{align}
The map $\bmap{ij}{k}{\rargs{a}}$ combines the partons $\mathsc{a}_i$ and $\mathsc{a}_j$ into a 
common ``mother'' parton $a_{\widetilde{\im\jm}}$, in the presence of the 
spectator $\mathsc{a}_k$ by defining a new flavour $f_{\widetilde{\im\jm}}$ and by 
redefining the particle momenta. The inverse map, $\rmap{\im\jm}{k}{\args{a}}$
determines the parton configuration of a real-emission subprocess from
a Born parton configuration and a related branching process 
$\widetilde{\im\jm},\tilde{k}\to ij,k$. The radiative variables $\Phi_{R|B}^{ij,k}$,
denoted $\Phi_1$ in Sec.~\ref{sec:catchy} for brevity,
are thereby employed to turn the $n$-parton momentum configuration into an 
$(n+1)$-parton momentum configuration.

The real-emission matrix elements, $\mc{R}(\rargs{a})$, can be approximated
in the soft and collinear limits by subtraction terms $\mc{D}^{\rm(S)}_{ij,k}(\rargs{a})$, 
which capture the universal singularity structure when two partons $i$ and $j$ become 
collinear, or one of them becomes soft in the presence of a spectator parton $k$.
\begin{equation}\label{eq:def_subterms}
  \mc{R}\,\overset{\substack{ij\;\text{collinear}\\ i,j\;\text{soft}}}{\longrightarrow}\mc{D}^{\rm(S)}_{ij,k}(\rargs{a})\;.
\end{equation}
These subtraction terms are related to the ones defined in Sec.~\ref{sec:catchy}
through $\mr{D}^{\rm(S)}\to\mc{L}(\rargs{a})\,\mc{D}^{\rm(S)}_{ij,k}(\rargs{a})$, 
with implicit notation of dipole indices, phase space and flavour configuration 
on the left hand side. They are not uniquely defined, but can be constructed, 
for example, using the Catani-Seymour method~\cite{Catani:1996vz,*Catani:2002hc}
or the FKS approach~\cite{Frixione:1995ms,*Frixione:1997np}.
Furthermore they can be restricted in phase space~\cite{Nagy:2003tz} or 
extended with an arbitrary finite function.
Similar terms can also be computed in a parton-shower approximation as
\begin{equation}\label{eq:factorisation_ps}
  \mc{R}\,\overset{ij\;\text{collinear}}{\longrightarrow}
    \mc{D}^{\rm(PS)}_{ij,k}(\rargs{a})\,=\;\mc{B}(\bmap{ij}{k}{\rargs{a}})\,
    \frac{S(\bmap{ij}{k}{\rargs{f\,}})}{S(\rargs{f\,})}\,
    \frac{1}{2\,p_ip_j}\,8\pi\,\alpha_s\,\mc{K}_{ij,k}(p_i,p_j,p_k)\;,
\end{equation}
where $\mc{K}_{ij,k}$ denote the parton-shower evolution kernels.
The parton-shower approximation is meaningful only in the collinear region, 
as it implements an exact factorisation of the colour structure into a Born part 
and an emission part. In many cases also a factorisation of the helicity structure
is assumed.

\subsection{From fixed-order to resummation}\label{sec:mcatnlo:construction}
Section~\ref{sec:catchy} only introduced expressions for the \emph{total cross section}
in the \MCatNLO and \POWHEG methods. A detailed discussion requires an analysis of 
the expectation value of arbitrary infra-red safe observables. 
The respective formulae are developed in the following.

Using the subtraction terms introduced in Eq.~\eqref{eq:def_subterms}, the expectation value 
of an observable $O$ is determined to next-to-leading order accuracy as
\begin{equation}\label{eq:master_nlo}
  \begin{split}
  \abr{O}^{\rm (NLO)}\,=&\;\sum_{\args{f\,}}\int\done\Phi_B(\args{p})\,
  \sbr{\,\mr{B}(\args{a})+\tilde{\mr{V}}(\args{a})+
    \sum_{\{\widetilde{\im\jm},\tilde{k}\}}\mr{I}_{\widetilde{\im\jm},\tilde{k}}^{\rm(S)}(\args{a})\,}\,O(\args{p})\\
  &\qquad+\sum_{\rargs{f\,}}\int\done\Phi_R(\rargs{p})\,
  \sbr{\,\mr{R}(\rargs{a})\,O(\rargs{p})-\sum_{\{ij,k\}}\mr{D}_{ij,k}^{\rm(S)}(\rargs{a})\,O(\bmap{ij}{k}{\rargs{p}})\,}\;.
  \end{split}
\end{equation}
where $\mr{I}^{\rm(S)}(\args{a})$ represent the subtraction terms $\mr{D}^{\rm(S)}(\rargs{a})$ 
integrated over the extra-emission phase space. Note that the configurations $\rargs{f\,}$, $\rargs{p}$ 
and $\rargs{a}$ on the second line each include one more particle.

This equation can be modified by adding and subtracting an additional 
arbitrary set of subtraction terms $\mr{D}^{\rm(A)}(\rargs{a})$
with the same kinematics mapping as $\mr{D}^{\rm(S)}(\rargs{a})$
\begin{equation}\label{eq:master_nlo_addsub}
  \begin{split}
  \abr{O}^{\rm (NLO)}\,=&\;\sum_{\args{f\,}}\int\done\Phi_B(\args{p})\,
  \bar{\mr{B}}^{\rm(A)}(\args{a})\,O(\args{p})\\
  &\qquad+\sum_{\rargs{f\,}}\int\done\Phi_R(\rargs{p})\,
  \sbr{\,\mr{R}(\rargs{a})\,O(\rargs{p})-\sum_{\{ij,k\}}\mr{D}^{\rm(A)}_{ij,k}(\rargs{a})\,O(\bmap{ij}{k}{\rargs{p}})\,}\;,
  \end{split}
\end{equation}
where the function $\bar{\mr{B}}^{\rm(A)}(\args{a})$ is defined as
\begin{equation}\label{eq:def_bbar_a}
  \begin{split}
    \bar{\mr{B}}^{\rm(A)}(\args{a})\;=&\,\mr{B}(\args{a})+\tilde{\mr{V}}(\args{a})+
      \sum_{\{\widetilde{\im\jm},\tilde{k}\}}\mr{I}_{\widetilde{\im\jm},\tilde{k}}^{\rm(S)}(\args{a})\\
    &\qquad+\sum_{\{\widetilde{\im\jm},\tilde{k}\}}
      \sum_{f_i=q,g}\int\done\Phi_{R|B}^{ij,k}\;
      \sbr{\,\mr{D}^{\rm(A)}_{ij,k}(\rmap{\im\jm}{k}{\args{a}})
        -\mr{D}^{\rm(S)}_{ij,k}(\rmap{\im\jm}{k}{\args{a}})\,}\;,
  \end{split}
\end{equation}
and where $\done\Phi_{R|B}^{ij,k}$ represents an integral over the radiative phase space. 

When combining fixed-order calculations with resummation, the task is to define a unique 
starting condition for the parton shower. As was argued in~\cite{Frixione:2002ik}, it is
not possible to process the terms $\mr{R}(\rargs{a})$ and $\mr{D}^{\rm(A)}_{ij,k}(\rargs{a})$
in Eq.~\eqref{eq:master_nlo_addsub} separately because this would lead to double-counting.
Instead, the problem is solved if the observables on the second line of 
Eq.~\eqref{eq:master_nlo_addsub} are both assumed to depend on the momentum configuration $\rargs{p}$. 
This leads to
\begin{equation}\label{eq:master_nlo_addsub_hmod}
  \begin{split}
  \abr{O}^{\rm (NLO)}\,=&\;\sum_{\args{f\,}}\int\done\Phi_B(\args{p})\,
  \bar{\mr{B}}^{\rm(A)}(\args{p})\,O(\args{p})\\
  &\qquad+\sum_{\rargs{f\,}}\int\done\Phi_R(\rargs{p})\,
  \sbr{\,\mr{R}(\rargs{a})-\sum_{\{ij,k\}}\mr{D}^{\rm(A)}_{ij,k}(\rargs{a})\,}\,O(\rargs{p})
  +\abr{O}^{\rm(corr)}\;,
  \end{split}
\end{equation}
introducing the correction term
\footnote{ The dependence of the observable $O$ on the kinematics of the partonic final state
  makes the need for a correction term $\abr{O}^{\rm(corr)}$ manifest. This term integrates 
  to zero in Eq.~\eqref{eq:total_xsec_split_trivial}, as $O$=1 and the phase-space dependence
  is trivial.}
\begin{equation}\label{eq:nlo_correction}
  \abr{O}^{\rm(corr)}\,=\;\sum_{\rargs{f\,}}\int\done\Phi_R(\rargs{p})\,
    \sum_{\{ij,k\}}\mr{D}^{\rm(A)}_{ij,k}(\rargs{a})\,
    \sbr{\,\vphantom{\int}O(\rargs{p})-O(\bmap{ij}{k}{\rargs{p}})\,}\;.
\end{equation}

The essence of both the \MCatNLO and \POWHEG methods is to generate Eq.~\eqref{eq:nlo_correction}
by using a Sudakov branching algorithm, which is formulated in terms of an evolution variable 
$t$, where $t\varpropto2\,p_ip_j$. Let a corresponding form factor be defined as
\begin{equation}\label{eq:nbp_ps}
  \bar{\Delta}^{\rm(A)}(t;\args{a})\,=\;\prod_{\{\widetilde{\im\jm},\tilde{k}\}}
  \bar{\Delta}^{\rm(A)}_{\widetilde{\im\jm},\tilde{k}}(t;\args{a})\;,
\end{equation}
where
\begin{equation}\label{eq:nbp_ijk}
  \begin{split}
  \bar{\Delta}^{\rm(A)}_{\widetilde{\im\jm},\tilde{k}}(t;\args{a})\,=&\;
  \exp\left\{-\sum_{\mathsc{f}_i=q,g}
    \int\done\Phi_{R|B}^{ij,k}\;\Theta(t(\Phi_{R|B}^{ij,k})-t)\,
  \right.\\&\qquad\qquad\qquad\times\left.
    \frac{1}{S_{ij}}\,
    \frac{S(\rmap{\im\jm}{k}{\mathsc{f}_i;\args{f\,}})}{S(\args{f\,})}\,
    \frac{\mr{D}^{\rm(A)}_{ij,k}(\rmap{\im\jm}{k}{\mathsc{f}_i,\Phi_{R|B}^{ij,k};\args{a}})}{
      \mr{B}(\args{a})}\,\right\}\;.
  \end{split}
\end{equation}
The ratio of symmetry factors, including $S_{ij}$, is explained in detail in~\cite{Hoeche:2010pf}. 
It accounts for the way in which the phase space is successively filled by a parton shower.

It can then be proven~\cite{Frixione:2002ik,Nason:2004rx,*Frixione:2007vw,Hoeche:2010pf},
that the following formula for the expectation value of an infrared safe observable
reproduces Eq.~\eqref{eq:master_nlo_addsub_hmod} to $\order(\alpha_s)$:
\begin{align}\label{eq:master_nlomc}
  \abr{O}^{\rm(NLOMC)}\,=&\;\sum_{\args{f\,}}\int\done\Phi_B(\args{p})\,
  \bar{\mr{B}}^{\rm(A)}(\args{a})\left[\vphantom{\Bigg)_{\int}^{\int}}\right.\,
    \underbrace{\bar{\Delta}^{\rm(A)}(t_0;\args{a})}_{\text{unresolved}}\,
    O(\args{p})\nnb\\
  &\qquad+\,
    \sum_{\{\widetilde{\im\jm},\tilde{k}\}}
    \sum_{\mathsc{f}_i=q,g}\int\done\Phi_{R|B}^{ij,k}\;\Theta(t(\Phi_{R|B}^{ij,k})-t_0)\,\;
    O(\rmap{\im\jm}{k}{\args{p}})\nnb\\
  &\qquad\qquad\times\,
    \underbrace{\frac{1}{S_{ij}}\,
    \frac{S(\rmap{\im\jm}{k}{\mathsc{f}_i;\args{f\,}})}{S(\args{f\,})}\,
    \frac{\mr{D}_{ij,k}^{\rm(A)}(\rmap{\im\jm}{k}{\mathsc{f}_i,\Phi_{R|B}^{ij,k};\args{a}})}{
      \mr{B}(\args{a})}\;
    \bar{\Delta}^{\rm(A)}(t;\args{a})}_{\text{resolved, singular}}\,
    \left.\vphantom{\Bigg)_{\int}^{\int}}\right]\nnb\displaybreak[3]\\
  &+\;\sum_{\rargs{f\,}}\int\done\Phi_R(\rargs{p})\,
    \underbrace{\sbr{\,\mr{R}(\rargs{a})-\sum_{ij,k}\mr{D}^{\rm(A)}_{ij,k}(\rargs{a})\,}}_{\text{resolved, non-singular}}\,
    O(\rargs{p})\;.
\end{align}
NLO Monte-Carlo events reproducing Eq.~\eqref{eq:master_nlomc} can be generated
in the following way:
\begin{itemize}
\item  A seed event is produced according to either the first or the second
  line of Eq.~\eqref{eq:master_nlo_addsub_hmod},\\
  not including the correction factor $\abr{O}^{\rm(corr)}$.
\item If the second line is chosen ($\mb{H}$-event) the event has real-emission kinematics and is kept as-is.\\
  This generates the ``resolved, non-singular'' term of Eq.~\eqref{eq:master_nlomc}
\item If the first line is chosen ($\mb{S}$-event), 
the event has Born-like kinematics and is processed through a one-step Sudakov branching algorithm described 
by Eqs.~\eqref{eq:nbp_ps} and~\eqref{eq:nbp_ijk}.
The necessary techniques for that case will be detailed in Sec.~\ref{sec:techniques_mcatnlo}.
An emission might or might not occur, which is represented
by the ``resolved, singular'' and ``unresolved'' terms of Eq.~\eqref{eq:master_nlomc}, respectively.\end{itemize}

\subsection{The \texorpdfstring{\MCatNLO}{MC@NLO} and \texorpdfstring{\POWHEG}{POWHEG} methods}
\label{sec:construct_mcatnlo}
Equation~\eqref{eq:master_nlomc} becomes particularly simple when
\begin{equation}\label{eq:def_mcatnlo}
  \mr{D}^{\rm(A)}_{ij,k}\to\mr{D}^{\rm(S)}_{ij,k}\;.
\end{equation}
In this special case the integral over the radiative phase space in 
Eq.~\eqref{eq:def_bbar_a} is zero. \changed{We will refer to this as the \MCatNLO 
technique.} It can be thought of as a method which uses the Sudakov branching 
probability for subtraction \changed{or, conversely, which uses the subtraction 
kernels for parton evolution.
Possible event generation techniques to achieve this will be discussed in Sec.~\ref{sec:techniques_mcatnlo}.}

The \POWHEG method, in its original form, on the other hand, defines
\begin{equation}\label{eq:def_powheg}
  \mr{D}^{\rm(A)}_{ij,k}\to\rho_{ij,k}(\rargs{a})\,\mr{R}(\rargs{a})
  \qquad\text{where}\qquad
  \rho_{ij,k}(\rargs{a})=\frac{\mc{D}^{\rm(S)}_{ij,k}(\rargs{a})}{\sum_{mn,l}\mc{D}^{\rm(S)}_{mn,l}(\rargs{a})}\;,
\end{equation}
\changed{It} can be thought of as exponentiating the full radiative corrections 
into a Sudakov form factor. As such, it bears strong similarity to
matrix-element corrected \changed{power} showers. 
This aspect has been discussed extensively in~\cite{Hoeche:2010pf,Hoeche:2010kg}.

One can construct a mixed scheme, where $\mr{D}_{ij,k}^{\rm(A)}$ is defined as
\begin{equation}\label{eq:def_powheg_zhsplit}
  \mr{D}^{\rm(A)}_{ij,k}\to\rho_{ij,k}(\rargs{a})\,\sbr{\,\mr{R}(\rargs{a})-\mr{R}^{(r)}(\rargs{a})\,}\;,
\end{equation}
with $\mr{R}^{(r)}$ an arbitrary infrared-finite part of the real-emission cross section
and $\rho_{ij,k}$ given by Eq.~\eqref{eq:def_powheg}. This method \changed{leads to the 
original \MCatNLO prescription, proposed in~\cite{Frixione:2002ik}, if 
$\mr{D}^{\rm(A)}_{ij,k}=\mr{D}^{\rm(PS)}_{ij,k}$. It was also used in~\cite{Alioli:2008gx}
to deal with the problem of radiation zeros in the $W^\pm$-production process in the 
\POWHEG approach.}

Note that all the above choices of $\mr{D}_{ij,k}^{\rm(A)}$
fulfil the requirement that the kinematics and flavour mapping
be identical to the one in $\mr{D}_{ij,k}^{\rm(S)}$, 
cf.\ Eq.~\eqref{eq:master_nlo_addsub}.

\subsection{Event-generation techniques}
\label{sec:techniques_mcatnlo}
In the parton-shower approximation, Eq.~\eqref{eq:nbp_ijk} would read
\begin{equation}\label{eq:nbp_ps_ijk}
  \begin{split}
  \Delta_{\widetilde{\im\jm},\tilde{k}}(t',t'';\args{a})\,=&\;
  \exp\left\{-\sum_{\mathsc{f}_i=q,g}
    \int_{t'}^{t''}\frac{\done t}{t}\,\int_{z_{\rm min}}^{z_{\rm max}}\done z\,
    \int_0^{2\pi}\frac{\done\phi}{2\pi}\,J_{ij,k}(t,z,\phi)\,
  \right.\\&\qquad\qquad\qquad\times\left.
    \frac{1}{S_{ij}}\,\frac{\alpha_s}{2\pi}\,
    \mc{K}_{ij,k}(t,z,\phi)\,
    \frac{\mc{L}(\rmap{\im\jm}{k}{\mathsc{f}_i,t,z,\phi;\args{a}};t)}
         {\mc{L}(\args{a\,};t)}\,\right\}\;,
  \end{split}
\end{equation}
where $z$ is called the splitting variable of the parton-shower model and
$J_{ij,k}$ is the Jacobian factor associated with the transformation of 
$\done\Phi_{R|B}^{ij,k}$ into $\done t\,\done z\,\done\phi$~\cite{Hoeche:2010pf}.

It is well known how to generate emissions according to Eq.~\eqref{eq:nbp_ps_ijk}.
The task of a proper implementation of \MCatNLO and \POWHEG is, however, to employ 
Eq.~\eqref{eq:nbp_ijk}. \changed{In our formulation of \MCatNLO, i.e.\ with 
$\mr{D}^{\rm(A)}_{ij,k}=\mr{D}^{\rm(S)}_{ij,k}$, this can involve the exponentiation
of subtraction terms which are negative, due to subleading colour configurations in the 
real-emission matrix elements, cf.\ Sec.~\ref{sec:subtleties}.
This leads to a form factor larger than one, which cannot be interpreted in terms 
of a no-branching probability. In this section we will show how to generate such
form factors efficiently in a Markov chain Monte Carlo. The respective algorithm 
is the basis for our implementation of the \MCatNLO method. It was outlined 
in~\cite{Hoeche:2009xc} and reformulated in~\cite{Platzer:2011dq}.} 
As it needs to be modified slightly in the current context, it is briefly discussed here again.
The method is based on an extension of the veto algorithm~\cite{Sjostrand:2006za}.

\subsubsection*{The veto algorithm}
Let $t$ be the parton-shower evolution variable and $f(t)$ the splitting kernel
$\mc{K}$, integrated over the splitting variable $z$.\footnote{ For simplicity,
  the existence of only one splitting function is assumed, i.e.\ that there is no flavour change 
  of the splitter during the evolution. The extension to flavour changing splittings 
  is straightforward, but it would unnecessarily complicate the notation at this point.}
The differential probability for generating a branching at scale $t$, when starting
from an upper evolution scale $t'$ is then given by
\begin{equation}\label{eq:va_prob}
  \mc{P}(t,t')\,=\;f(t)\,\exp\cbr{-\int_t^{t'}\done\bar{t}\,f(\bar{t})}\;.
\end{equation}
A new scale $t$ is therefore found as
\begin{align}
  t\,=&\;F^{-1}\sbr{\,F(t')+\log \#\,}
  &&\text{where}
  &F(t)\,=\,\int_t\done \bar t\,f(\bar t)\;,
\end{align}
and where $\#$ is a random number between zero and one.
The key point of the veto algorithm is, that even if the integral $F(t)$ is unknown, one
can still generate events according to $\mc{P}$ using an overestimate $g(t)\ge f(t)$
with a known integral $G(t)$. Firstly, a value $t$ is generated as $t=G^{-1}\sbr{\,G(t')+\log \#\,}$.
Secondly, the value is accepted with probability $f(t)/g(t)$. 
A splitting at $t$ with $n$ intermediate rejections is then produced with probability
\begin{equation}\label{eq:va_mcprob}
  \begin{split}
  \mc{P}_n(t,t')\,=&\;\frac{f(t)}{g(t)}\,g(t)\,\exp\cbr{-\int_t^{t_1}\done\bar{t}\,g(\bar{t})}\\
    &\qquad\times\prod_{i=1}^n\sbr{\,\int_{t_{i-1}}^{t_{n+1}}\done t_i\rbr{1-\frac{f(t_i)}{g(t_i)}}g(t_i)\,
      \exp\cbr{-\int_{t_i}^{t_{i+1}}\done\bar{t}\,g(\bar{t})}}\;,
  \end{split}
\end{equation}
where $t_{n+1}=t'$ and $t_0=t$. The nested integrals in Eq.~\eqref{eq:va_mcprob} can be disentangled, and summing 
over $n$ leads to the exponentiation of the factor $g(t)-f(t)$, such that Eq.~\eqref{eq:va_prob}
is reproduced~\cite{Sjostrand:2006za}.

\subsubsection*{Analytic weights}
The purpose here is to introduce an additional overestimate $h(t)$.
The additional weight $g(t)/h(t)$ is then applied analytically rather than using a hit-or-miss method.
This leads to the following expression for the differential
probability to generate an emission at $t$ with $n$ rejections between $t$ and $t'$
\begin{equation}\label{eq:wva_mcprob}
  \begin{split}
  \mc{P}_n(t,t')\,=&\;\frac{f(t)}{g(t)}\,h(t)\,\exp\cbr{-\int_t^{t_1}\done\bar{t}\,h(\bar{t})}\\
    &\qquad\times\prod_{i=1}^n\sbr{\,\int_{t_{i-1}}^{t_{n+1}}\done t_i\rbr{1-\frac{f(t_i)}{g(t_i)}}h(t_i)\,
      \exp\cbr{-\int_{t_i}^{t_{i+1}}\done\bar{t}\,h(\bar{t})}}\;.
  \end{split}
\end{equation} 
Note that the function $h(t)$ and the ratio $f(t)/g(t)$ must be positive definite.
Generating events according to Eq.~\eqref{eq:wva_mcprob} is then straightforward.
In order to recover the desired distribution, the following analytic weight needs to be applied
\begin{equation}\label{eq:wva_analytic}
  w(t,t_1,\ldots,t_n)\,=\;\frac{g(t)}{h(t)}\,\prod_{i=1}^n\frac{g(t_i)}{h(t_i)}\frac{h(t_i)-f(t_i)}{g(t_i)-f(t_i)}\;.
\end{equation}
Here the term $g(t)/h(t)$ is due to the acceptance of the emission.
The product, which is needed for an exponentiation of $h(t)-f(t)$ 
instead of $g(t)-f(t)$, runs over all correction weights for rejected steps.
Equation~\eqref{eq:wva_analytic} can lead to negative weights, which reflect the fact
that sub-leading colour configurations are taken into account and that the a-priori
density $h(t)$ might underestimate $f(t)$.

\subsubsection*{Implementation of the \texorpdfstring{\MCatNLO}{MC@NLO} method}
\changed{
In order to implement the \MCatNLO method with $\mr{D}^{\rm(A)}_{ij,k}=\mr{D}^{\rm(S)}_{ij,k}$
we need to identify the function $f$ above with the $(z,\phi)$-integrated subtraction terms 
$\mr{D}^{\rm(S)}_{ij,k}$, divided by the Born contribution. A convenient choice of 
the function $h$ will be the $(z,\phi)$-integral of the parton-shower evolution kernels
$\mr{D}^{\rm(PS)}_{ij,k}$, divided by the Born contribution. We are then free to choose
the auxiliary function $g$ on a point-by-point basis, but a convenient way is to define
$g=C\,f$, where $C$ is a constant larger than one. This guarantees that both acceptance 
and rejection term are generated in sufficient abundance to reduce Monte-Carlo errors.
}
In the remainder of this paper, the functions $f$, $g$ and $h$ are thus schematically 
identified as follows:
\begin{equation}\label{eq:def_mcatnlo_mc_weight}
  f\to \mr{D}^{\rm(A)}\;,\\
  h\to \mr{D}^{\rm(PS)}\;,\\\text{and}\\
  g\,=\;C\,f\;,\\
\end{equation}
where $C\approx 2$ is a constant.

\changed{By exponentiating $\mr{D}^{\rm(S)}_{ij,k}$ in this manner, we can suppress
the integral over the radiative phase space in Eq.~\eqref{eq:def_bbar_a}
and thus substantially simplify the event generation.}

%% file: text/results.tex
\section{Analysis of perturbative uncertainties}
\label{sec:perturbative}

In the following the uncertainties discussed in Sec.~\ref{sec:subtleties} are investigated
in detail. The event generator \Sherpa~\cite{Gleisberg:2003xi,*Gleisberg:2008ta}
sets the framework for this study, including its automated \MCatNLO 
implementation, the matrix-element generator \Amegic~\cite{Krauss:2001iv}, 
an automated implementation~\cite{Gleisberg:2007md} of the Catani-Seymour 
dipole subtraction method~\cite{Catani:1996vz} and the parton shower model 
described in~\cite{Schumann:2007mg,Hoeche:2009xc}. The CTEQ6.6 PDF 
set~\cite{Nadolsky:2008zw} is used together with the corresponding parametrisation of the running coupling.
All analyses are carried out with the help of Rivet~\cite{Buckley:2010ar}.

\subsection{Higgs-boson production in gluon fusion}
\label{sec:ggh}
The production of a Higgs boson at the LHC with 7~TeV centre-of-mass energy 
serves as a first example. This process amplifies some of the effects discussed in Sec.~\ref{sec:subtleties}
because the radiating partons are gluons. There are thus no valence PDF's involved at Born level, which
allows to test fixed-order effects and resummation in a relatively clean setting.
Results presented in this section do not include non-perturbative or multiple parton scattering effects. 
A detailed study of the \POWHEG algorithm and its associated uncertainties has already
been presented in~\cite{Hoeche:2010pf}. In the following, the focus will therefore mainly lie
on \changed{our implementation of} the \MCatNLO method, but comparisons with \POWHEG will be made at the appropriate junctures.

\begin{figure}[t]
  \centering
  \subfloat[][]{
    \begin{minipage}{0.49\textwidth}
      \begin{picture}(0,210)
	\put(0.0,0.0){\includegraphics[width=\textwidth]{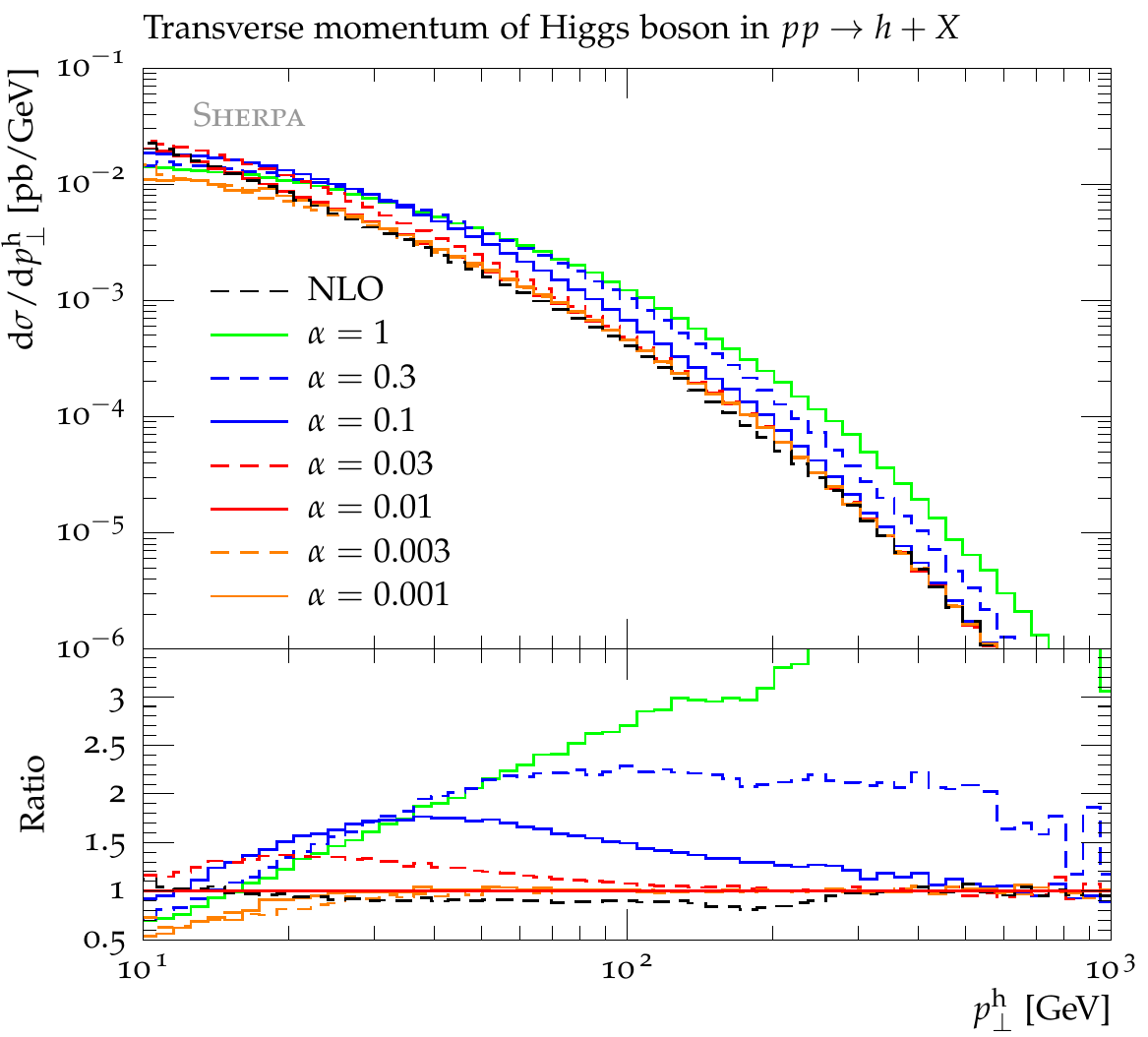}}
	\put(128,127){\includegraphics[width=0.4\textwidth]{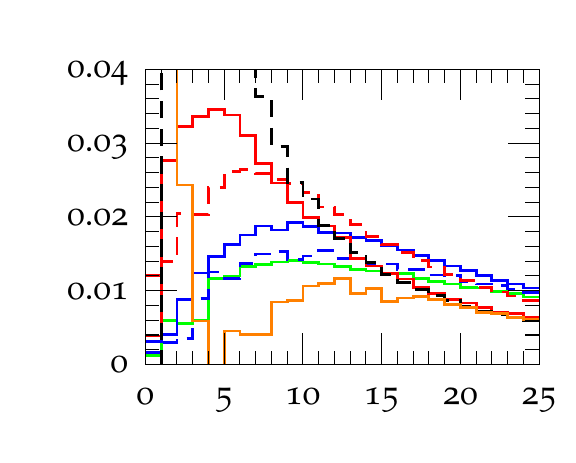}}
      \end{picture}
    \end{minipage}
    \label{fig:ggh:hpt_a}}
  \subfloat[][]{
    \begin{minipage}{0.49\textwidth}
      \begin{picture}(0,210)
	\put(0.0,0.0){\includegraphics[width=\textwidth]{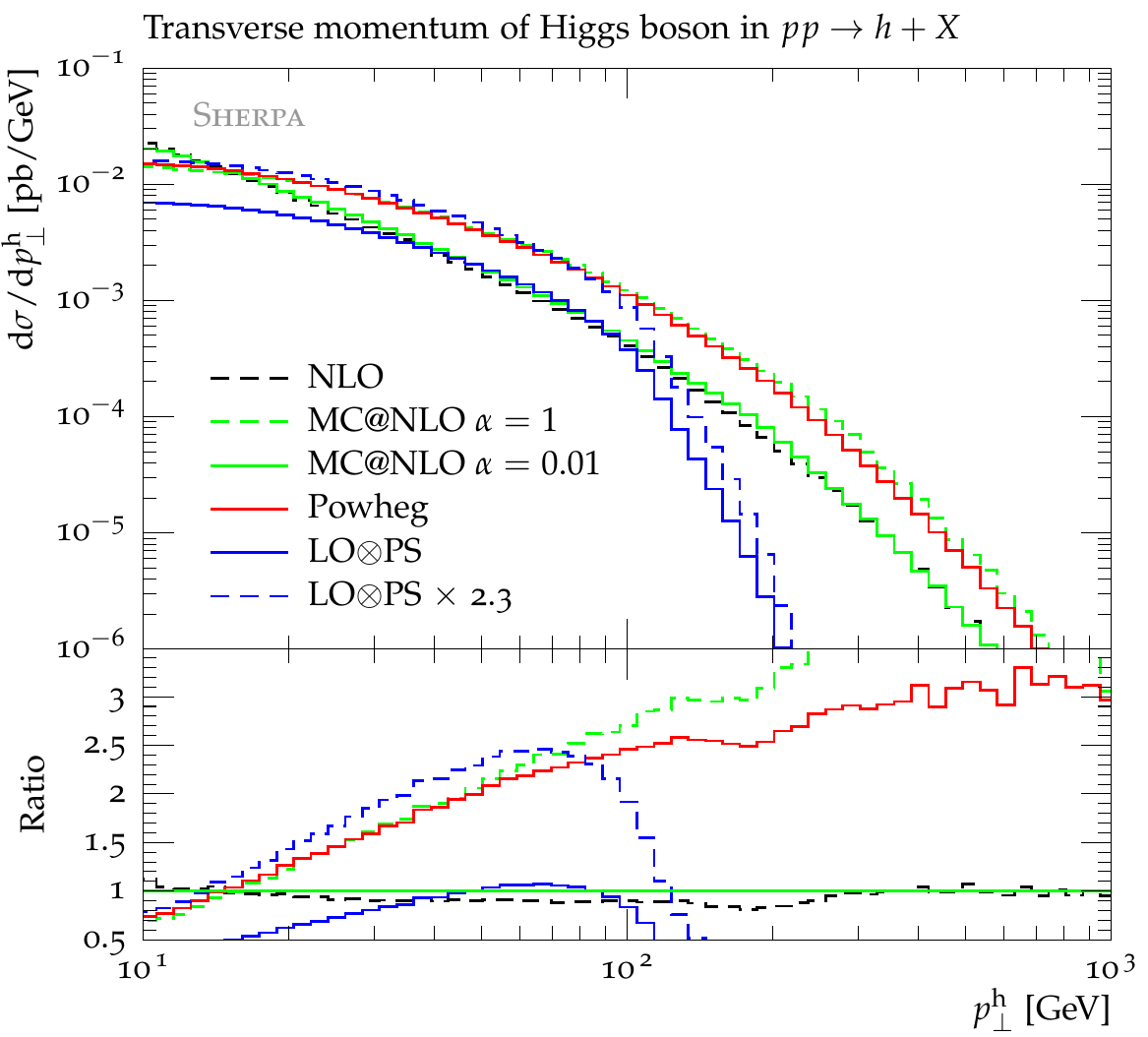}}
	\put(128,127){\includegraphics[width=0.4\textwidth]{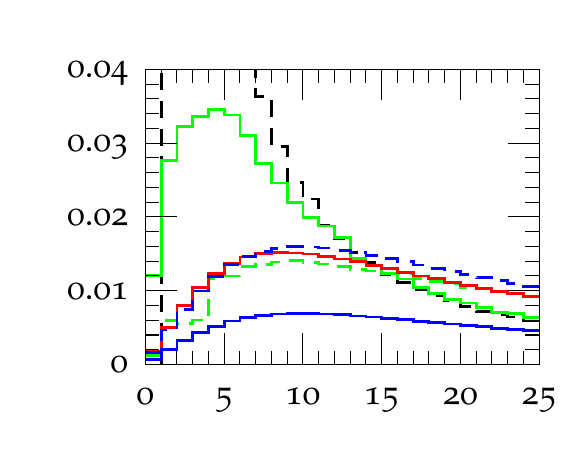}}
      \end{picture}
    \end{minipage}
    \label{fig:ggh:hpt_b}}
  \caption{Transverse momentum of the Higgs boson in inclusive Higgs boson production ($m_h=120$ GeV)
           at $E_{\rm cms}=7$~TeV. The variation of MC@NLO
           predictions with varying $\alpha_{\rm cut}$ (denoted $\alpha$ in the legend)
           is shown in Fig.~\protect\subref{fig:ggh:hpt_a},
           while Fig.~\protect\subref{fig:ggh:hpt_b} compares the \MCatNLO, \POWHEG and \LOPS methods.}
  \label{fig:ggh:hpt}
\end{figure}

\begin{figure}[t]
  \begin{center}\includegraphics[width=0.48\textwidth]{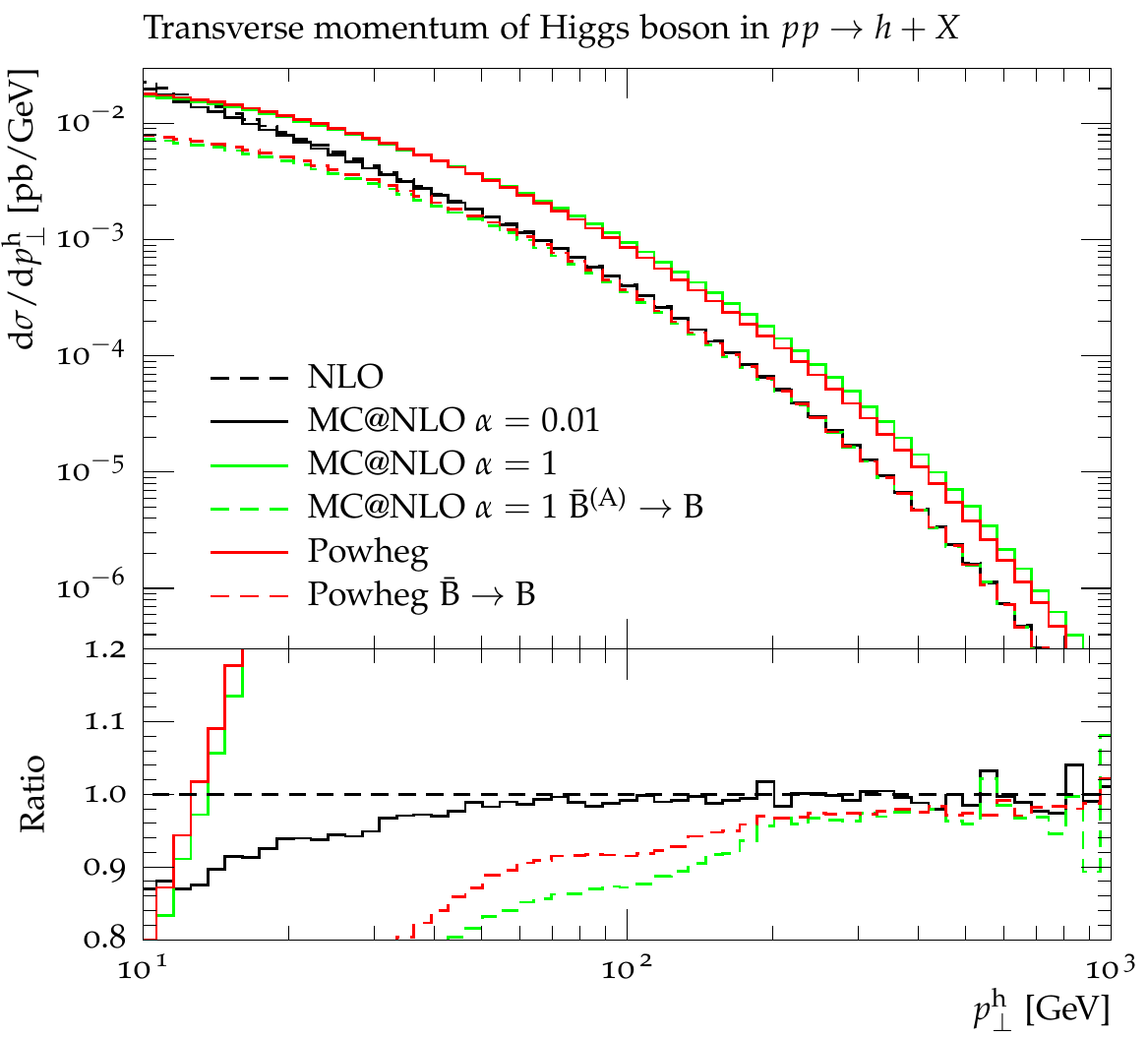}\end{center}\vspace*{-3mm}
  \caption{Transverse momentum of the Higgs boson (left) 
	   in inclusive Higgs boson production ($m_h=120$ GeV) at 
           $E_{\rm cms}=7$~TeV. The solid lines show the full \MCatNLO (green) 
           and \POWHEG (red) results, while the corresponding dashed lines show 
           the results where the prefactor of the resummed $\mathbb{S}$ events, 
           $\bar{\mr{B}}$ and $\bar{\mr{B}}^\mr{(A)}$ respectively, have been 
	   replaced by $\mr{B}$. In all cases only the first emission is taken 
	   into account. The black dashed line shows the
	   NLO fixed-order prediction. 
	  }
  \label{fig:ggh:bbbar:hpt_j1pt}
\end{figure}

\begin{figure}[t]
  \centering
  \subfloat[][]{
    \includegraphics[width=0.49\textwidth]{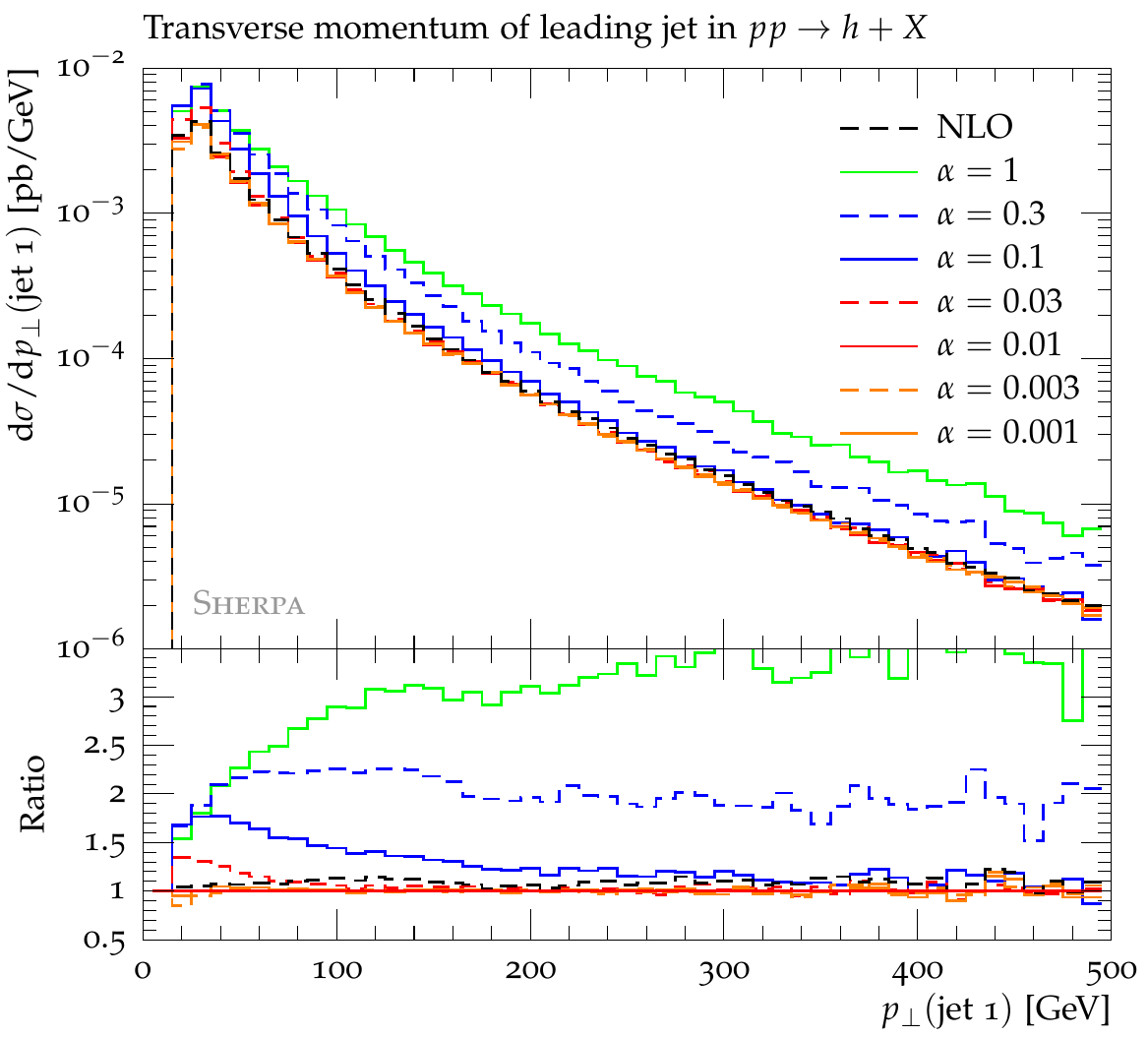}
    \label{fig:ggh:j1pt_a}}
  \subfloat[][]{
    \includegraphics[width=0.49\textwidth]{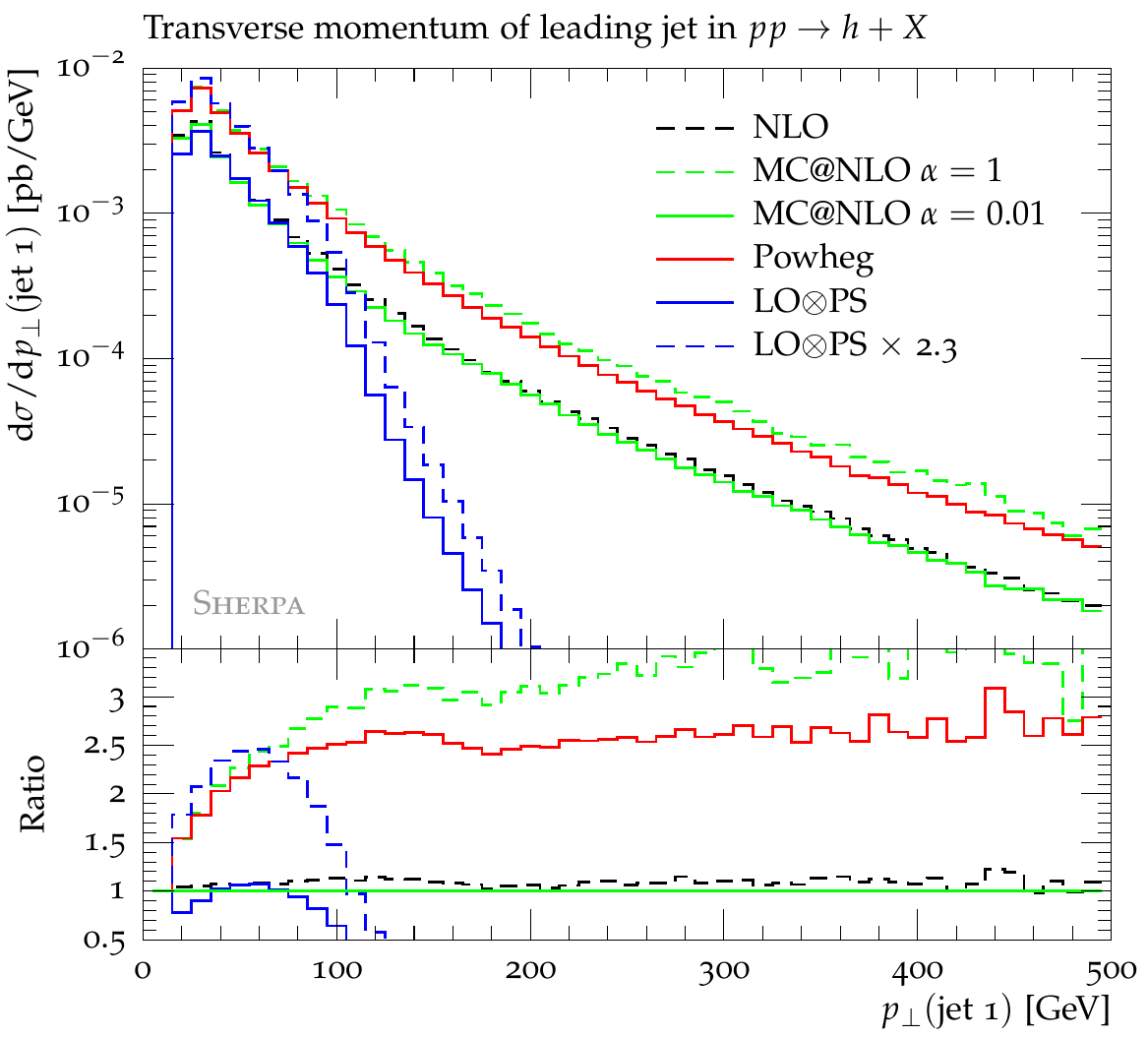}
    \label{fig:ggh:j1pt_b}}
  \caption{Transverse momentum of the leading jet in inclusive Higgs-boson production ($m_h=120$ GeV) 
           at $E_{\rm cms}=7$~TeV. See Fig.~\ref{fig:ggh:hpt} for details.}
  \label{fig:ggh:j1pt}
\end{figure}

\begin{figure}[t!]
  \centering
  \subfloat[][]{
    \includegraphics[width=0.49\textwidth]{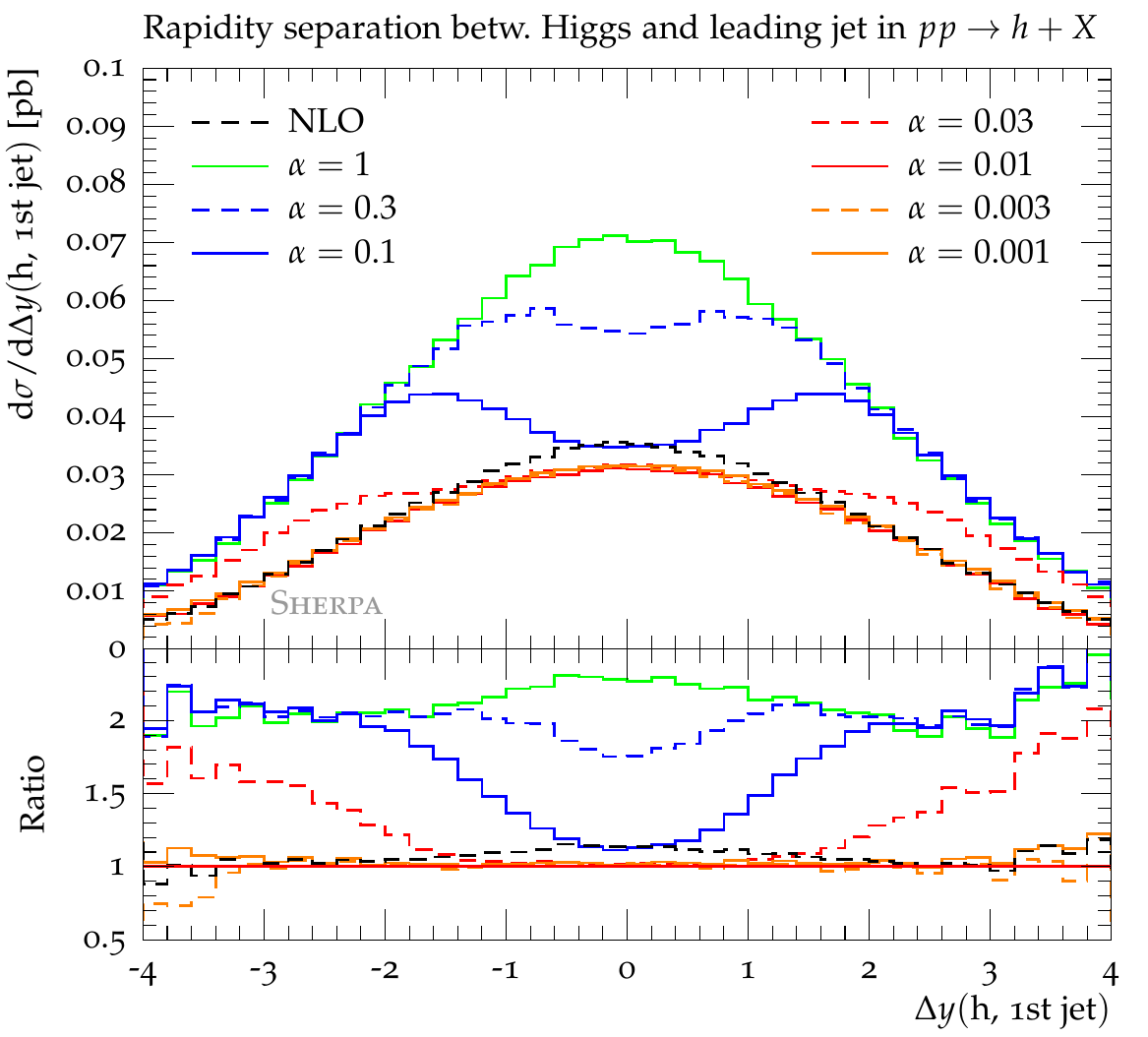}
    \label{fig:ggh:hjdy_a}}
  \subfloat[][]{
    \includegraphics[width=0.49\textwidth]{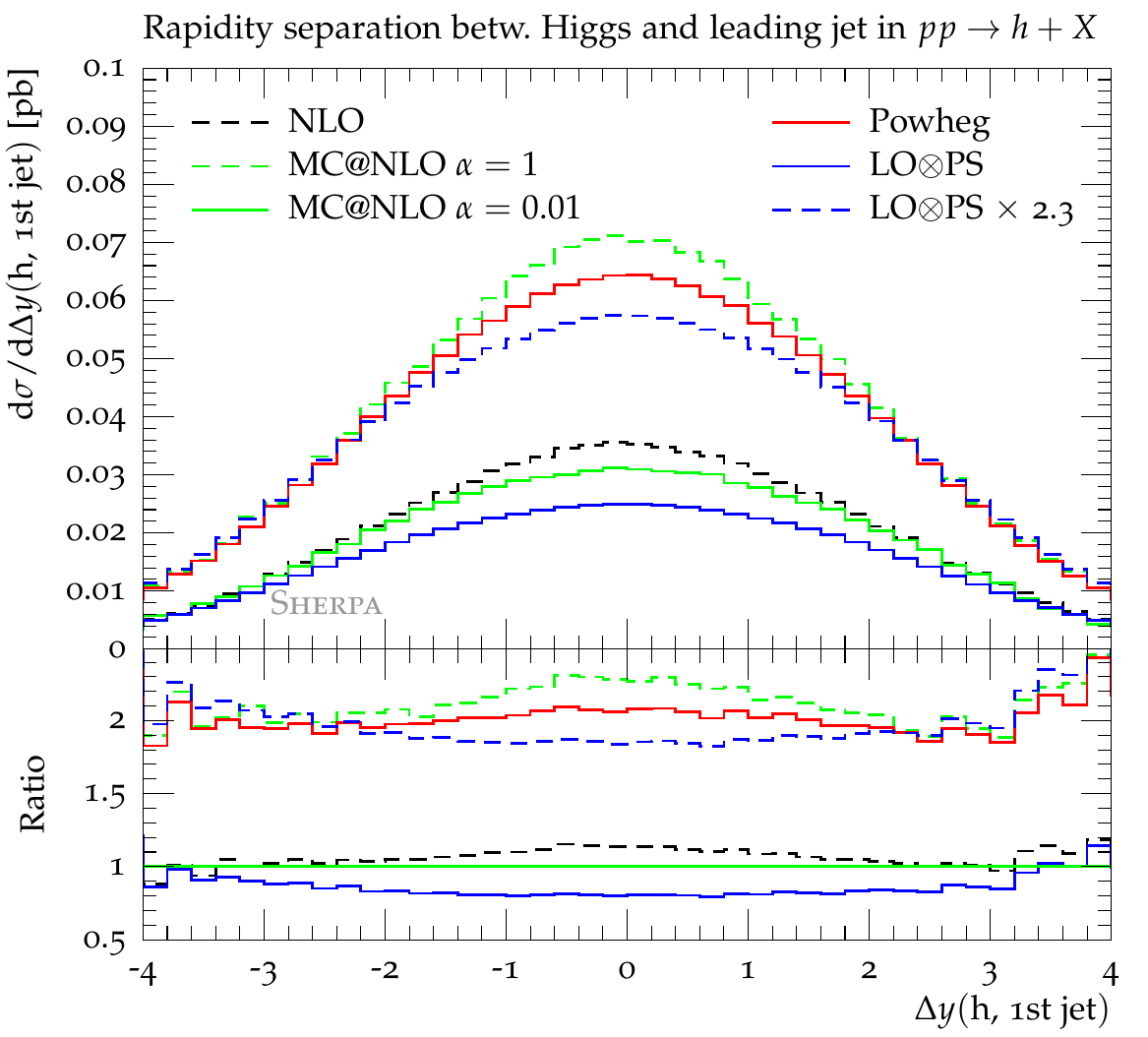}
    \label{fig:ggh:hjdy_b}}
  \caption{Rapidity separation between Higgs-boson and leading jet in inclusive 
           Higgs-boson production ($m_h=120$ GeV) at $E_{\rm cms}=7$~TeV.
           See Fig.~\ref{fig:ggh:hpt} for details.}
  \label{fig:ggh:hjdy}
\end{figure}

\begin{figure}[tbp]
  \centering
  \subfloat[][]{
    \includegraphics[width=0.49\textwidth]{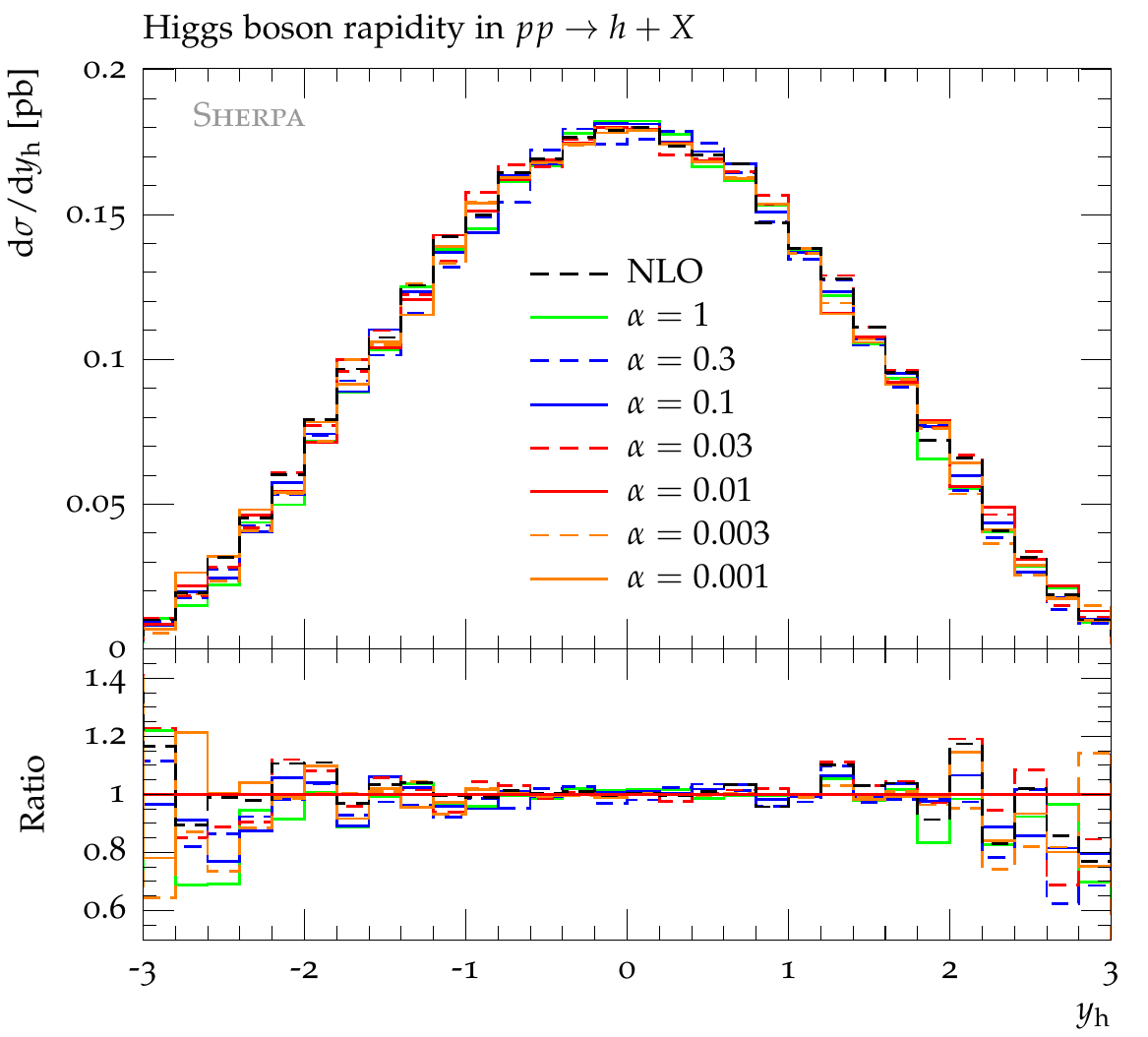}
    \label{fig:ggh:hy_a}}
  \subfloat[][]{
    \includegraphics[width=0.49\textwidth]{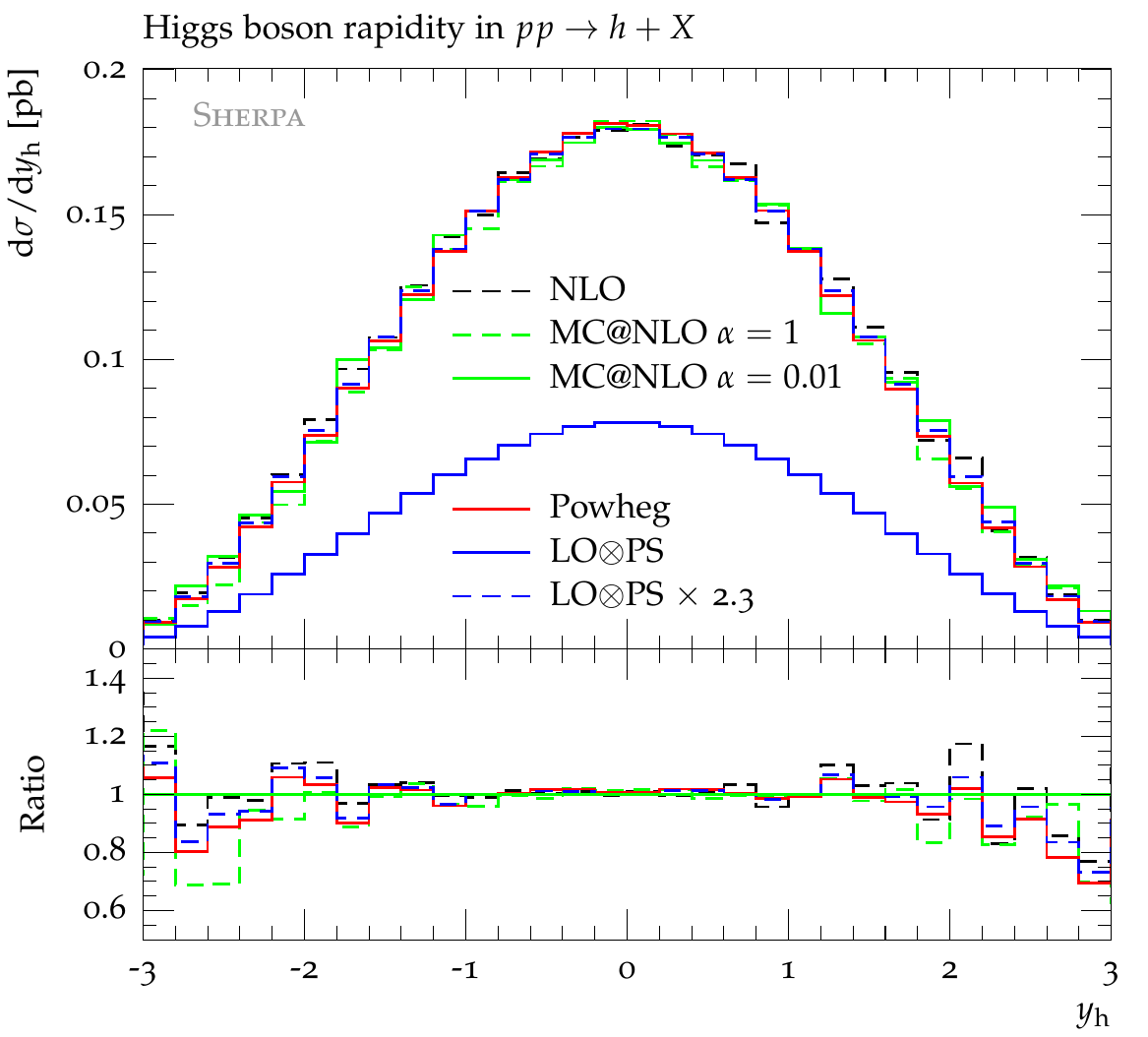}
    \label{fig:ggh:hy_b}}
  \caption{Prediction and uncertainties for the rapidity of the Higgs boson 
	   in inclusive Higgs boson production ($m_h=120$ GeV) at $E_{\rm cms}=7$~TeV. 
           See Fig.~\ref{fig:ggh:hpt} for details.}
  \label{fig:ggh:hy}
\end{figure}

\begin{figure}[tbp]
  \centering
      \includegraphics[width=0.49\textwidth]{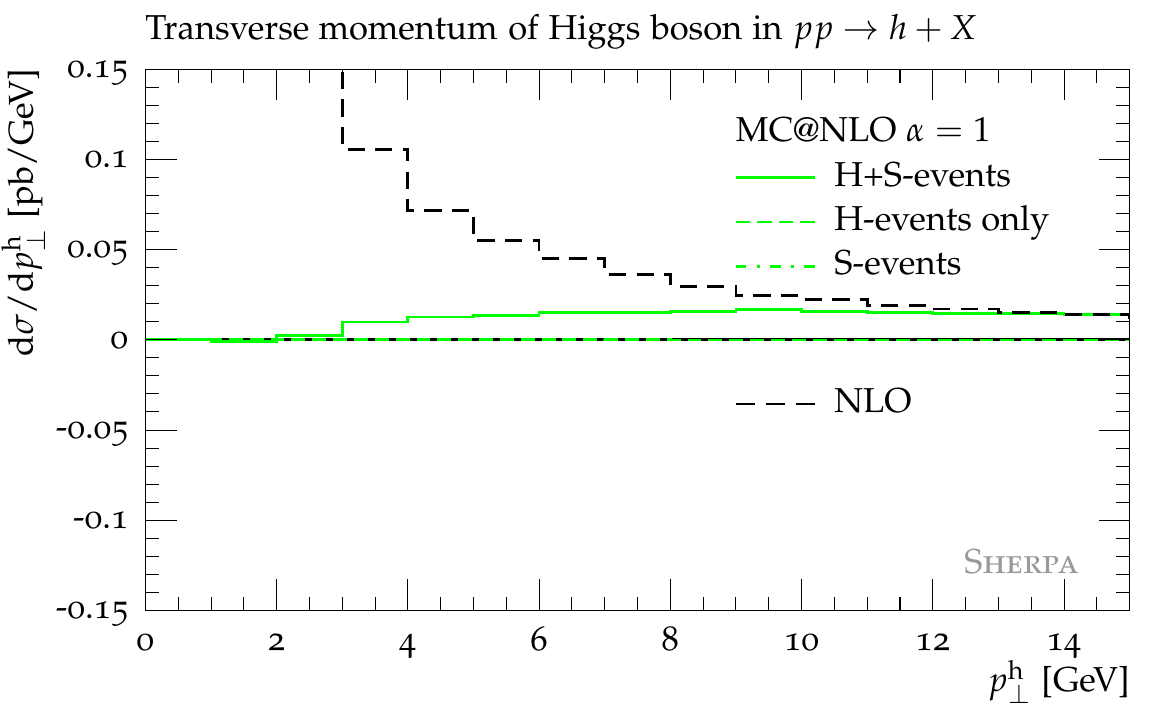}\\
      \includegraphics[width=0.49\textwidth]{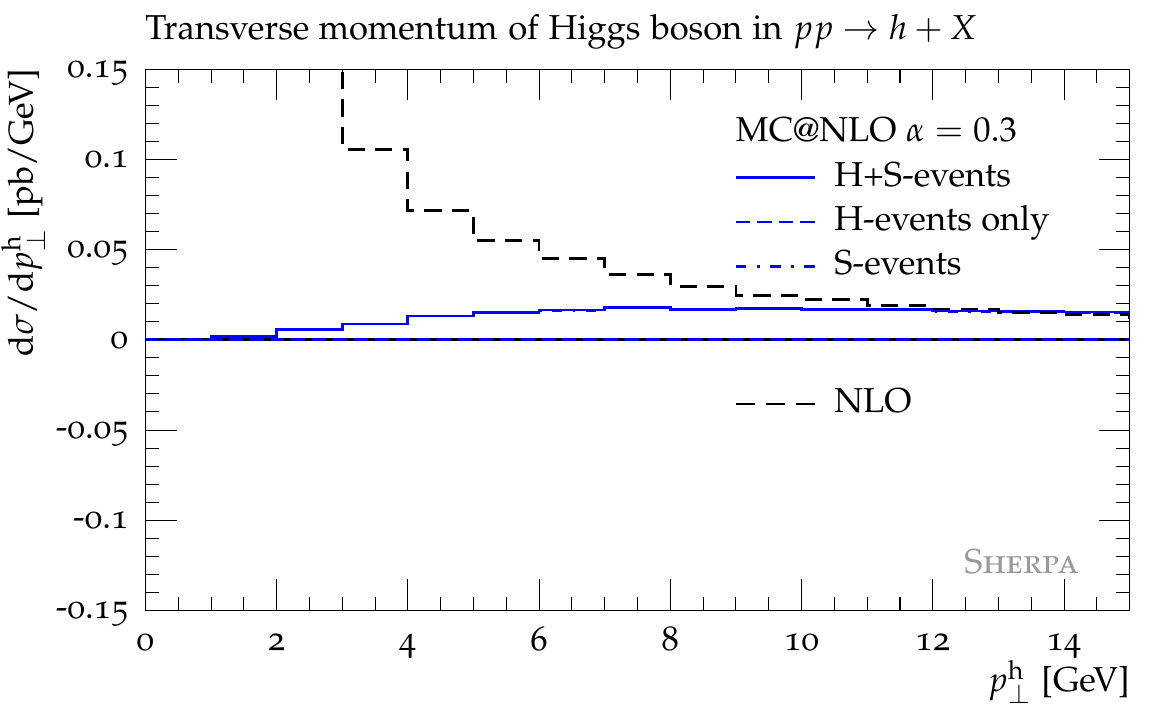}\hfill
      \includegraphics[width=0.49\textwidth]{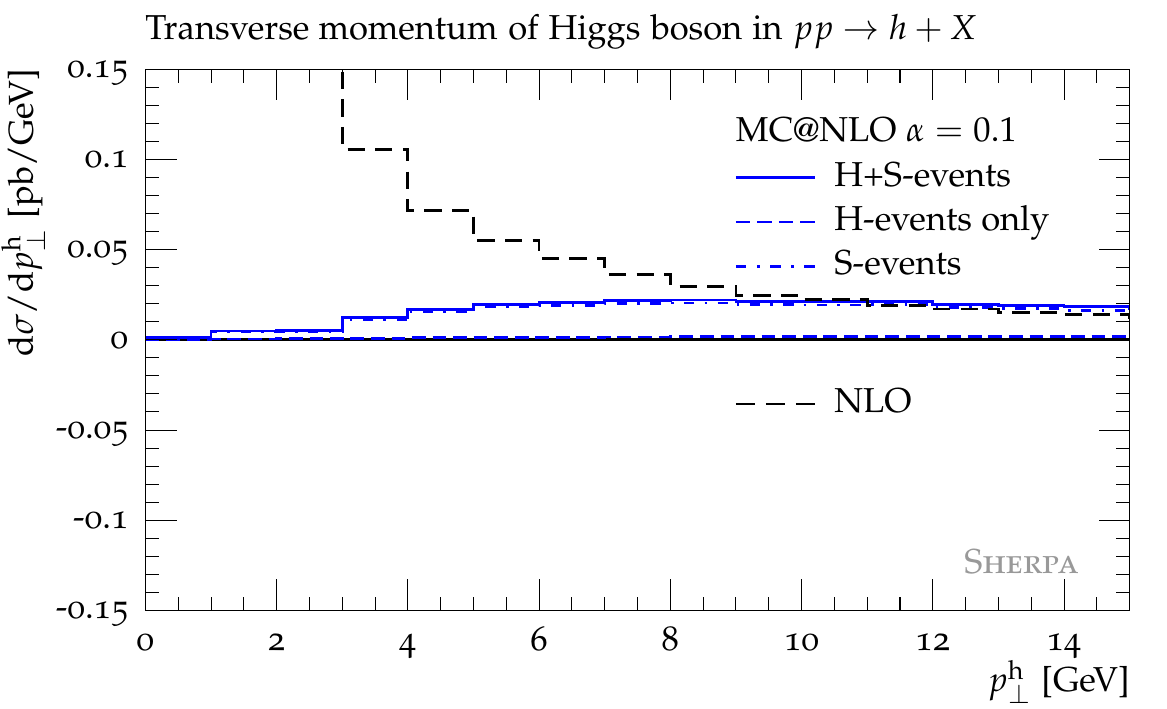}\\
      \includegraphics[width=0.49\textwidth]{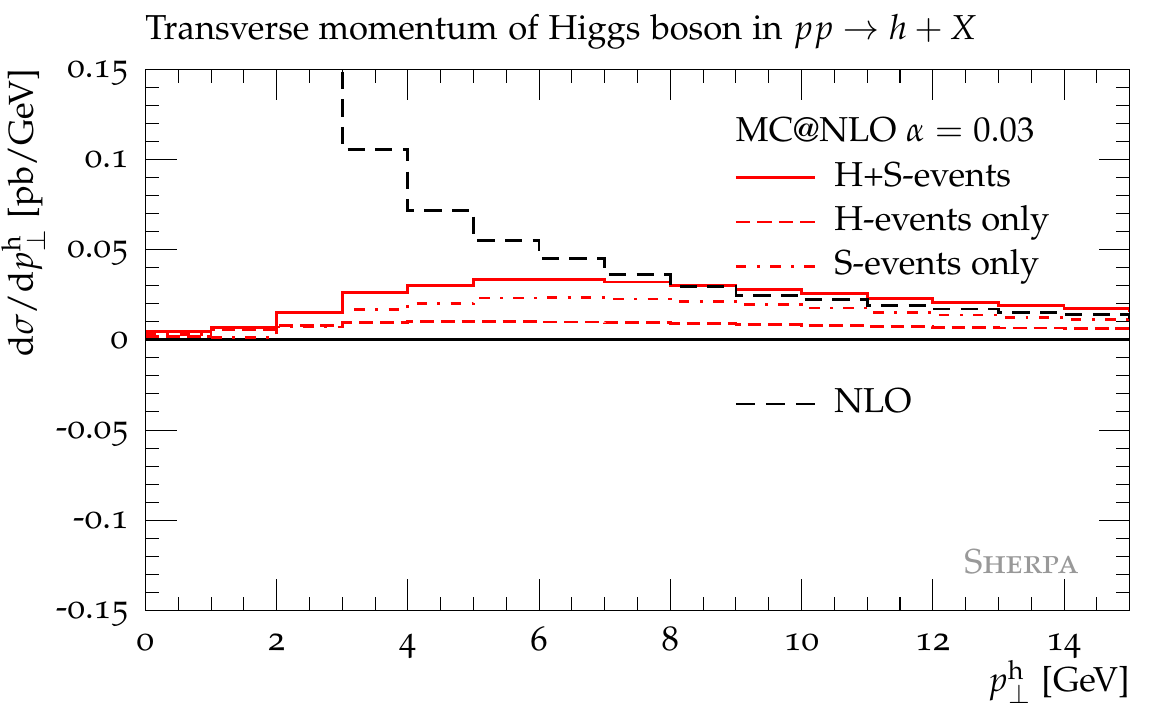}\hfill
      \includegraphics[width=0.49\textwidth]{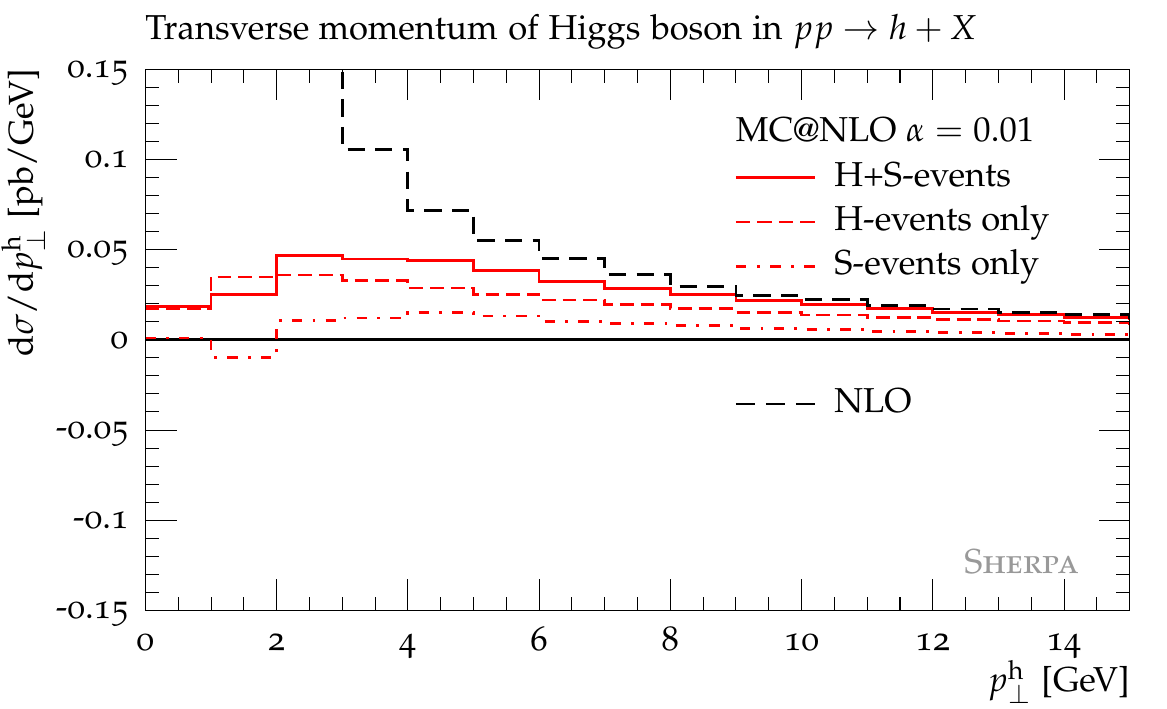}\\
      \includegraphics[width=0.49\textwidth]{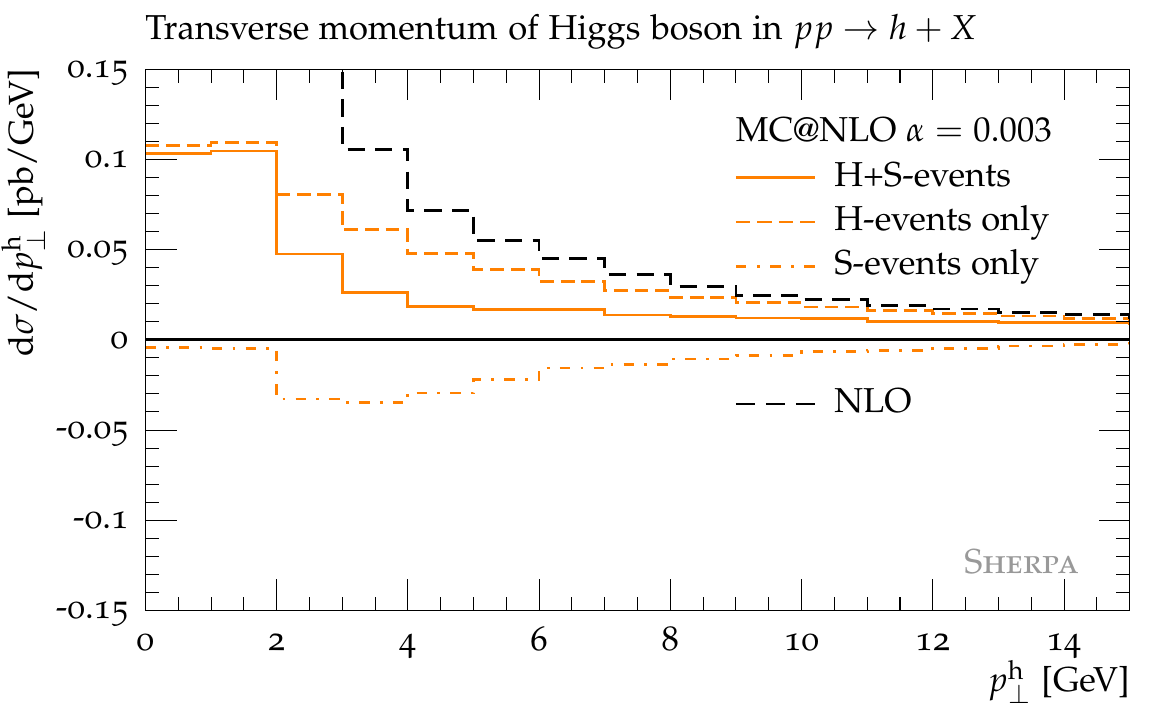}\hfill
      \includegraphics[width=0.49\textwidth]{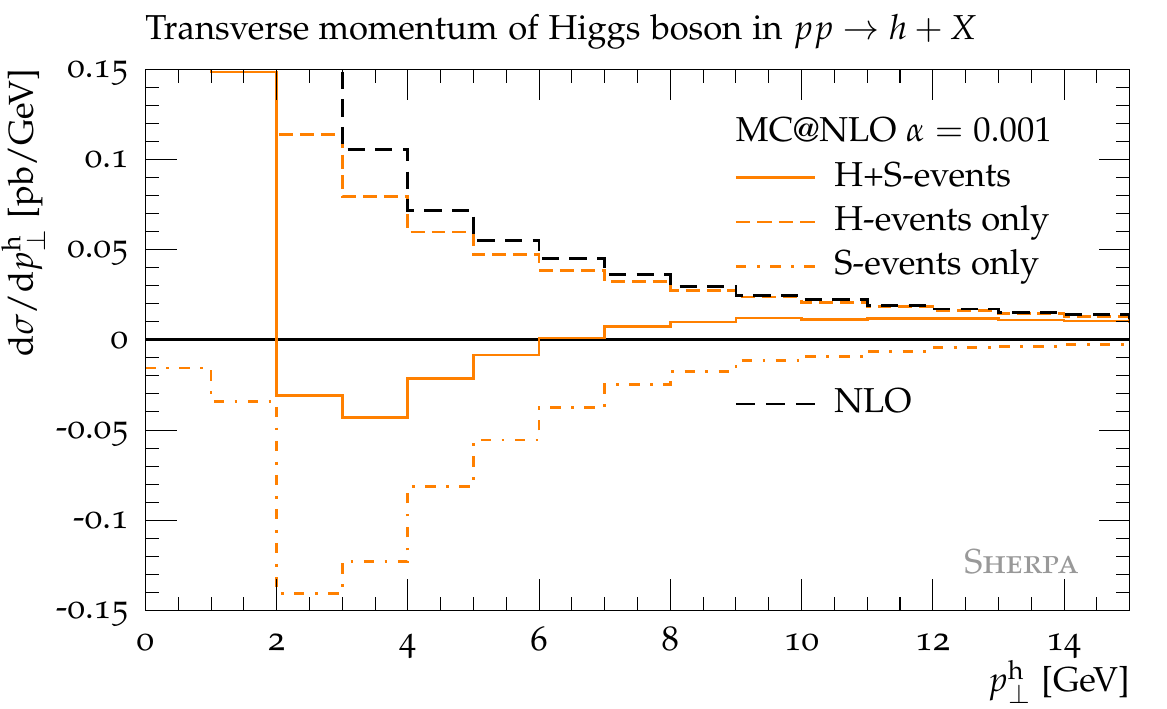}
  \caption{Higgs-boson transverse momentum for small values of $p_\perp^{\rm h}$ 
           in inclusive Higgs-boson production ($m_h=120$ GeV) at $E_{\rm cms}=7$~TeV.
           The contribution of $\mathbb{S}$- and $\mathbb{H}$-events in the \protect\MCatNLO method
           is displayed separately for different values of $\alpha_{\rm cut}$.
           For $\alpha_{\rm cut}=1$ and $\alpha_{\rm cut}=0.3$ the sample consist 
           primarily of $\mb{S}$-events with small additions of negatively and 
           positively valued $\mb{H}$-events, respectively.}
  \label{fig:ggh:hpt_soft_noCSS}
\end{figure}

\begin{figure}[tbp]
  \centering
      \includegraphics[width=0.49\textwidth]{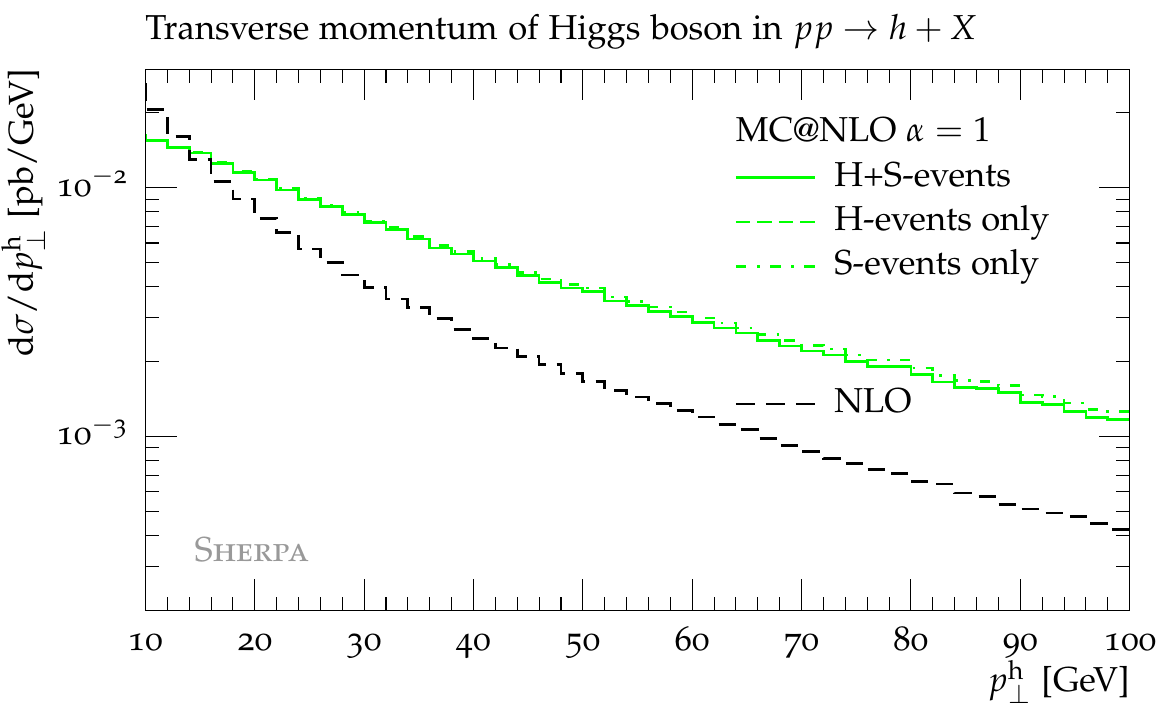}\\
      \includegraphics[width=0.49\textwidth]{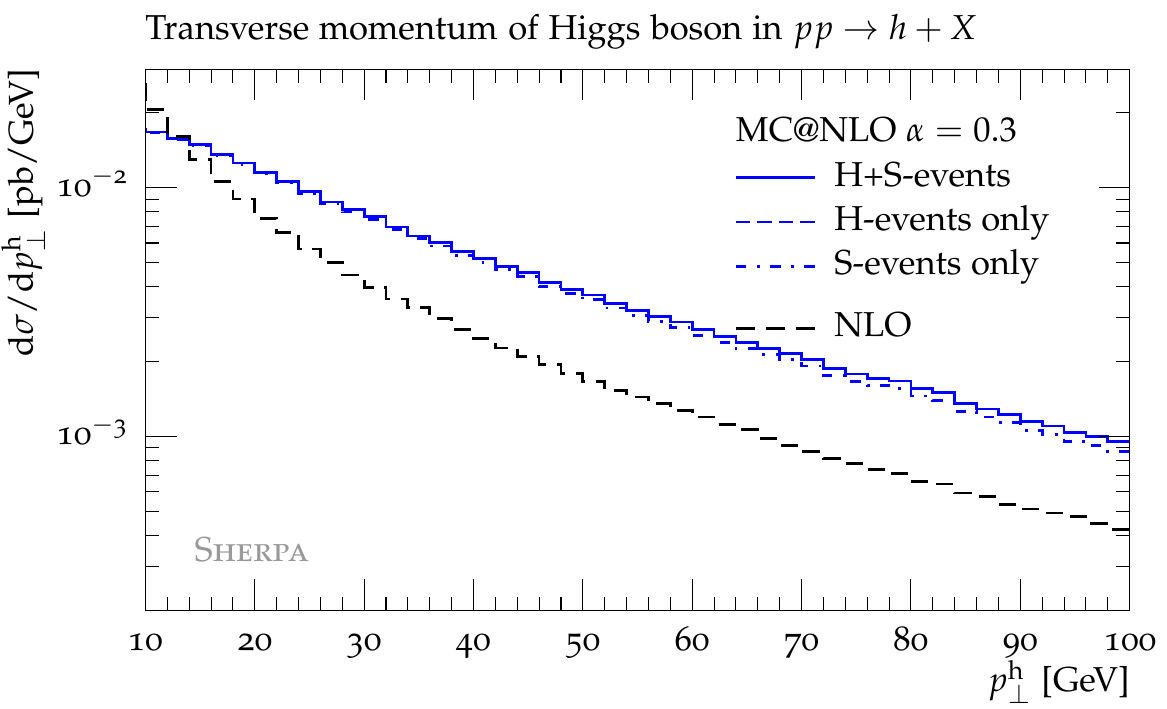}\hfill
      \includegraphics[width=0.49\textwidth]{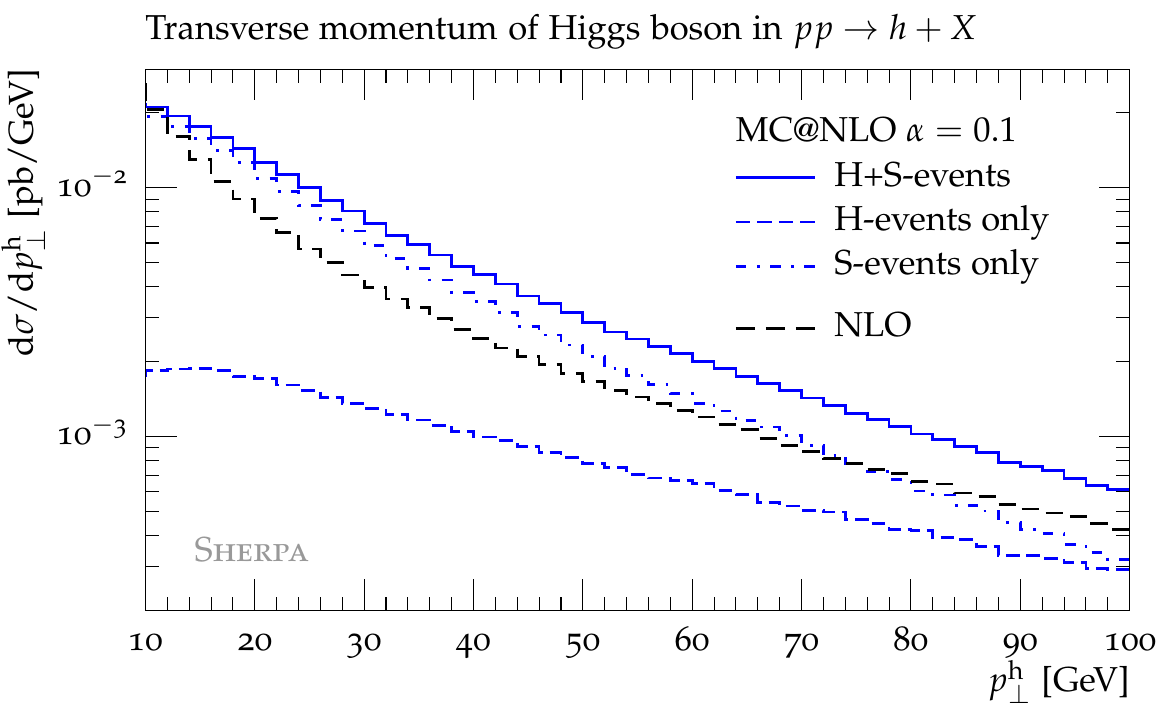}\\
      \includegraphics[width=0.49\textwidth]{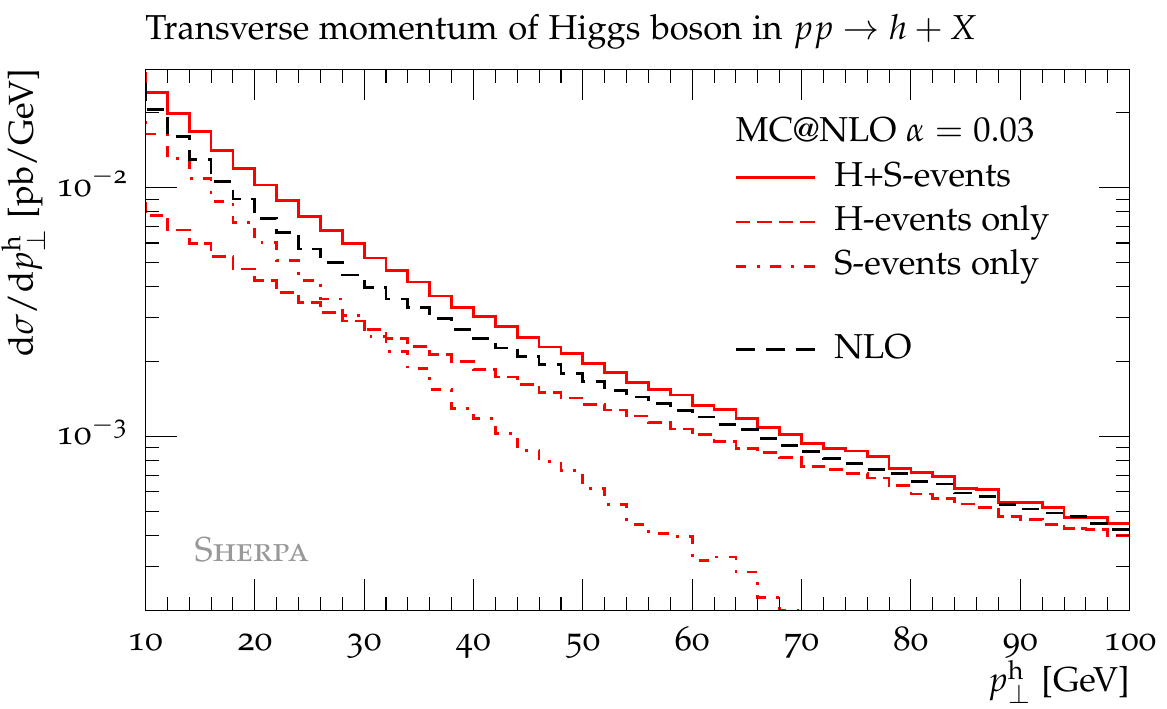}\hfill
      \includegraphics[width=0.49\textwidth]{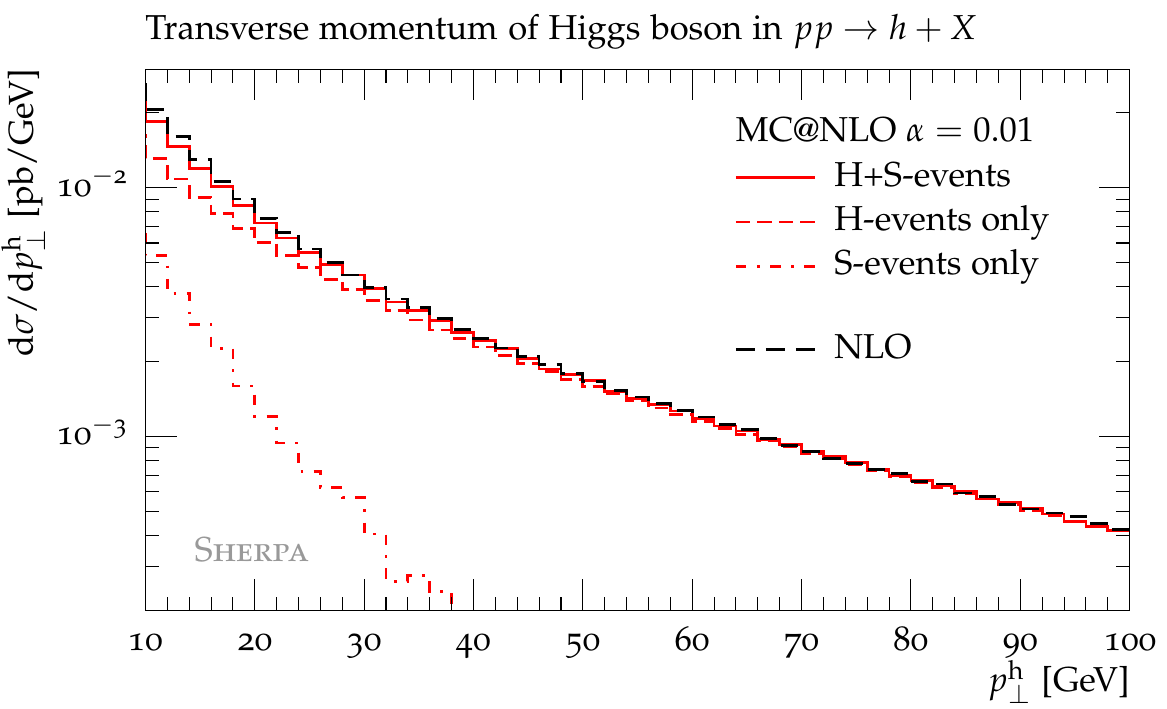}\\
      \includegraphics[width=0.49\textwidth]{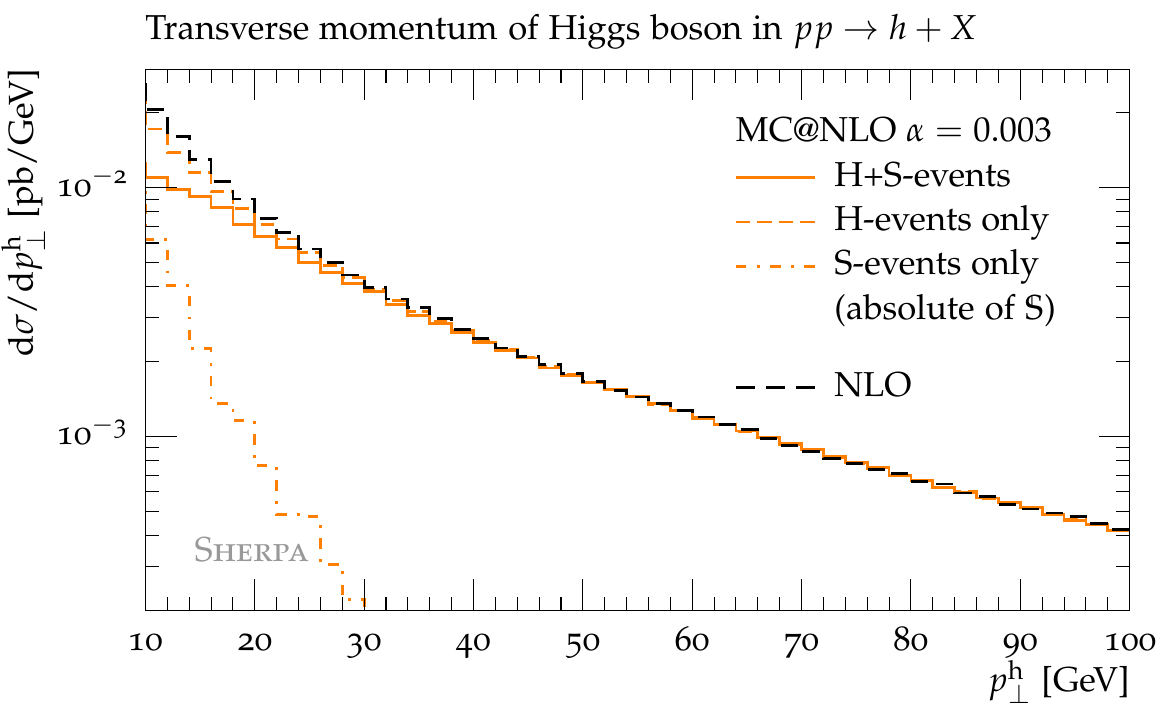}\hfill
      \includegraphics[width=0.49\textwidth]{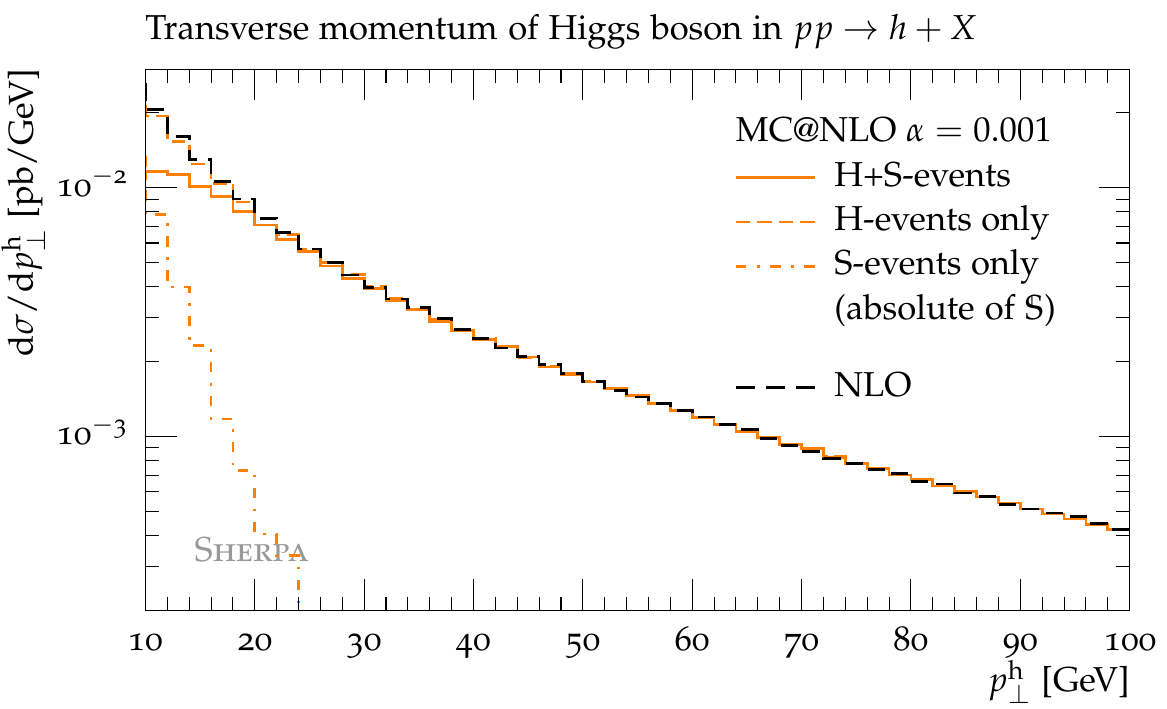}
  \caption{Higgs-boson transverse momentum for large values of $p_\perp^{\rm h}$ 
           in inclusive Higgs-boson production ($m_h=120$ GeV) at $E_{\rm cms}=7$~TeV.
           The contribution of $\mathbb{S}$- and $\mathbb{H}$-events in the \protect\MCatNLO method
           is displayed separately for different values of $\alpha_{\rm cut}$.
           For $\alpha_{\rm cut}=1$ and $\alpha_{\rm cut}=0.3$ the sample consist 
           primarily of $\mb{S}$-events with small additions of negatively and 
           positively valued $\mb{H}$-events, respectively.}
  \label{fig:ggh:hpt_hard_noCSS}
\end{figure}

For an unbiased comparison of the \MCatNLO and \POWHEG techniques, the events are analysed
with a minimal set of cuts. A Higgs-boson mass of $m_h=120$ GeV is assumed and two $\tau$-leptons 
with $|\eta|<3.5$ and $p_\perp>25$~GeV are required; jets are defined according to the inclusive
$k_\perp$-algorithm~\cite{Catani:1993hr,*Ellis:1993tq,*Cacciari:2005hq} 
with $R=0.7$ and $p_{\perp,\rm min}=20$~GeV. The renormalisation and 
factorisation scales are set to $\mu_F=\mu_R=m_h$. The effective coupling of the 
Higgs to gluons, mediated by a top-quark loop, is modeled through an effective Lagrangian 
\cite{Dawson:1990zj,*Djouadi:1991tka}. \changed{Both powers of $\alpha_s$ in the 
effective gluon-gluon-Higgs coupling are also evaluated at $m_h$.}

The comparison with fixed-order predictions and predictions from standard parton showers
(referred to as \LOPS in the following) presents a first crucial test 
of the \MCatNLO and \POWHEG methods. In this context, \Sherpa is also used
as a framework for NLO fixed-order event generation, enabling a comparison of all approaches 
with identical input parameters. 
In order to vary the amount of exponentiated real-emission corrections in \changed{our} \MCatNLO, 
which is governed by Eqs.~\eqref{eq:Powheg_split_trivial}-\eqref{eq:def_mod_Sudakov_trivial} 
or, more rigorous, Eqs.~\eqref{eq:nbp_ijk}-\eqref{eq:def_mcatnlo},
the phase-space restriction described in~\cite{Nagy:2003tz} is employed and 
the parameter $\alpha_{\rm cut}$ is varied. It has the effect of increasing
($\alpha_{\rm cut}\to 1$) or decreasing ($\alpha_{\rm cut}\to 0$) the phase space
for non-singular contributions in $\mr{D}^{(\mr{A})}$ by setting an upper bound
on the virtuality $\hat{t}$ of the splitting parton. 
The leading logarithm is then of the form $\alpha_s\log^2(\hat{t}/\alpha_{\rm cut}\,s)$, 
with $s$ the centre-of-mass energy of the colliding protons. 
Hence, for $\alpha_{\rm cut}\sim 1$ the leading logarithm contains the wrong 
argument $\alpha_{\rm cut}s\gg\mu_F^2$, clearly at odds with usual choices of the 
resummation scale and in violation of factorisation theorems. Similarly, 
the original \POWHEG method exponentiates radiative corrections throughout the real-emission 
phase space, corresponding to \changed{our} \MCatNLO with $\alpha_{\rm cut}=1$. 
\changed{Some} practical implementations of \POWHEG propose to 
suppress non-singular terms with a continuous function which tends to 
one in the singular limits and approaches zero in those regions of phase space 
where emissions become hard~\cite{Alioli:2008gx}. 
However, \changed{even in this approach} the volume of phase space where exponentiation 
is performed is left invariant, and only the influence of non-logarithmic terms is reduced.
The leading logarithm is thus still of the form $\alpha_s\log^2(\hat{t}/s)$.
The traditional \LOPS method directly implements 
the DGLAP resummation and thus uses the factorisation scale $\mu_F$ as a
phase space constraint, resulting in the correct logarithmic form $\alpha_s\log^2(\hat{t}/\mu_F^2)$.
\changed{It is clearly desirable to implement this very same constraint in our
\MCatNLO. This requires the computation of a new class of integrated subtraction terms
in the Catani-Seymour framework, which is beyond the scope of this work and will be
the topic of a forthcoming publication.}

The influence of $\alpha_{\rm cut}$ on the \MCatNLO predictions and its 
relation to the \POWHEG and \LOPS predictions are investigated for various 
observables in Figs.~\ref{fig:ggh:hpt}-\ref{fig:ggh:hy}.
Taking the $\order(\alpha_s)$ expansion of Eq.~\eqref{eq:master_nlomc} at face 
value leads to the expectation that results from \MCatNLO (for any $\alpha_{\rm cut}$) 
and \POWHEG should \changed{approximately} agree with fixed-order predictions
in the limit of hard, well separated partons. In the limit of 
soft/collinear radiation the common logarithmic structure of the resummation
in \MCatNLO, \POWHEG and \LOPS should lead to identical results in all three
approaches.

In practice, the transverse momentum spectrum of the Higgs boson, 
shown in Fig.~\ref{fig:ggh:hpt_a}, exhibits a large dependence on 
$\alpha_{\rm cut}$. This roots in the different composition of the 
distribution in terms of $\mb{H}$- and $\mb{S}$-events, which is depicted in 
Figs.~\ref{fig:ggh:hpt_soft_noCSS} and \ref{fig:ggh:hpt_hard_noCSS}. Neglecting 
subsequent parton shower emissions, $\mb{H}$-events above $\alpha_{\rm cut}$ 
coincide with those of the NLO calculation, while $\mb{S}$-events resum the 
leading logarithm of the real emission matrix element and occur with an 
additional enhancement $\bar{\mr{B}}^{\rm(A)}/{\rm B}$. This enhancement, 
though formally contributing to $\order(\alpha_s^2)$, can be numerically large. 
Thus, when allowing $\mb{S}$-type events to fill more than the phase space 
suited for the resummation of the DGLAP logarithms, their emission probability 
is not only distorted by the exponentiation but also by an artificial 
$K$-factor $\bar{\mr{B}}^{\rm(A)}/{\rm B}$, as compared to the NLO result. 
For smaller values of $\alpha_{\rm cut}$, the argument of the leading logarithm 
is closer to the one in DGLAP evolution, even though $\alpha_{\rm cut}$ has 
a functional form which makes a direct comparison difficult. It is thus observed
that the \MCatNLO distributions come close to the fixed-order result in the 
hard region, and the method behaves exactly as expected.

Further, excessively small values of 
$\alpha_{\rm cut}$ lead to spurious results, as the $\mb{S}$-event 
prefactor $\bar{\mr{B}}^{\rm(A)}/{\rm B}$ turns negative. This is exemplified in 
Fig.~\ref{fig:ggh:hpt_soft_noCSS}.  
Thus, those values of $\alpha_{\rm cut}$ are preferred which ensure that 
$\mb{S}$-events only fill the phase space suited for resummation and 
simultaneously avoid $\bar{\mr{B}}^{\rm(A)}$ turning negative. Defining such 
an allowed range of $\alpha_{\rm cut}$ introduces an unwanted process dependence 
into this implementation. Clearly a physically more meaningful 
definition of phase space constraints will ultimately lead to removing this 
uncertainty.

Comparing these results to the predictions of the \POWHEG and \LOPS 
methods, depicted in Fig.~\ref{fig:ggh:hpt_b}, one finds that \POWHEG 
behaves as \changed{our \MCatNLO implementation} with $\alpha_{\rm cut}=1$, suffering from a too large 
phase space for exponentiation. On the other hand, the resummation properties of 
the parton shower are retained. The difference in the actual Sudakov shape 
can entirely be attributed to the respective size and functional form of 
the phase-space boundaries. Further, within the region where the standard 
parton shower is able to fill the phase space (up to $|\hat{t}|=\mu_F^2$), 
the \LOPS result scaled by a global factor 
$K=\sigma_\text{NLO}/\sigma_\text{LO}=2.3$ is in good agreement with the \POWHEG 
and \MCatNLO with $\alpha_{\rm cut}=1$ distributions. This is a consequence 
of the simple form of the virtual corrections resulting in a local $K$-factor
$\bar{\mr{B}}^{\rm(A)}/{\rm B}$, that largely coincides with the global $K$-factor.

\changed{It was claimed in~\cite{Alioli:2008tz} that the difference between \POWHEG and \MCatNLO 
predictions for Higgs-boson production originates from the local $K$-factor $\bar{\mr{B}}/{\rm B}$ 
in \POWHEG. Figure~\ref{fig:ggh:bbbar:hpt_j1pt} shows explicitly that this claim is at odds with our findings. 
It displays predictions from a standard \POWHEG simulation and a modified one, where we replace 
$\bar{\mr{B}}\to{\rm B}$. The same analysis is repeated with \MCatNLO at $\alpha=1$, except that in this case 
we replace $\bar{\mr{B}}^{\rm(A)}\to{\rm B}$. It is manifest that the original and the modified 
\POWHEG results agree, except for a global $K$-factor, as was observed in~\cite{Alioli:2008tz}. 
However, the same effect is found for \MCatNLO at $\alpha=1$. In contrast, both the modified \POWHEG 
and the modified \MCatNLO result differ in shape from the NLO prediction at high $p_T$,
while the \MCatNLO result with $\alpha=0.01$ matches the NLO result exactly. 
The deviations between \POWHEG and \MCatNLO, which were observed in~\cite{Alioli:2008tz} 
can therefore {\em not} be attributed to the local $K$-factor $\bar{\mr{B}}\to{\rm B}$. 
Instead they must originate from the unrestricted phase space in \POWHEG.}

Similar observations are made for the transverse momentum of the hardest 
jet of Fig~\ref{fig:ggh:j1pt}. Here, the difference in the emission rate 
for $\mb{H}$- and $\mb{S}$-events, as discussed earlier, is plainly visible. 
Again, \MCatNLO with $\alpha_{\rm cut}=0.01$ agrees with the fixed-order 
NLO prediction for large jet transverse momenta. Also \POWHEG and \MCatNLO 
with $\alpha_{\rm cut}=1$ give similar results, overestimating the emission 
rate by more than 250\%.

This overestimation also feeds into the rapidity separation between the Higgs boson
and the first jet in Fig.~\ref{fig:ggh:hjdy}. The predictions of \MCatNLO 
with $\alpha_{\rm cut}=0.01$ and \MCatNLO with $\alpha_{\rm cut}=1$ differ 
by a factor of two. The effects of decreasing $\alpha_{\rm cut}$ are
examined in Fig.~\ref{fig:ggh:hjdy_a}. With decreasing $\alpha_{\rm cut}$, the rapidity difference distribution
develops a pronounced dip, which becomes broader and flattens out until the fixed-order result 
is approached for $\alpha_{\rm cut}\approx\,$0.01 and stabilises for even 
smaller $\alpha_{\rm cut}$. Again, this is explained by the different 
composition of the samples in terms of $\mb{H}$- and $\mb{S}$-events.
\changed{Similar findings were reported in~\cite{Mangano:2006rw}, where $t\bar{t}$-production
was analysed and the \MCatNLO method was compared to tree-level merging.
The authors concluded that a dip in the leading jet rapidity spectrum,
present only in the \MCatNLO prediction, was most likely an artifact 
of the incomplete phase-space coverage in \Herwig's parton shower.
While the dead zones of this parton shower are clearly different from the 
dead zones generated through $\alpha_{\rm cut}$ in our approach, the resulting effect 
is surprisingly similar. It suggests that the dip, which is not present 
in either NLO fixed-order or tree-level merged results, must be attributed 
to exponentiation uncertainties.}
Fig.~\ref{fig:ggh:hjdy_b} shows a comparison with \POWHEG and \LOPS. 
\POWHEG predictions agree well with those from \changed{our} \MCatNLO with $\alpha_{\rm cut}=1$ 
and from \LOPS with a global $K$-factor, while \MCatNLO with $\alpha_{\rm cut}=0.01$ 
agrees with standard \LOPS.

Fig.~\ref{fig:ggh:hy} shows the rapidity spectrum of the Higgs boson,
which can be defined at Born level and is thus described at NLO accuracy.
No significant variation is observed in this spectrum when the parameter $\alpha_{\rm cut}$
is varied, as expected. One can thus conclude that both the \MCatNLO and \POWHEG techniques
are consistently implemented in the event generator \Sherpa.

\changed{Summarising the above results, we find that
the perturbative uncertainties associated with the \POWHEG method are large, due to the
unrestricted phase space available for exponentiation. It is important to stress that
this problem is in principle solved by the \MCatNLO method and that we simply use
our implementation of \MCatNLO to quantify the effect.}

\subsection{Higgs-boson production in association with a jet}
\label{sec:gghj}
\begin{figure}[t]
  \centering
  \includegraphics[width=0.49\textwidth]{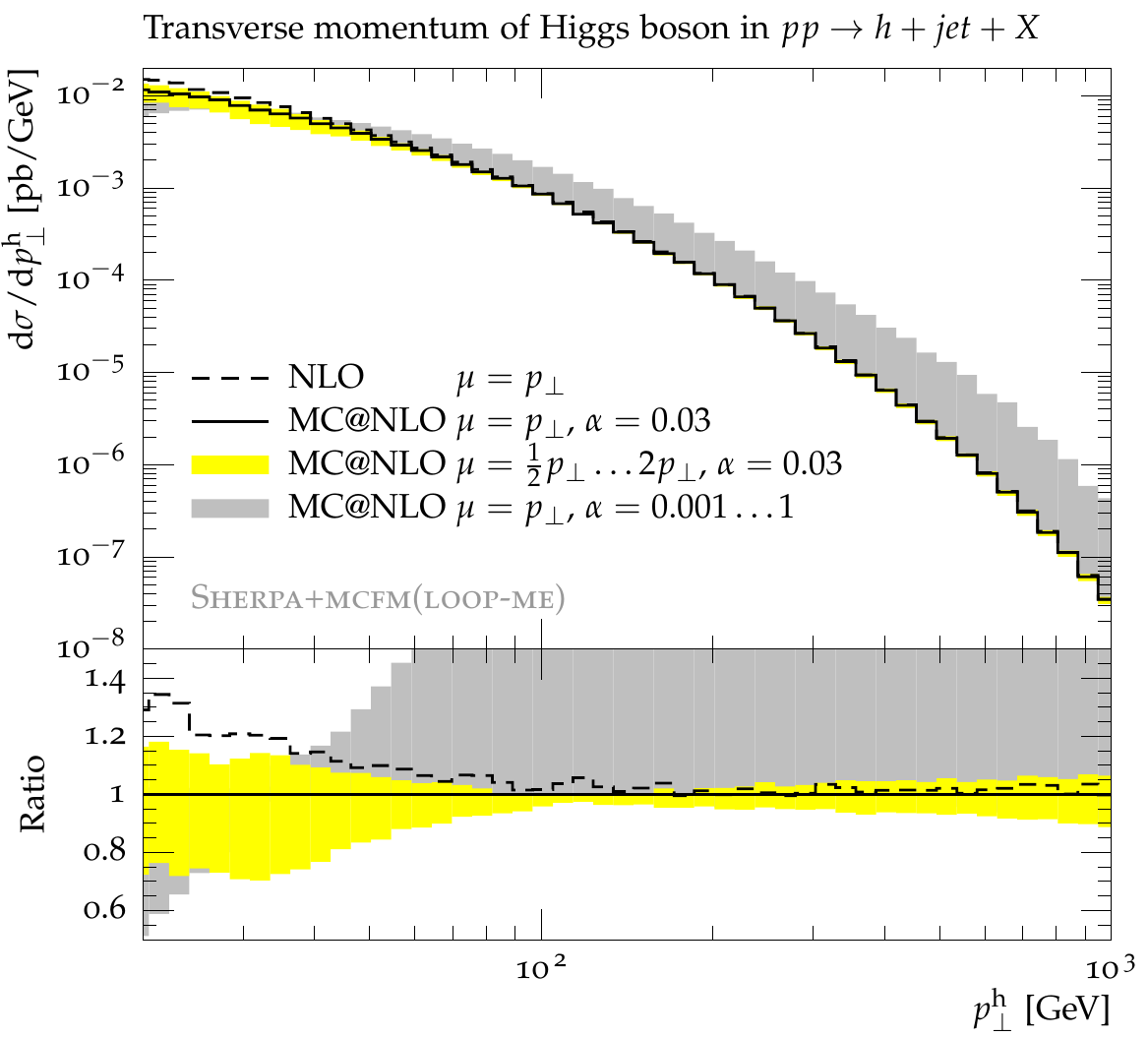}
  \includegraphics[width=0.49\textwidth]{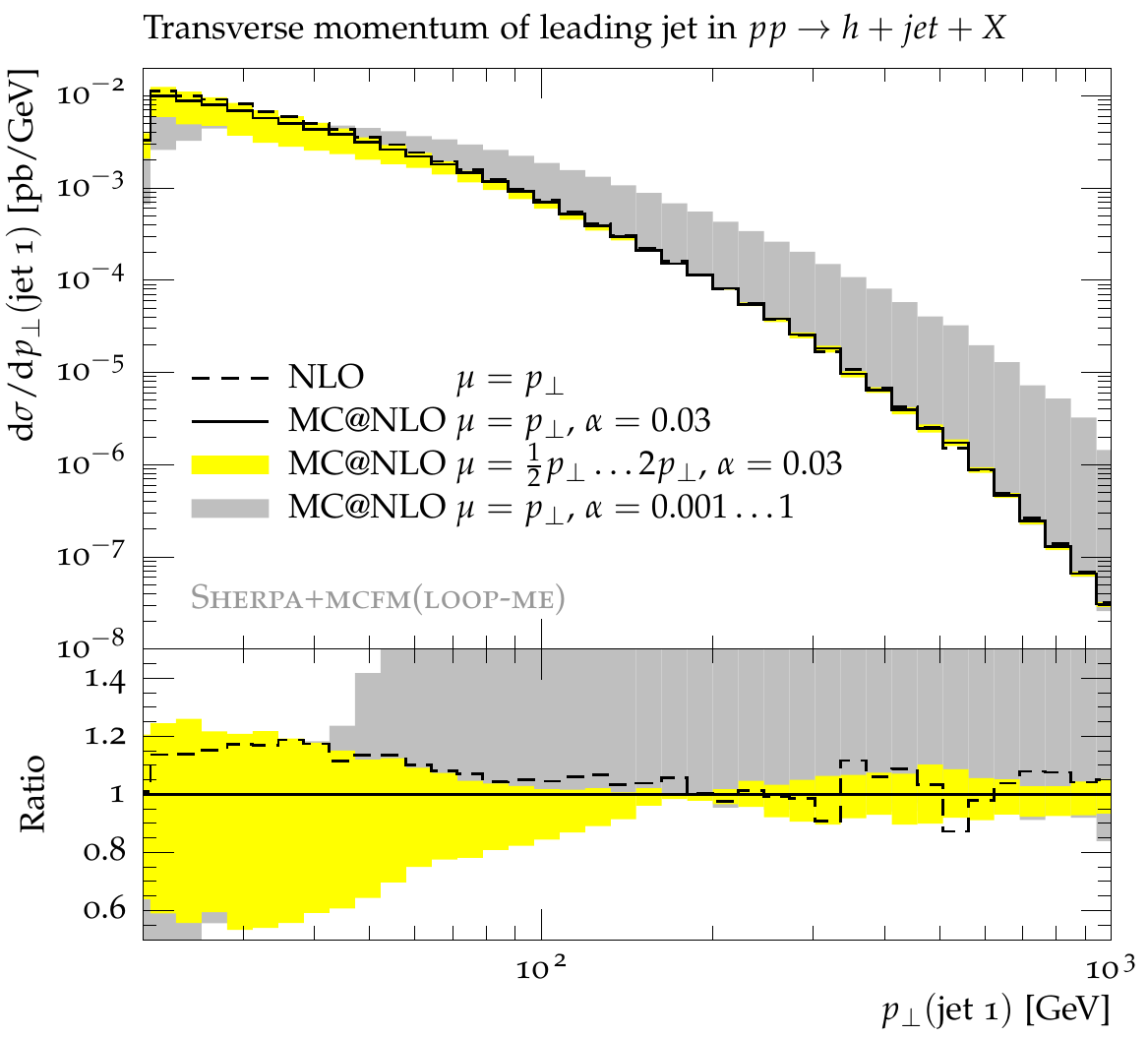}
  \caption{Prediction and uncertainties for the transverse momentum of the 
	   Higgs boson (left) and the leading jet (right) in Higgs boson 
	   plus jet events ($m_h=120$ GeV) at $E_{\rm cms}=7$~TeV.}
  \label{fig:gghj:hpt_j1pt}
\end{figure}
\begin{figure}[t]
  \includegraphics[width=0.49\textwidth]{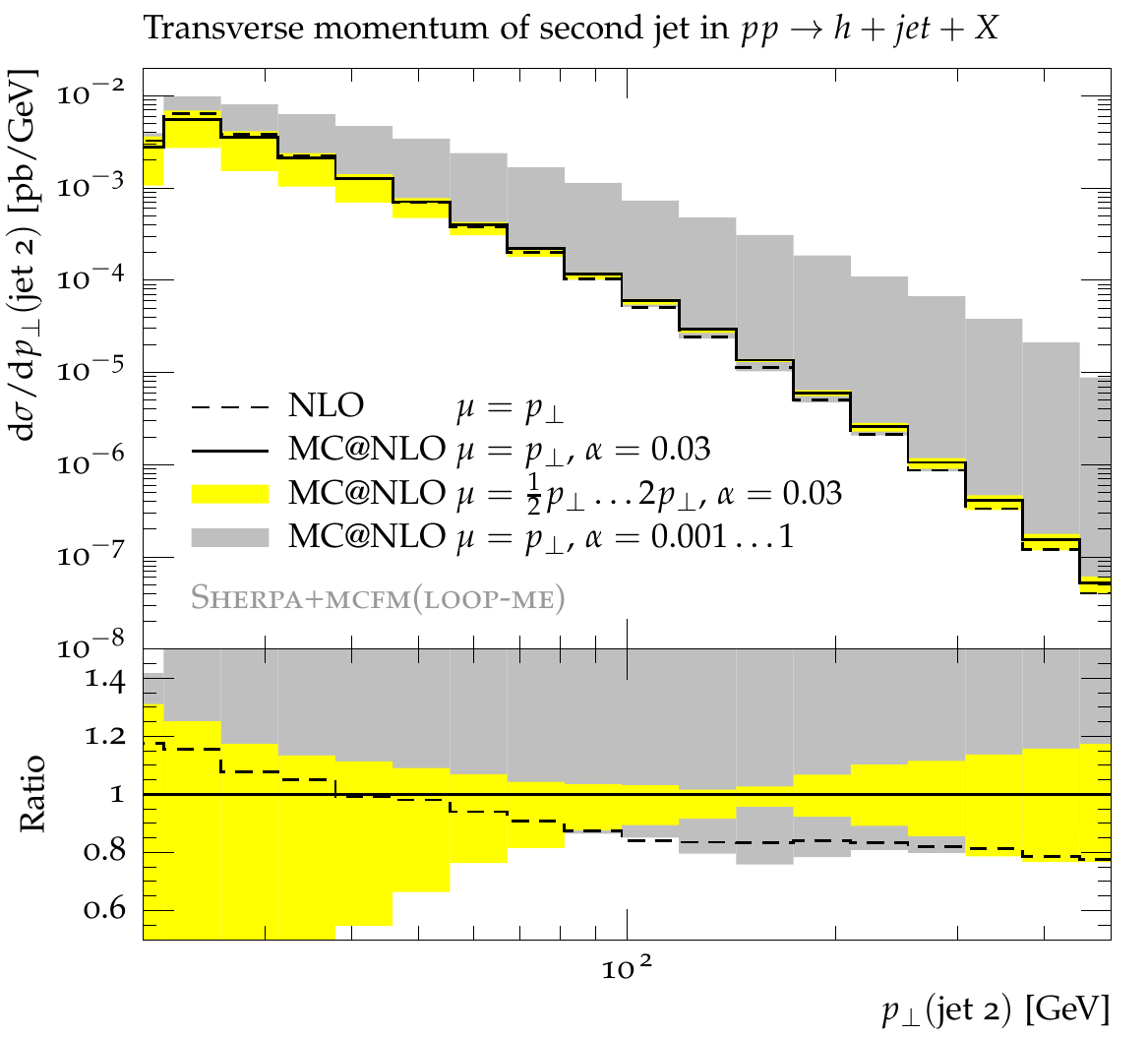}
  \includegraphics[width=0.49\textwidth]{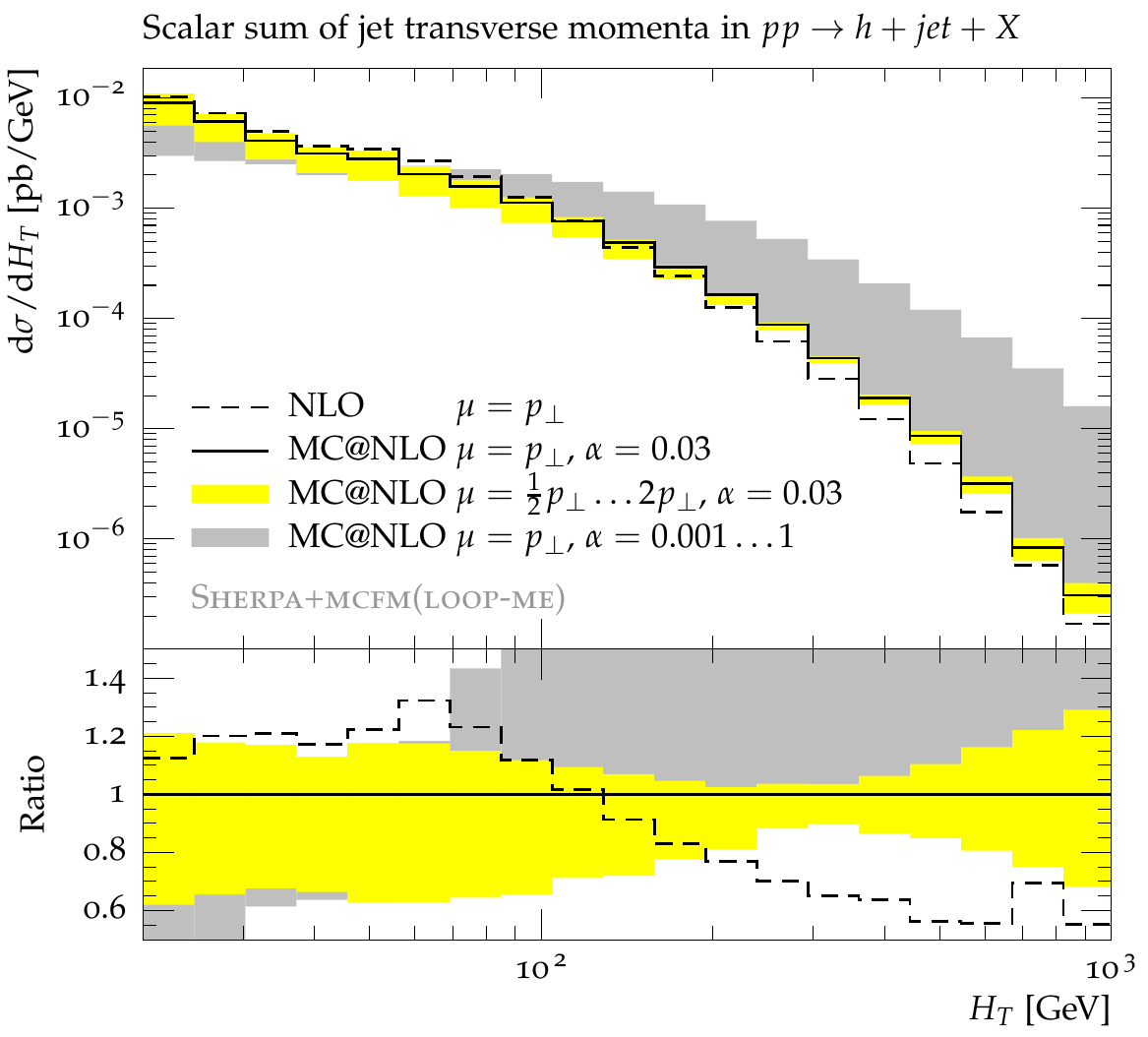}
  \caption{Predictions and uncertainties for transverse momentum of the 
	   second hardest jet (left) and the scalar sum of all jet transverse 
	   momenta (right) in Higgs boson plus jet events ($m_h=120$ GeV)
           at $E_{\rm cms}=7$~TeV.}
  \label{fig:gghj:j2pt_jHT}
\end{figure}
\begin{figure}[t!]
  \centering
  \includegraphics[width=0.49\textwidth]{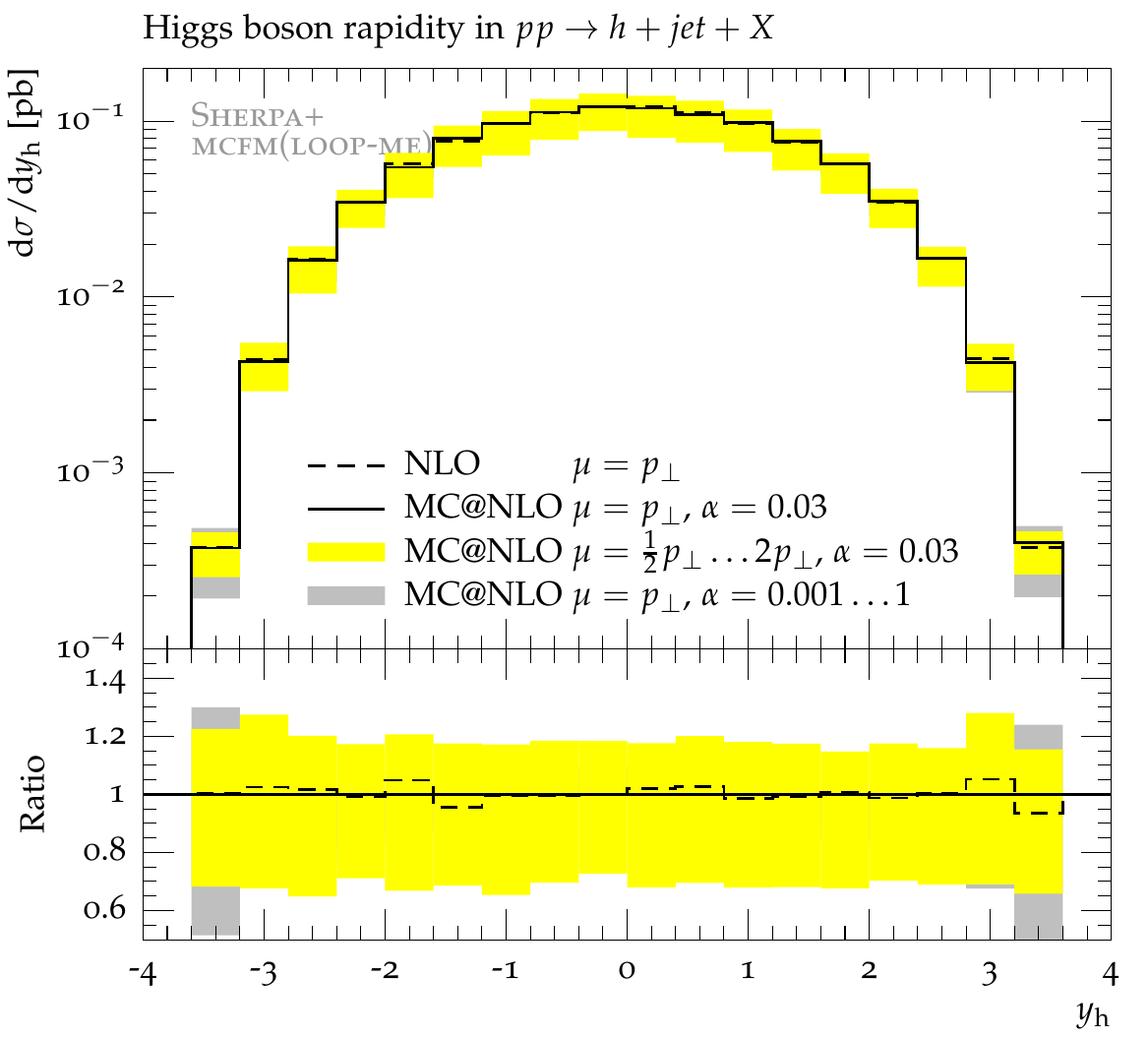}
  \includegraphics[width=0.49\textwidth]{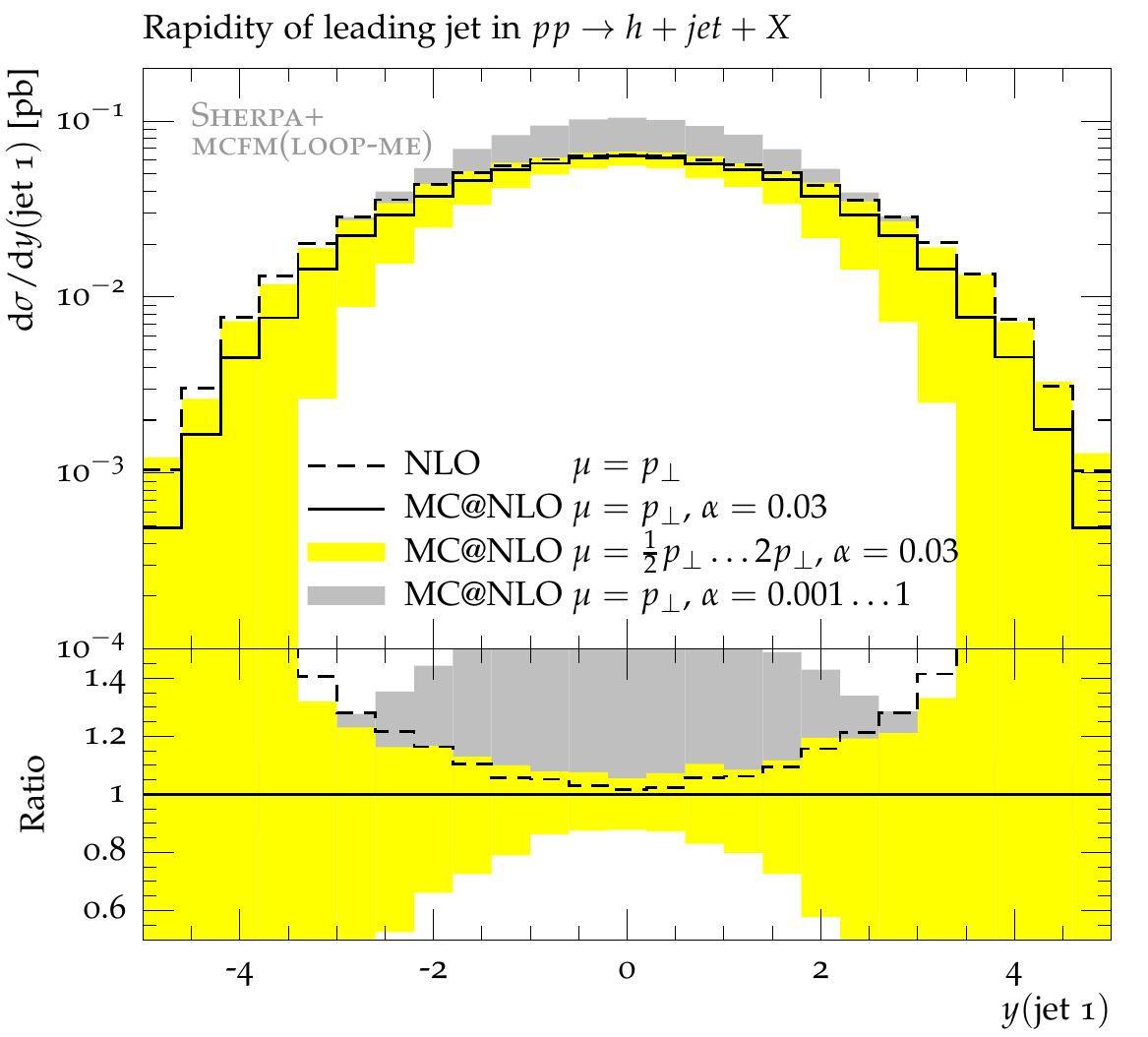}
  \caption{Prediction and uncertainties for the rapidity of the 
	   Higgs boson (left) and the leading jet (right) in Higgs boson 
	   plus jet events ($m_h=120$ GeV) at $E_{\rm cms}=7$~TeV.}
  \label{fig:gghj:hy_j1y}
\end{figure}
\begin{figure}[t]
  \centering
  \includegraphics[width=0.49\textwidth]{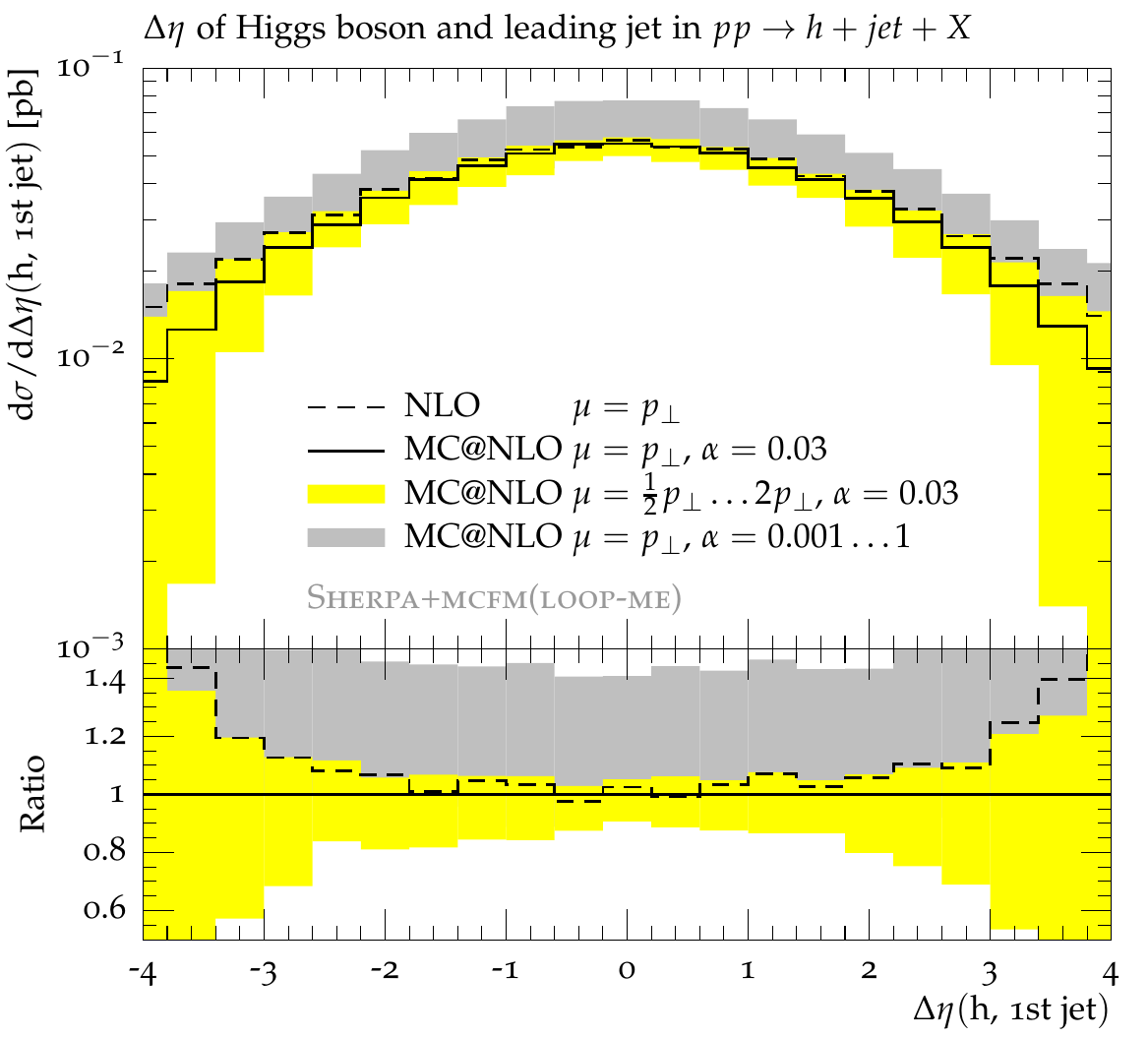}
  \includegraphics[width=0.49\textwidth]{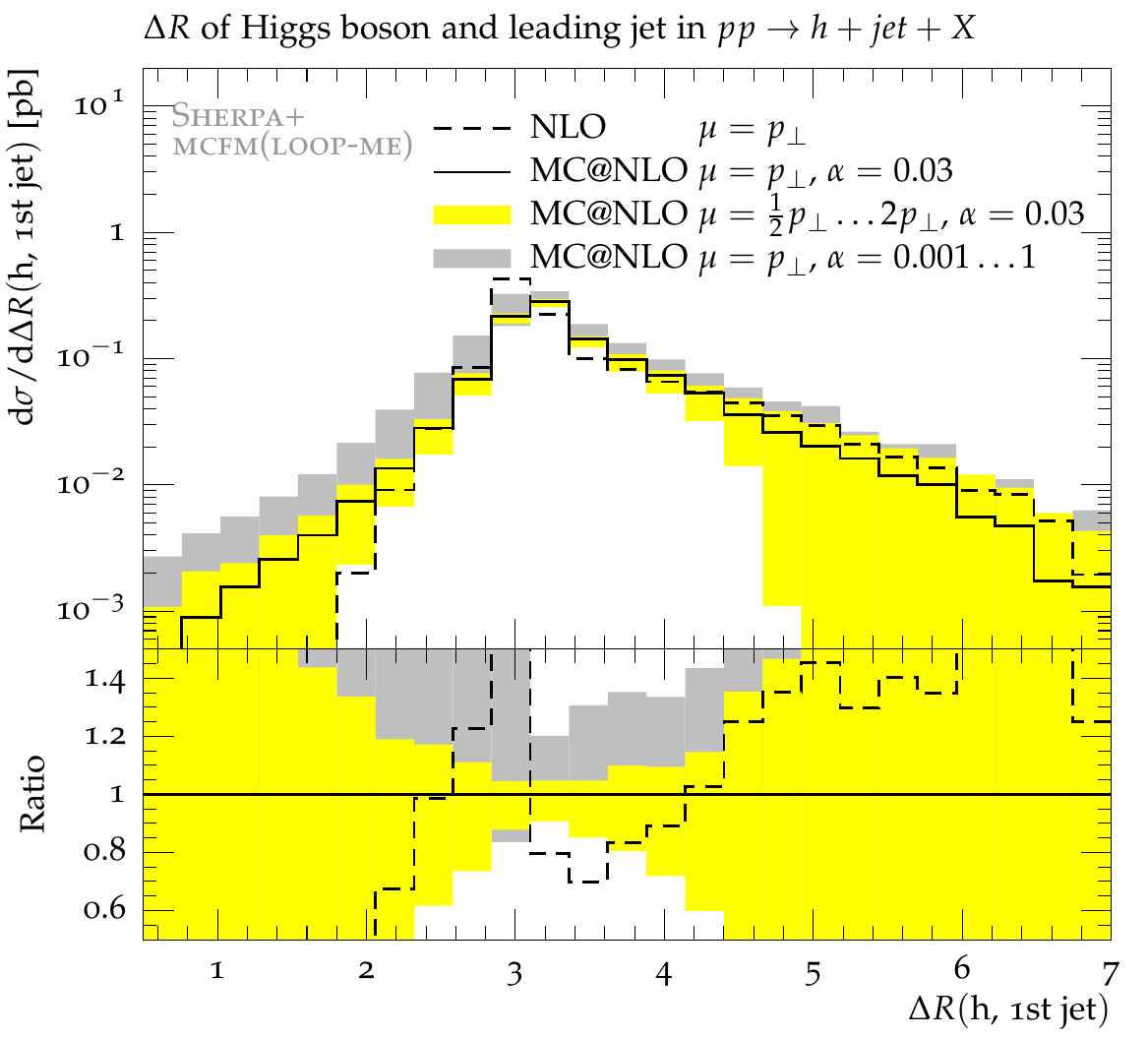}
  \caption{Predictions and uncertainties for the pseudorapidity (left) and 
	   angular (right) separation of the Higgs boson and the leading jet 
	   in Higgs boson plus jet events ($m_h=120$ GeV) at $E_{\rm cms}=7$~TeV.}
  \label{fig:gghj:hj1deta_hj1dR}
\end{figure}
\begin{figure}[t!]
  \includegraphics[width=0.49\textwidth]{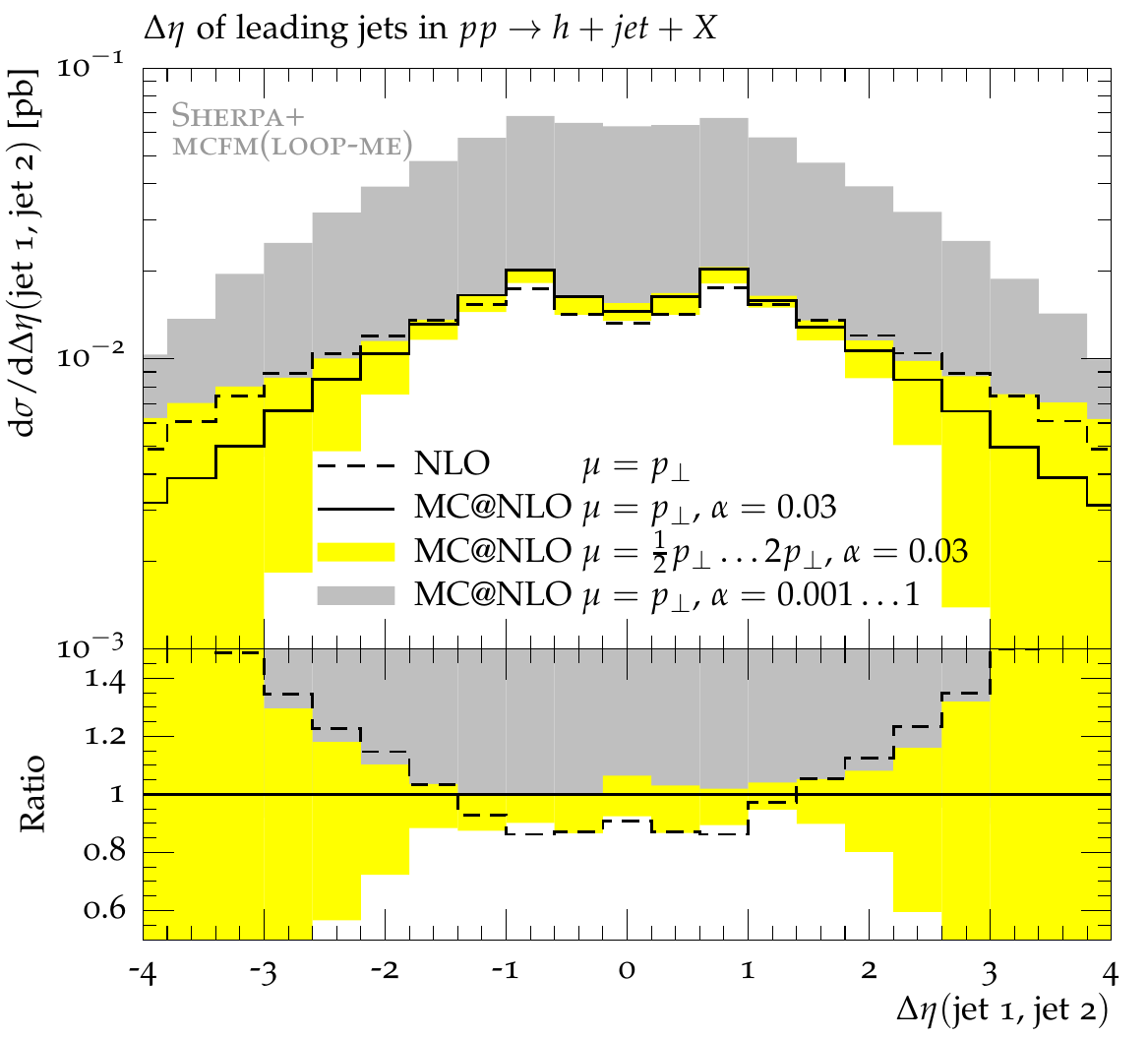}
  \includegraphics[width=0.49\textwidth]{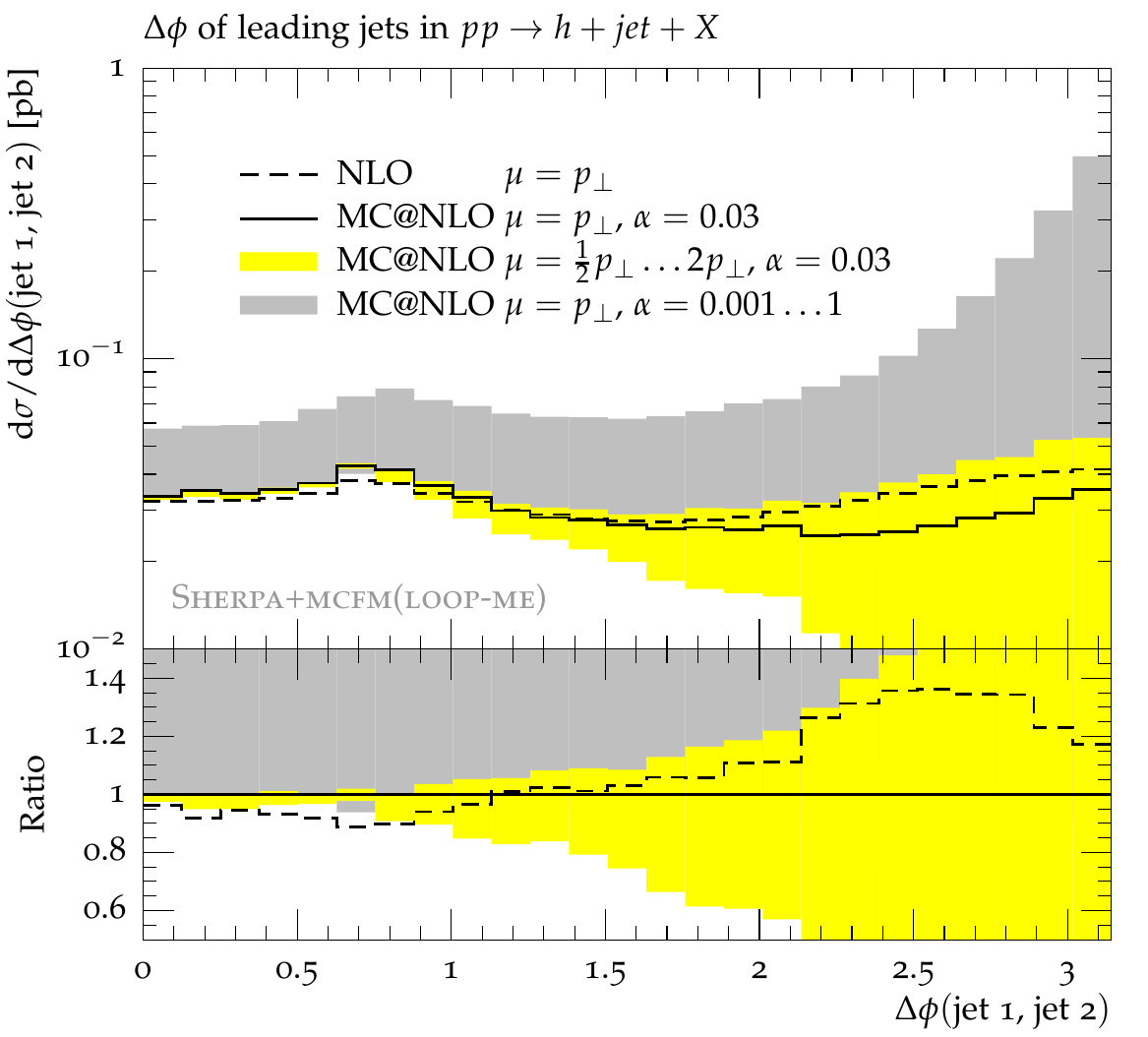}
  \caption{Predictions and uncertainties for the pseudorapidity (left) and 
	   azimuthal (right) separation of the two hardest jets in Higgs boson 
	   plus jet events ($m_h=120$ GeV) at $E_{\rm cms}=7$~TeV.}
  \label{fig:gghj:j1j2deta_j1j2dphi}
\end{figure}

This section presents predictions for the production of a Higgs boson 
in association with at least one jet. This process has not yet been investigated
using either the \MCatNLO or the \POWHEG approach. The Higgs mass is set to 
$m_h=120$ GeV. Jets at parton-level
(i.e.\ before parton shower emissions take place) are defined using the 
inclusive $k_\perp$ algorithm~\cite{Catani:1993hr,*Ellis:1993tq,*Cacciari:2005hq} 
with $R=0.5$ and $p_{\perp,\rm min}=10$~GeV. \changed{The independence of the 
results of the precise value of generation cut has been checked by varying 
$p_{\perp,\rm min}=5\ldots 15$~GeV.} Again, the effective coupling 
of the Higgs to gluons, mediated by a top-quark loop, is modeled through 
an effective Lagrangian \cite{Dawson:1990zj,*Djouadi:1991tka}, and the 
virtual matrix elements here are computed by \MCFM~\cite{MCFM,*Ravindran:2002dc}. 

Both the renormalisation and factorisation scales are set to the transverse 
momentum of the hardest parton-level jet in the event. \changed{The two powers of 
$\alpha_s$ in the effective gluon-gluon-Higgs coupling are evaluated at a fixed 
scale $m_h$, independent of the renormalisation scale $\mu$.}
The reason for this choice is that scale variations of the effective coupling do
not reflect a systematic uncertainty of the calculation at hand. They should be 
compensated by higher-order corrections to the effective coupling (i.e.\ diagrams 
with gluons attached to the top quarks in the loop), and will not be compensated 
by the one-loop corrections in the $h$+jet matrix elements that we use. Thus,
while the respective scale uncertainty is part of the systematics of 
the final cross section prediction, its proper assessment must come 
from a different calculation.

Our central prediction chooses
$\alpha_{\rm cut}=0.03$ to minimise the uncontrolled exponentiation 
of non-logarithmic terms and distortions of the Sudakov shape of the Higgs boson's 
transverse momentum spectrum. The predictions presented here include simulation of 
hadronisation~\cite{Winter:2003tt,*Krauss:2010xy} and multiple parton interactions
(MPI)~\cite{Sjostrand:1987su,*Alekhin:2005dx} as well as hadron decays~\cite{Krauss:2010xx}
and higher-order QED corrections to the $h\to \tau\tau$ decay~\cite{Schonherr:2008av}.

Uncertainty bands correspond to the variation of the renormalisation and factorisation
scales in the range $1/2\mu\ldots 2\mu$ as well as to the variation of $\alpha_{\rm cut}$
in a range from 1 to 0.001. \changed{Scales are varied only in the matrix elements, 
not in the parton shower.} As before, only minimal cuts are applied in the analysis: 
two $\tau$-leptons with $|\eta|<3.5$ and $p_{\perp}>25$~GeV are required. Jets are 
defined using the inclusive $k_\perp$-algorithm with $R=0.7$ and $p_{\perp,\rm min}=20$~GeV.
Their minimum transverse momentum requested in the analysis ensures that the parton-level
event selection is inclusive enough to guarantee coverage of the full phase space.

Figure~\ref{fig:gghj:hpt_j1pt} shows the transverse momentum spectra of the Higgs boson
and the leading jet. Grey bands indicate the uncertainty due to the choice of
$\alpha_{\rm cut}$, which will be referred to as exponentiation uncertainty in the following.
Yellow bands show the scale uncertainty. It is manifest, that the exponentiation uncertainty
is largest in the region of large transverse momenta, a fact that seems counter-intuitive
at first. Despite the transverse momentum distributions being described at NLO, they can 
apparently be altered significantly by additional emissions in the parton shower. Taking
a closer look at Eqs.~\eqref{eq:Powheg_split_trivial}-\eqref{eq:def_mod_Sudakov_trivial},
this fact becomes clear:
\begin{itemize}
\item The proof of next-to-leading order accuracy of both the \POWHEG and the \MCatNLO formulae
  is based on an expansion of the Sudakov form factors to first order. Higher-order
  corrections do play an important role, however, especially in regions of the phase space
  where parton emission is logarithmically enhanced. This region is drastically 
  enlarged if the upper integration limit of the exponentiated real emission is 
  shifted from $\mu_F$ to $s=E_\text{CMS}^2$, as is the case in \POWHEG and 
  \MCatNLO with $\alpha_{\rm cut}=1$ (cf.\ Sec.~\ref{sec:ggh}). This change alters the shape of the emission 
  term in Eq.~\eqref{eq:Powheg_split_trivial} compared to the emission term
  in Eq.~\eqref{eq:total_xsec_split_trivial}. This finding is reinforced by the fact 
  that in all distributions investigated, the predictions for $\alpha_{\rm cut}=0.03$ 
  coincide with those for $\alpha_{\rm cut}=0.001$.
\item The strong coupling is evaluated at a scale of the order of the relative 
  transverse momentum between two partons in the emission term in 
  Eq.~\eqref{eq:total_xsec_split_trivial}, while it is evaluated
  at a different scale (here the transverse momentum with respect to the beam) 
  in Eq.~\eqref{eq:split_bbar}. This leads to an additional distortion
  of the spectrum of the real-emission term, which is formally a sub-leading effect, but
  which becomes important in the enlarged logarithmically enhanced regions of the emission phase space.
\end{itemize}
The above effects are amplified in Fig.~\ref{fig:gghj:j2pt_jHT}, as the observables there,
the transverse momentum of the second jet and the scalar sum of all jets in the event, 
are described at leading-order accuracy only.

In contrast, Figs.~\ref{fig:gghj:hy_j1y} and~\ref{fig:gghj:hj1deta_hj1dR} show 
a rather mild dependence on $\alpha_{\rm cut}$ in the shape of the distributions. 
Nonetheless, a change in the normalisation can be observed, which is explained 
by the fact that due to larger emission rates with increasing $\alpha_{\rm cut}$, 
the amount of radiative events is increased. Their kinematic distribution, 
however, does not seem to differ significantly. In fact, the large scale dependence 
observed in Figs.~\ref{fig:gghj:hpt_j1pt} and 
\ref{fig:gghj:j2pt_jHT} is amplified here, especially for observables related 
to the hardest jet, indicating that, although $\mu=p_\perp$ is an appropriate 
scale choice for this process, the canonical variation to $\mu=\tfrac{1}{2}p_\perp$ 
leads to unphysical behaviour for jet rapidities beyond $|\eta|\approx 3.4$. 

As a consequence of the strongly varying jet rates with varying $\alpha_{\rm cut}$,
the pseudorapidity separation between first and second hardest jet suffers from
large uncertainties. It is displayed in Fig.~\ref{fig:gghj:j1j2deta_j1j2dphi}. 
While the shape of the distribution is mostly unaffected, the normalisation varies
strongly. The azimuthal separation between the two leading jets indicates how the
exponentiation uncertainty affects the relative position of both jets in phase space.
Compared to the pure next-to-leading order result, the back-to-back situation
is amplified by a factor $\sim 10$ in the \MCatNLO prediction. In such events,
the Higgs-boson is likely to have been produced at $p_\perp\to 0$,
consistent with the observation of a depletion of events in the first bins of the 
Higgs transverse momentum spectrum for $\alpha_{\rm cut}=1$, 
cf.\ Fig.~\ref{fig:gghj:hpt_j1pt}. 

\subsection{\texorpdfstring{$W$}{W}-boson production in association with a jet}
\label{sec:pheno:nlo}

In the following, $W[\to e\nu]$+jet production in $pp$ collisions at 7~TeV is studied.
As for the case of Higgs-boson production, the \MCatNLO implementation is validated
against a fixed-order calculation. In addition the scale uncertainties are
assessed by varying factorisation and renormalisation scales
in \MCatNLO by a factor of two in both directions.

To make the validation as meaningful as possible effects from 
parton showering beyond the first emission in $\mb{S}$-events, hadronisation and 
multiple parton scattering are not included in these parton level (PL) studies.
Events are generated requiring at least one jet with
$p_{\perp,\rm min}=10$~GeV, defined according to the inclusive 
$k_\perp$-algorithm~\cite{Catani:1993hr,*Ellis:1993tq,*Cacciari:2005hq} with $R=0.5$.
The transverse momentum of that hardest jet also sets the central value of factorisation
and renormalisation scales. The level of exponentiation is fixed by $\alpha_{\rm cut}=0.03$.

The total generated cross sections of the NLO calculation,
$\sigma_{\text{NLO}}=(4695 \pm 12)$~pb, and \MCatNLO,
$\sigma_{\text{\MCatNLO}}=(4680 \pm 30)$~pb, agree within
statistical uncertainties.
The total generated cross section in \MCatNLO varies between
$\sigma(\mu/2)=(4950 \pm 40)$~pb and $\sigma(\mu\cdot 2)=(4510 \pm 23)$~pb,
i.e. up to 5.8\% around the central value.

\begin{figure}[t]
  \centering
  \includegraphics[width=0.45\textwidth]{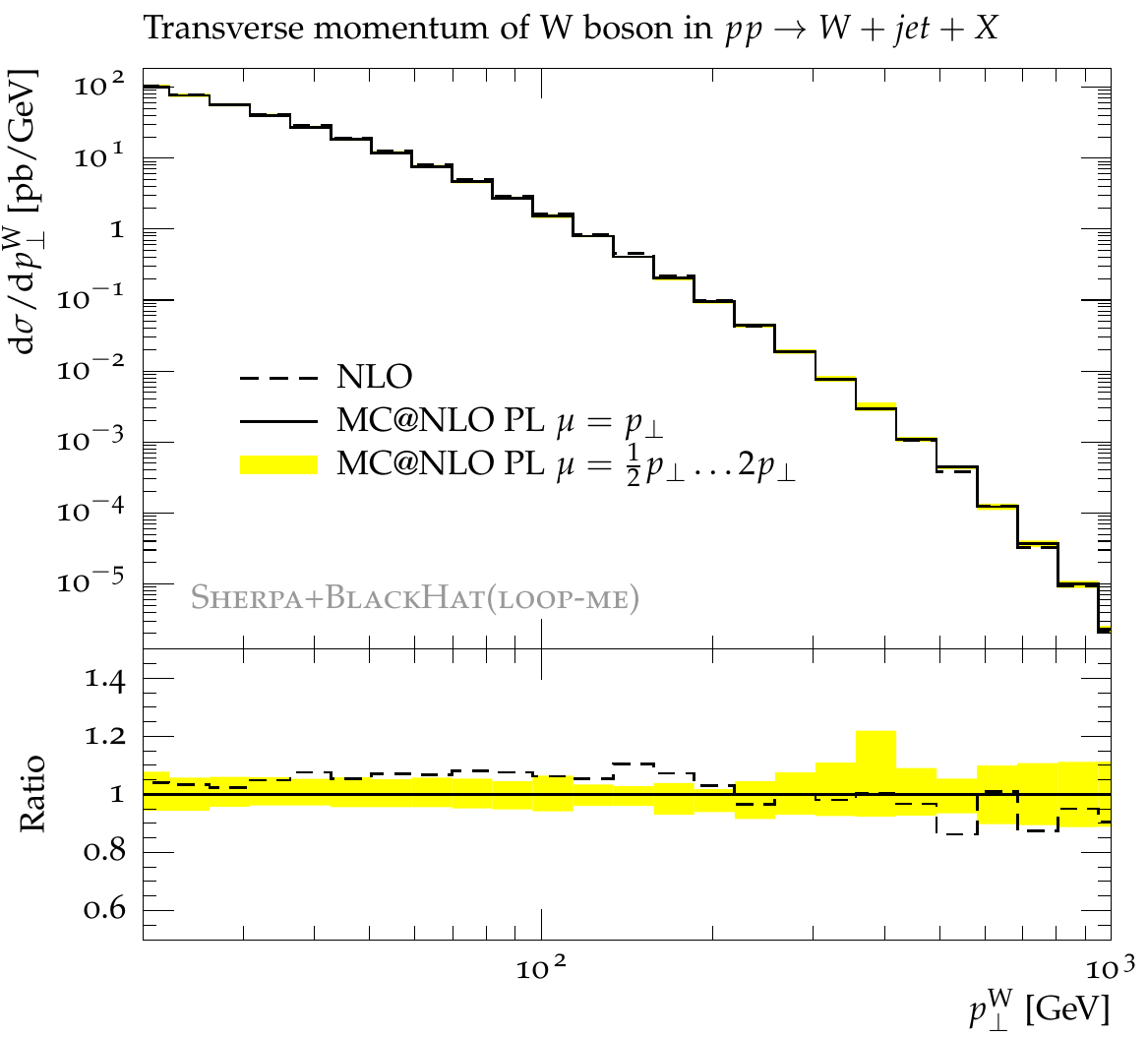}
  \includegraphics[width=0.45\textwidth]{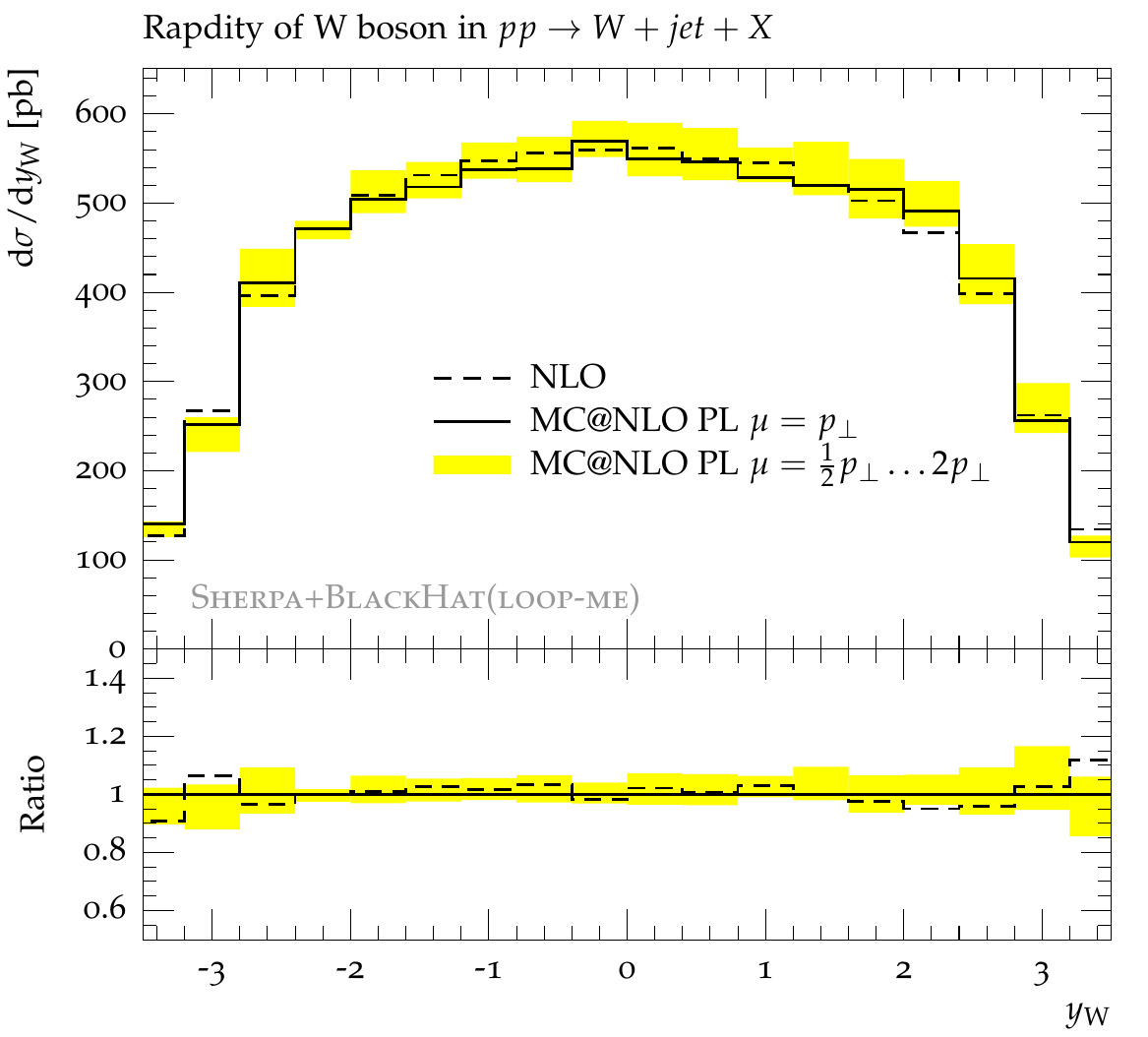}
  \caption{Transverse momentum and rapidity of the $W$-boson in $W[\to e\nu]+j$ production at the LHC.}
  \label{fig:fixed:w}
\end{figure}

\begin{figure}[t!]
  \centering
  \includegraphics[width=0.45\textwidth]{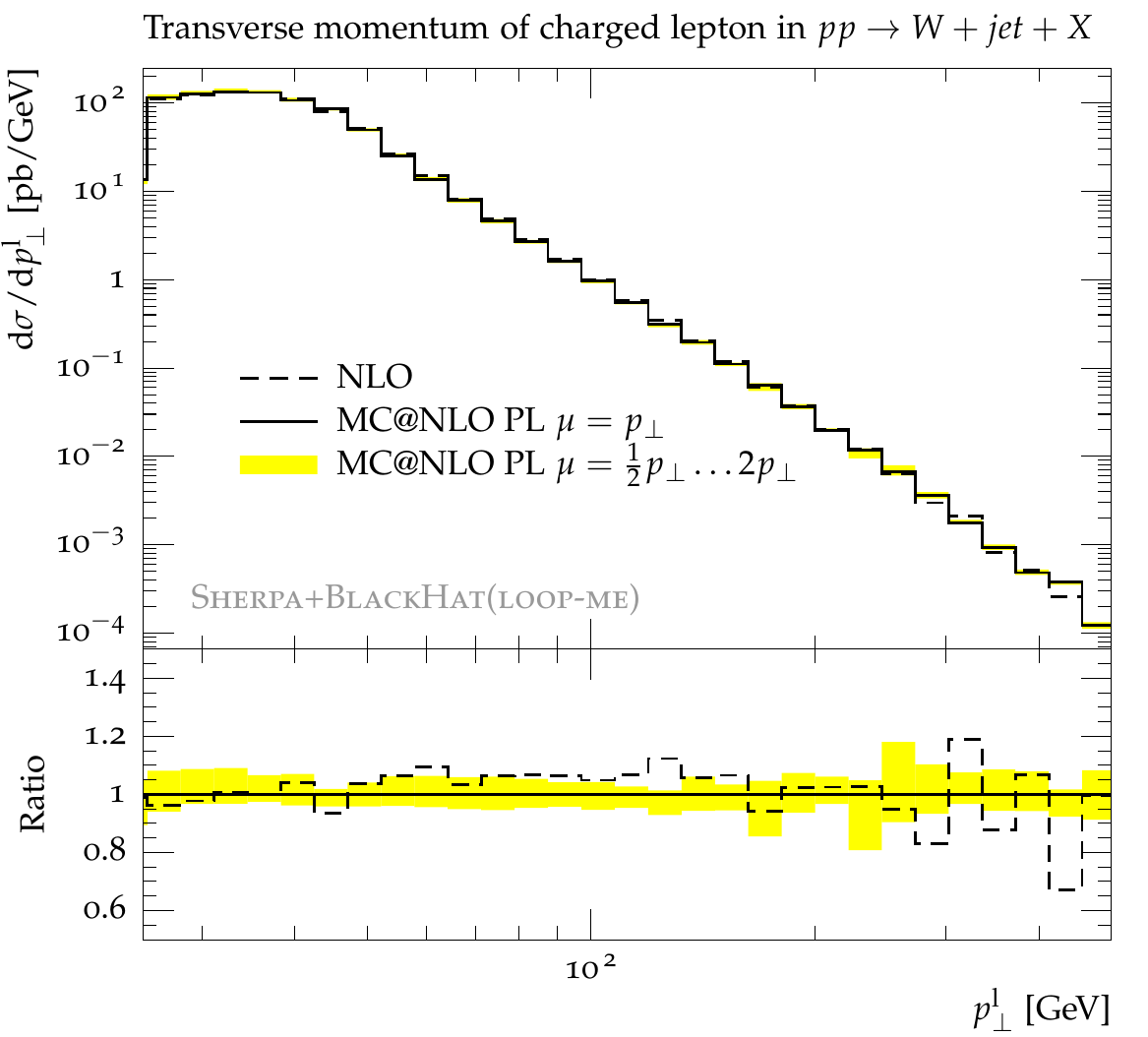}\nolinebreak
  \includegraphics[width=0.45\textwidth]{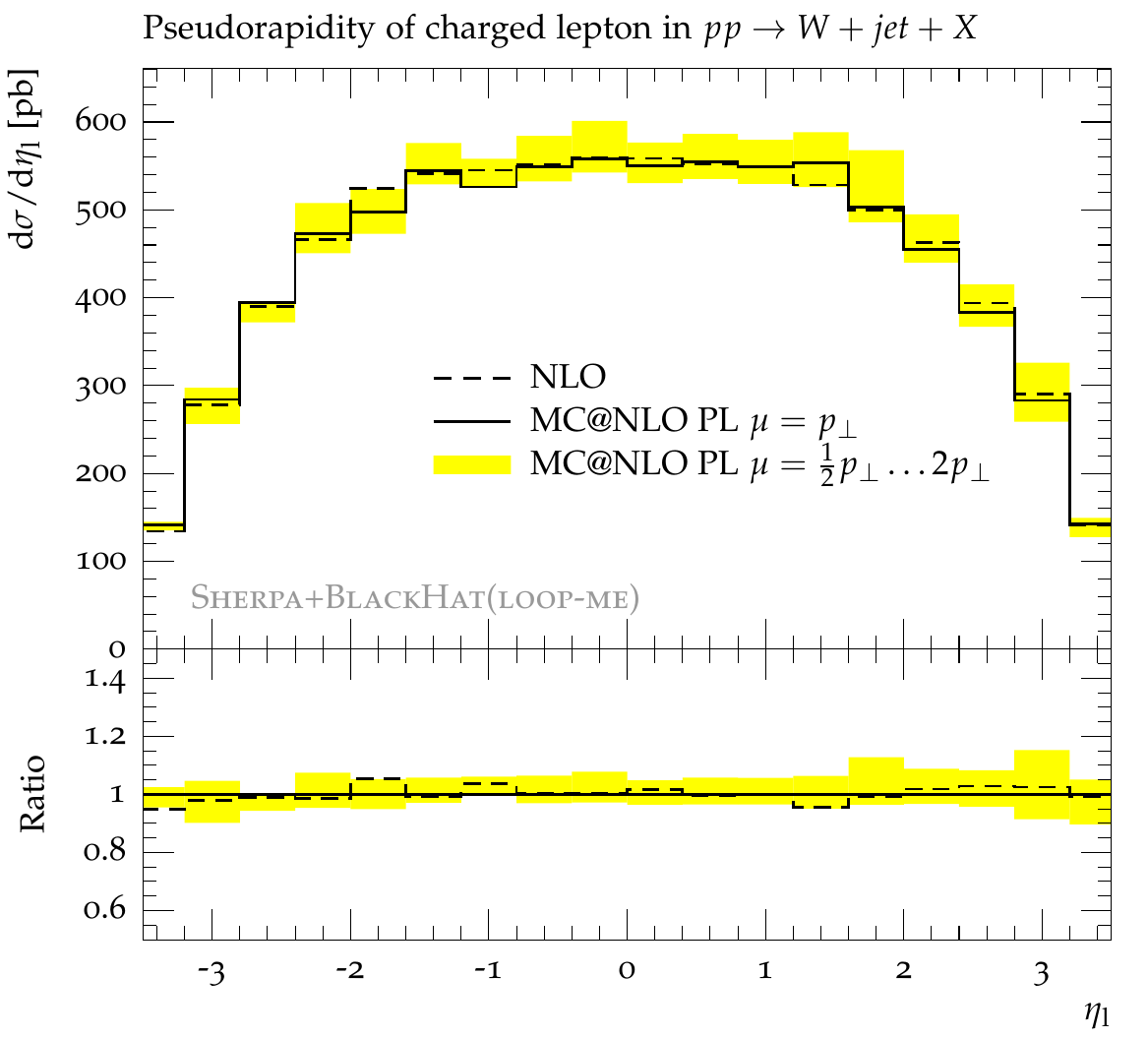}
  \caption{Transverse momentum and pseudorapidity of the electron in $W[\to e\nu]+j$ production at the LHC.}
  \label{fig:fixed:wlept}
\end{figure}

Properties of the $W$-boson, which is reconstructed
from the truth neutrino and the lepton, are displayed in Fig.~\ref{fig:fixed:w}.
Agreement between fixed-order and \MCatNLO results in both the rapidity 
and transverse momentum spectra is found.
Agreement is also found for the transverse momentum and pseudorapidity
distributions of the electron, which are shown in Fig.~\ref{fig:fixed:wlept}.

\begin{figure}[t!]
  \centering
  \includegraphics[width=0.45\textwidth]{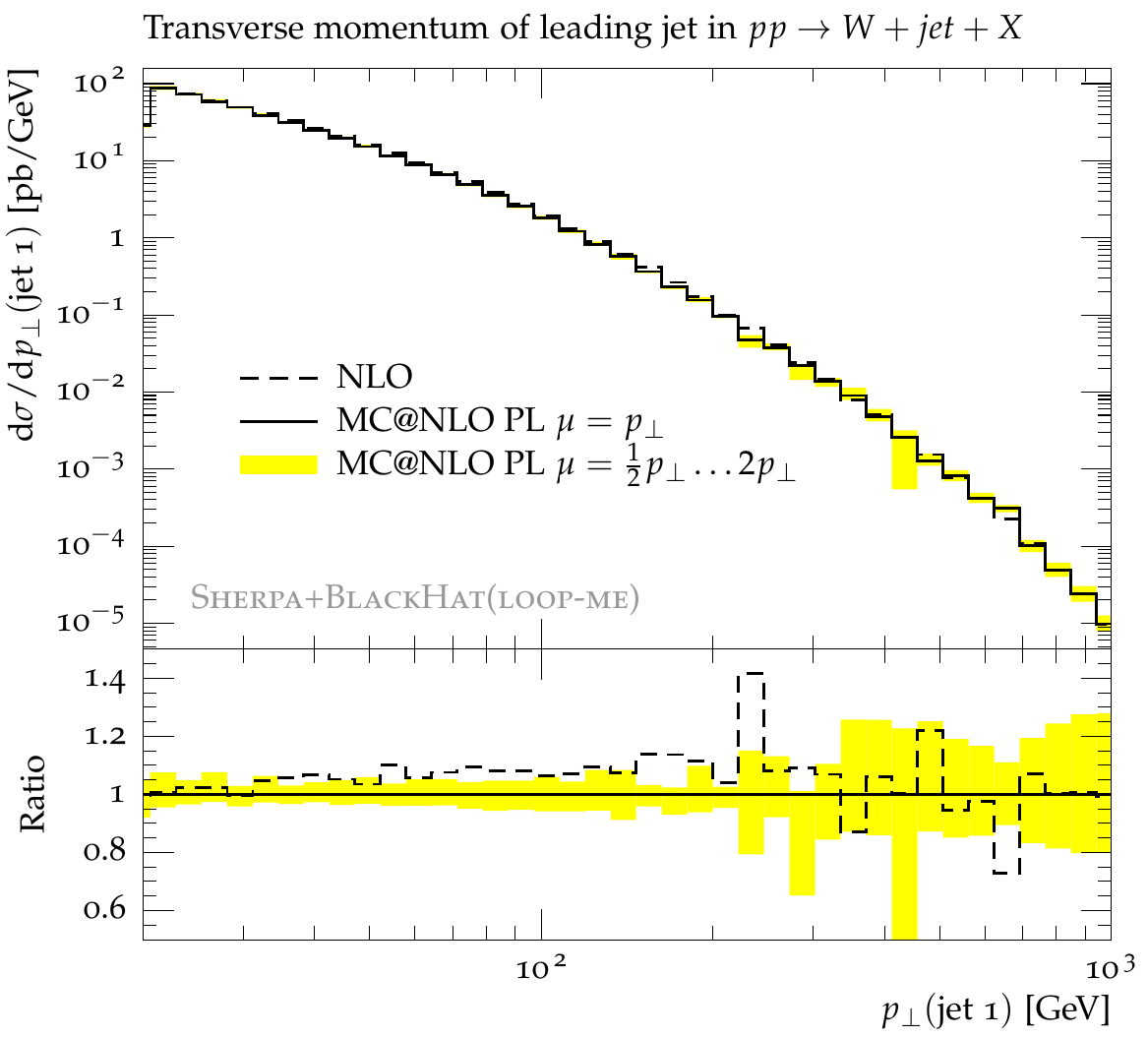}\nolinebreak
  \includegraphics[width=0.45\textwidth]{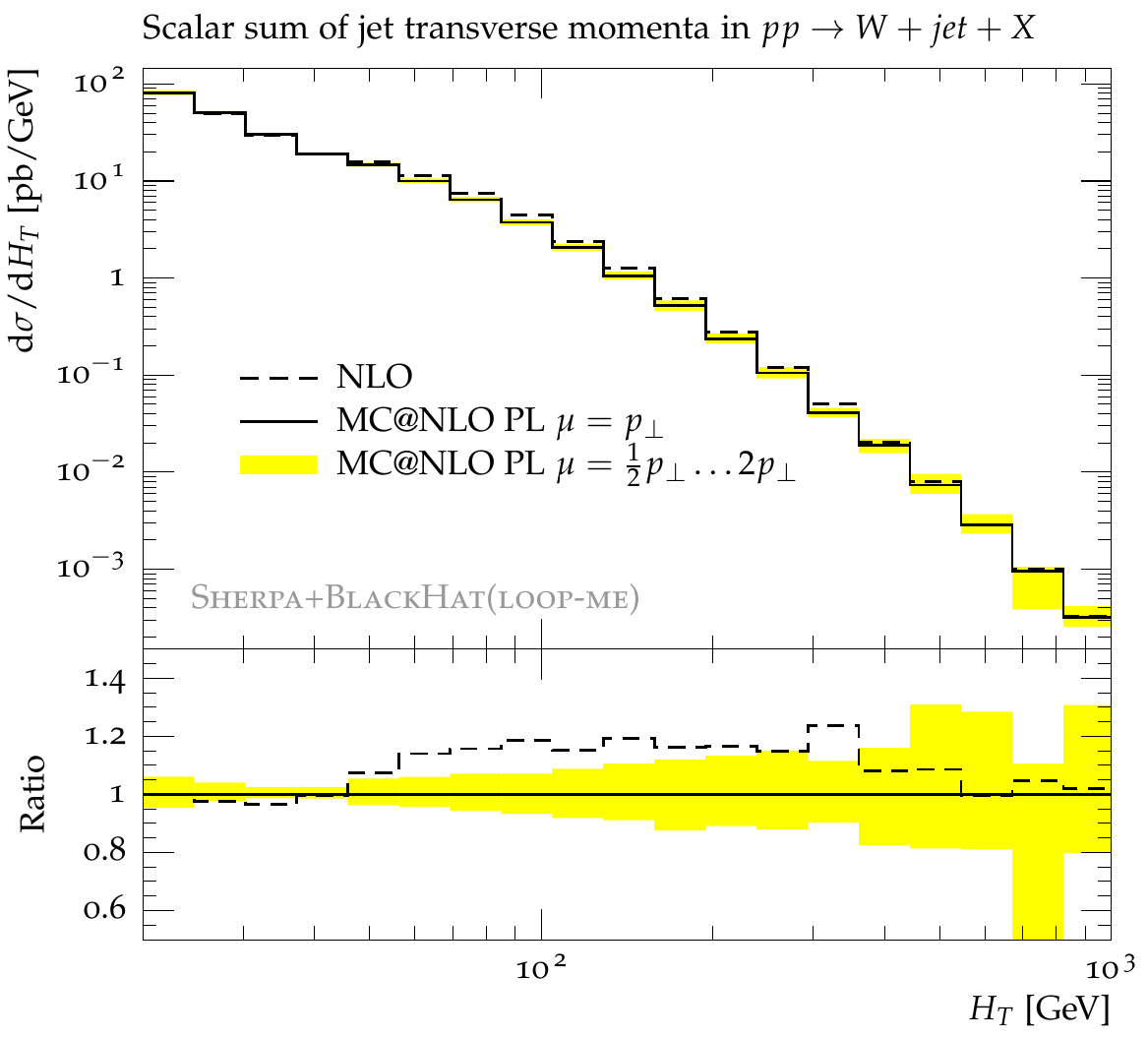}
  \caption{Transverse momentum of the leading jet and scalar sum of all jet transverse momenta
    in $W[\to e\nu]+j$ production at the LHC.}
  \label{fig:fixed:wjets}
\end{figure}

In the event analysis jets are defined according to the inclusive $k_\perp$ 
algorithm~\cite{Catani:1993hr,*Ellis:1993tq,*Cacciari:2005hq}
with $R=0.7$ and requiring $p_\perp>20$~GeV. Properties of these jets are
displayed in Fig.~\ref{fig:fixed:wjets}.
The transverse momentum of the leading jet, an observable described at
next-to-leading order accuracy, shows good agreement between both approaches.
The scalar sum of jet transverse momenta ($H_T$) is only described at
leading-order accuracy and thus suffers from larger exponentiation
uncertainties. This leads to a disagreement between the
fixed-order result and \MCatNLO, up to the level of $20\%$.

No distribution is affected significantly by scale variations beyond the change
in total rate. This indicates that the functional form of the scale choice
described at the beginning of this section works well for this process.

\section{Non-perturbative uncertainties and comparison to data}
\label{sec:nonperturbative}

In this section the \MCatNLO method is further studied using $W$+jet and $Z$+jet production
as a testing ground. Emphasis is now placed on the investigation of non-perturbative
effects and associated systematic variations.

\subsection{Analysis of non-perturbative effects}

\begin{figure}[t]
  \centering
  \includegraphics[width=0.45\textwidth]{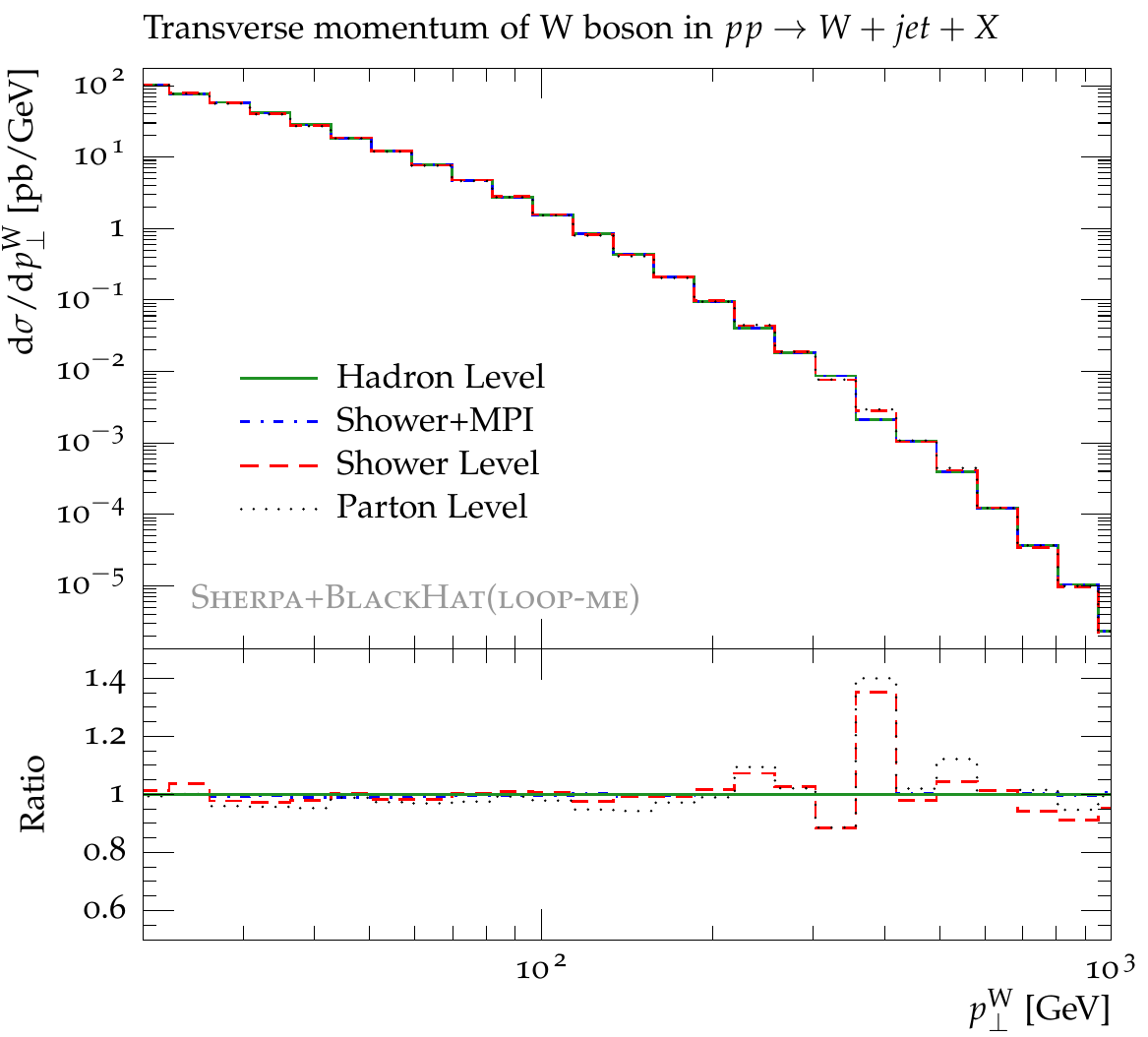}
  \includegraphics[width=0.45\textwidth]{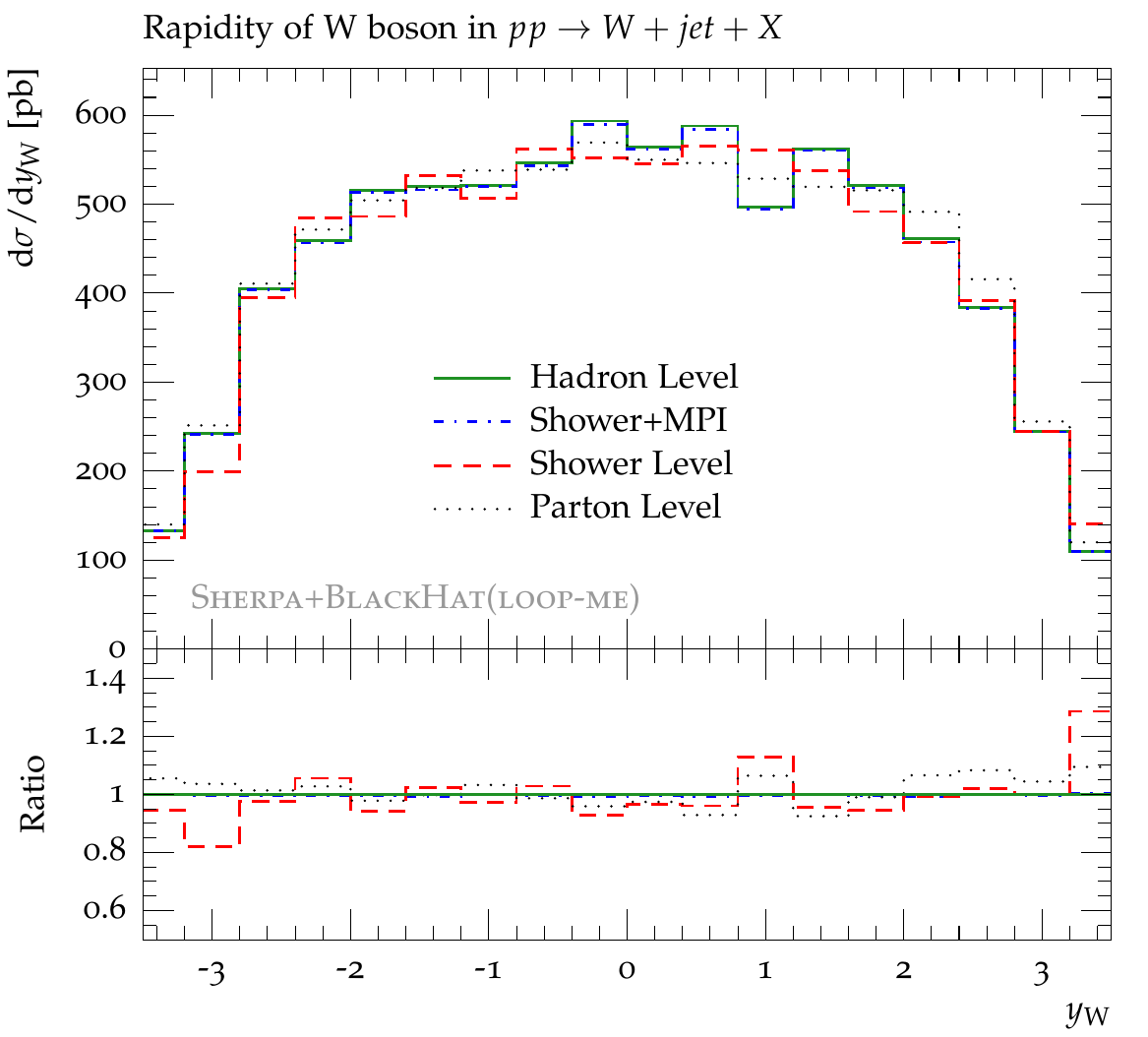}
  \caption{Transverse momentum and rapidity spectrum of the $W$-boson
    in $W[\to e\nu]+j$ production at the LHC.}
    \label{fig:nonpert:w}
\end{figure}

The \MCatNLO method allows to generate fully hadronised events as an input for detector 
simulation or for direct comparison to measurements at the particle level. A question that 
naturally arises is, whether the theoretical uncertainties of the full \MCatNLO simulation 
are then dominated by the perturbative or non-perturbative effects. We do not attempt 
to judge on this question here as it is not obvious on which grounds they can be compared
at all, but rather point out that both non-perturbative corrections
and non-perturbative uncertainties are modest compared to the intrinsic uncertainties of
the parton-level result, which were investigated in the previous section.

\begin{figure}[p]
  \centering
  \includegraphics[width=0.45\textwidth]{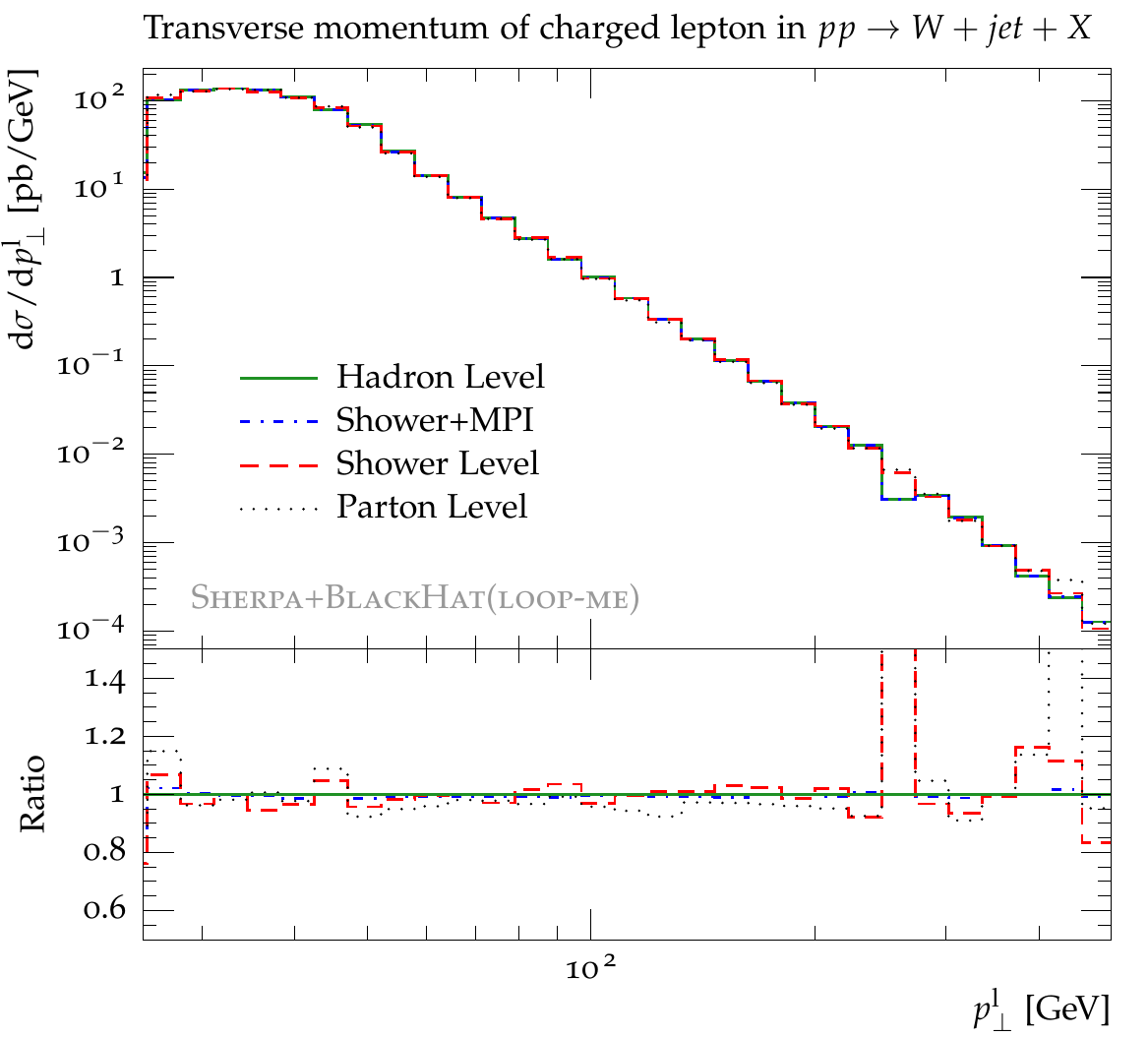}\nolinebreak
  \includegraphics[width=0.45\textwidth]{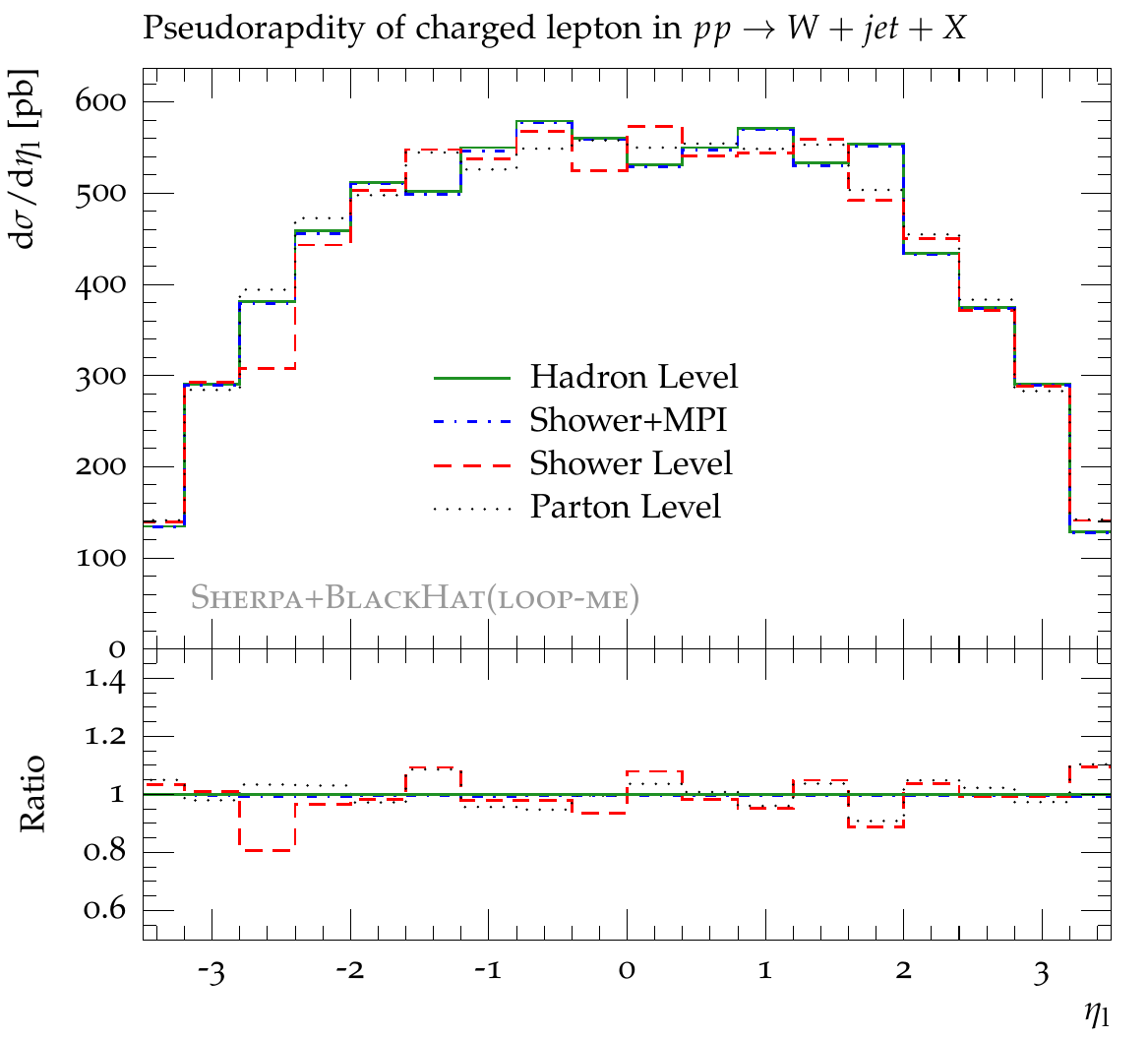}
  \caption{Transverse momentum and pseudorapidity spectrum of the electron 
    in $W[\to e\nu]+j$ production at the LHC.}
  \label{fig:nonpert:wlept}
\end{figure}

\begin{figure}[p]
  \centering
  \includegraphics[width=0.45\textwidth]{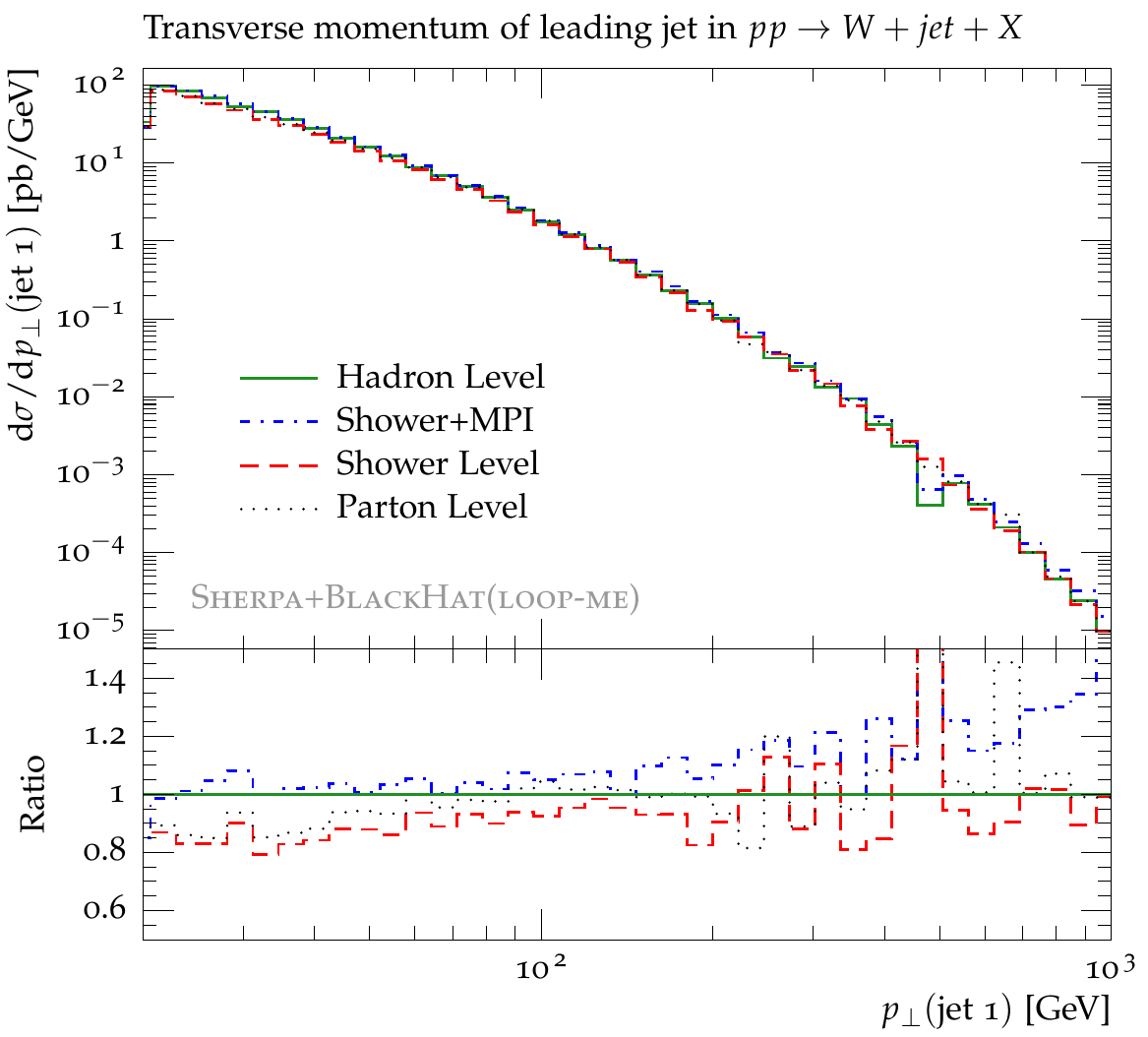}\nolinebreak
  \includegraphics[width=0.45\textwidth]{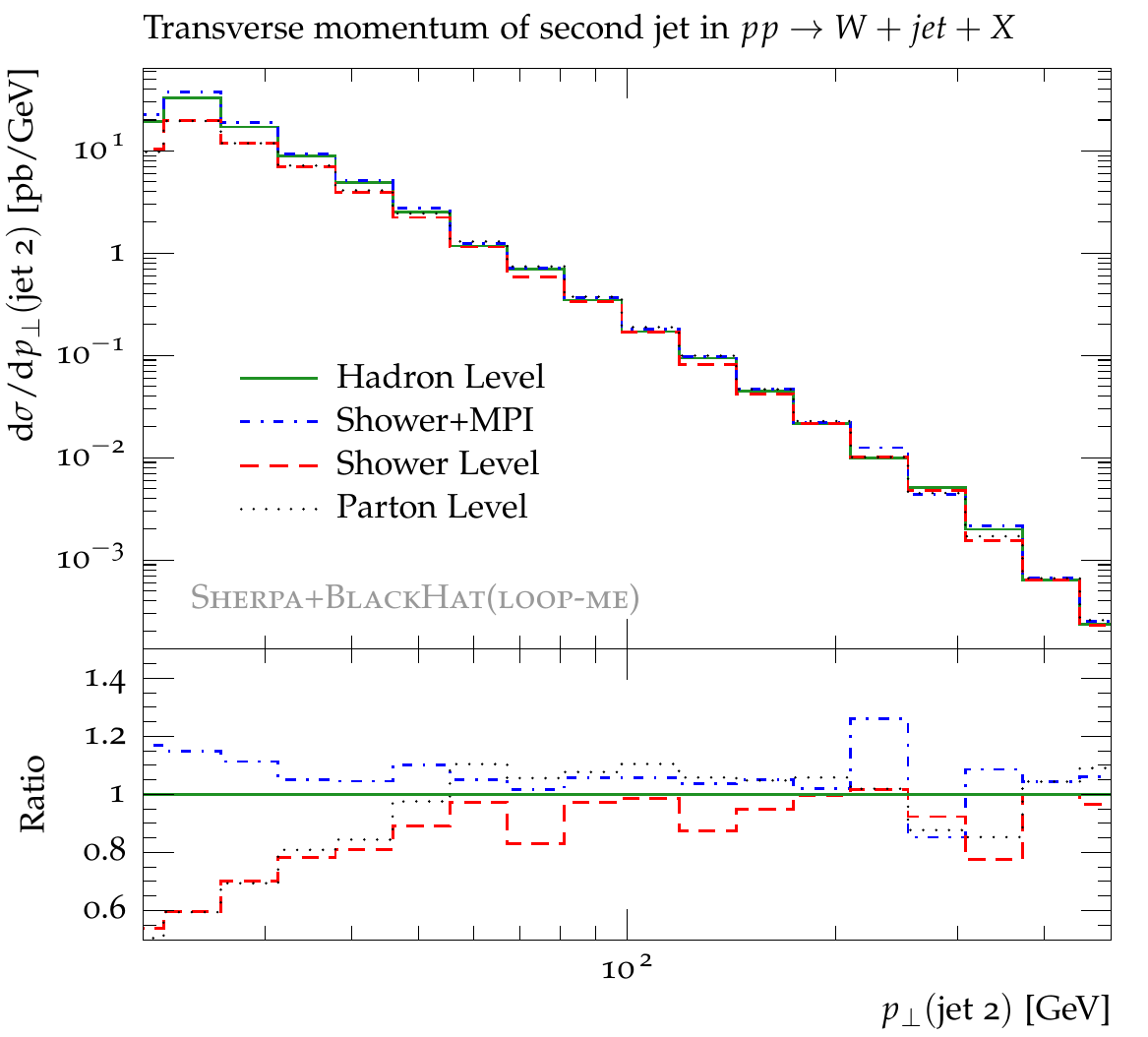}\\
  \includegraphics[width=0.45\textwidth]{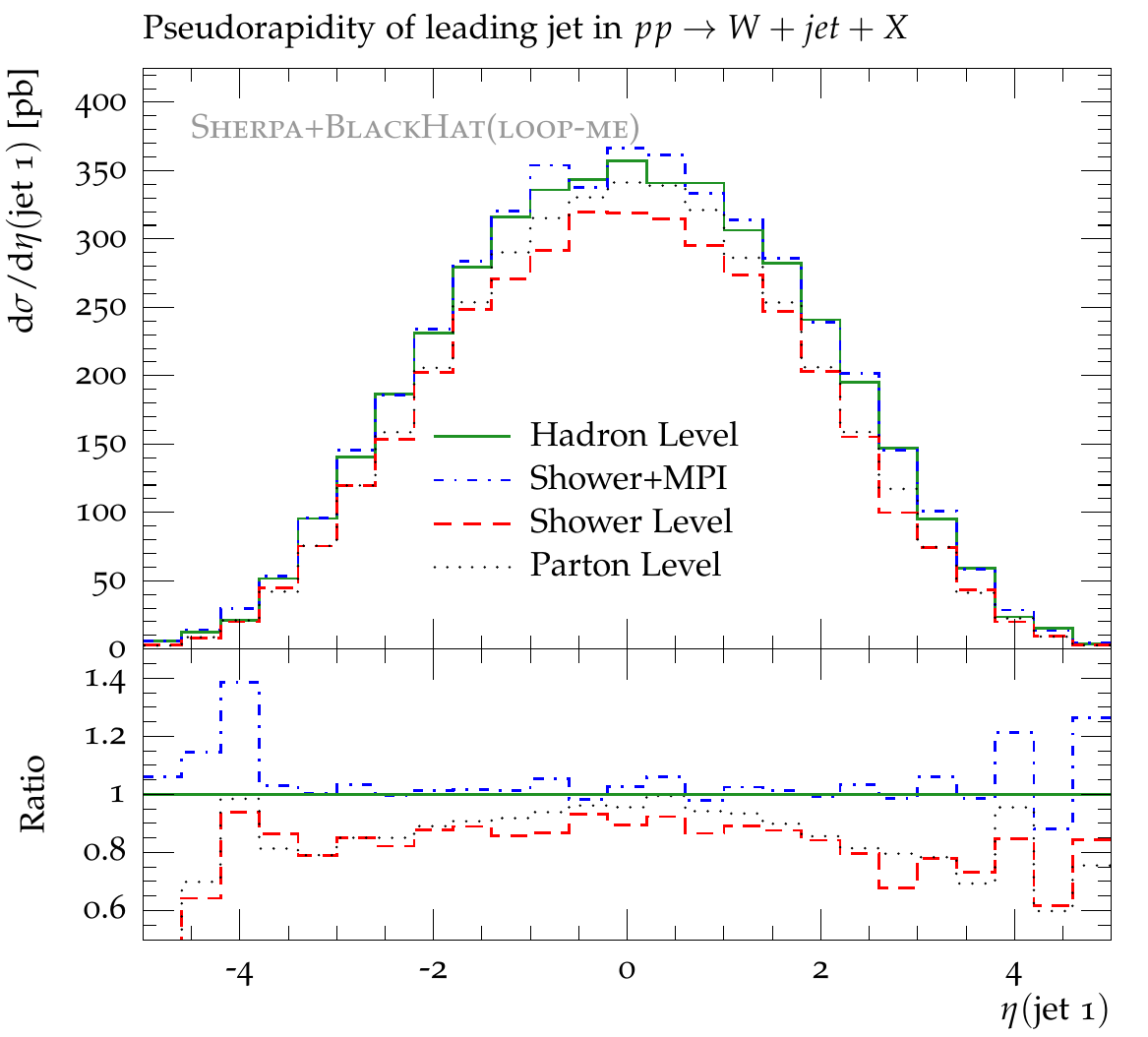}\nolinebreak
  \includegraphics[width=0.45\textwidth]{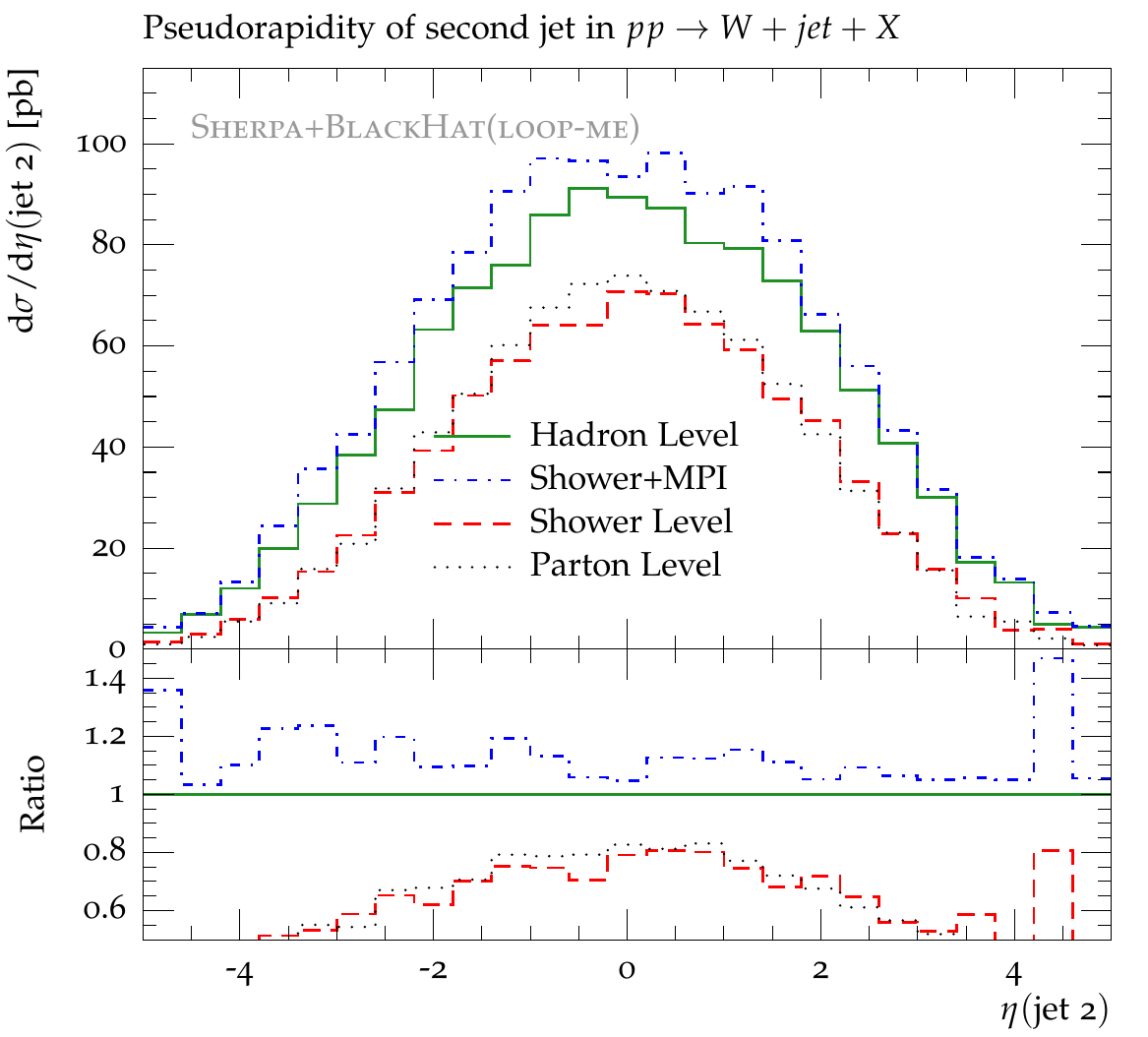}\\
  \caption{Transverse momenta and pseudorapidities of the two
    leading jets in $W[\to e\nu]+j$ production at the LHC.
  \label{fig:nonpert:wjetpt}}
\end{figure}

\begin{figure}[p]
  \centering
  \includegraphics[width=0.45\textwidth]{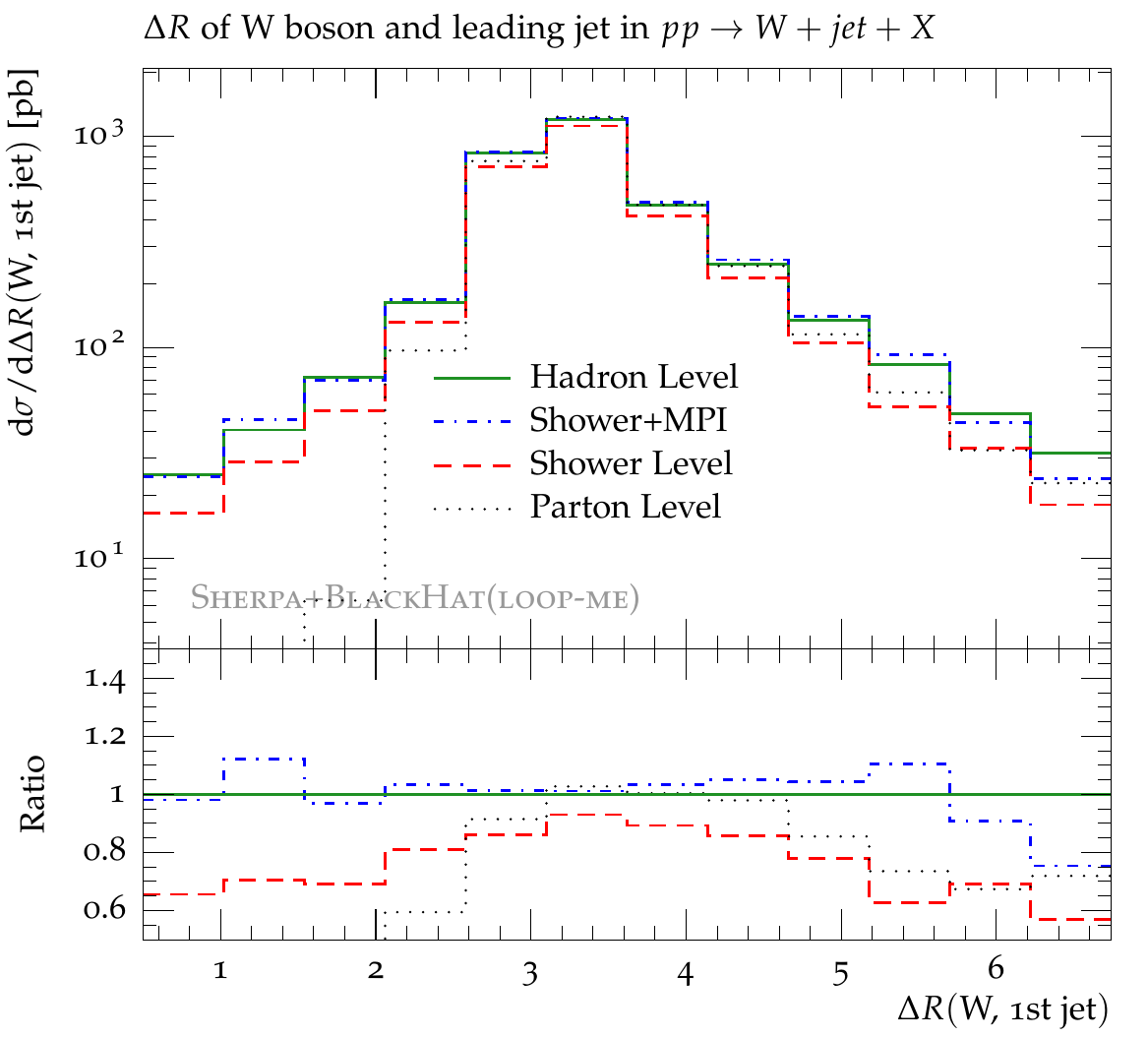}\nolinebreak
  \includegraphics[width=0.45\textwidth]{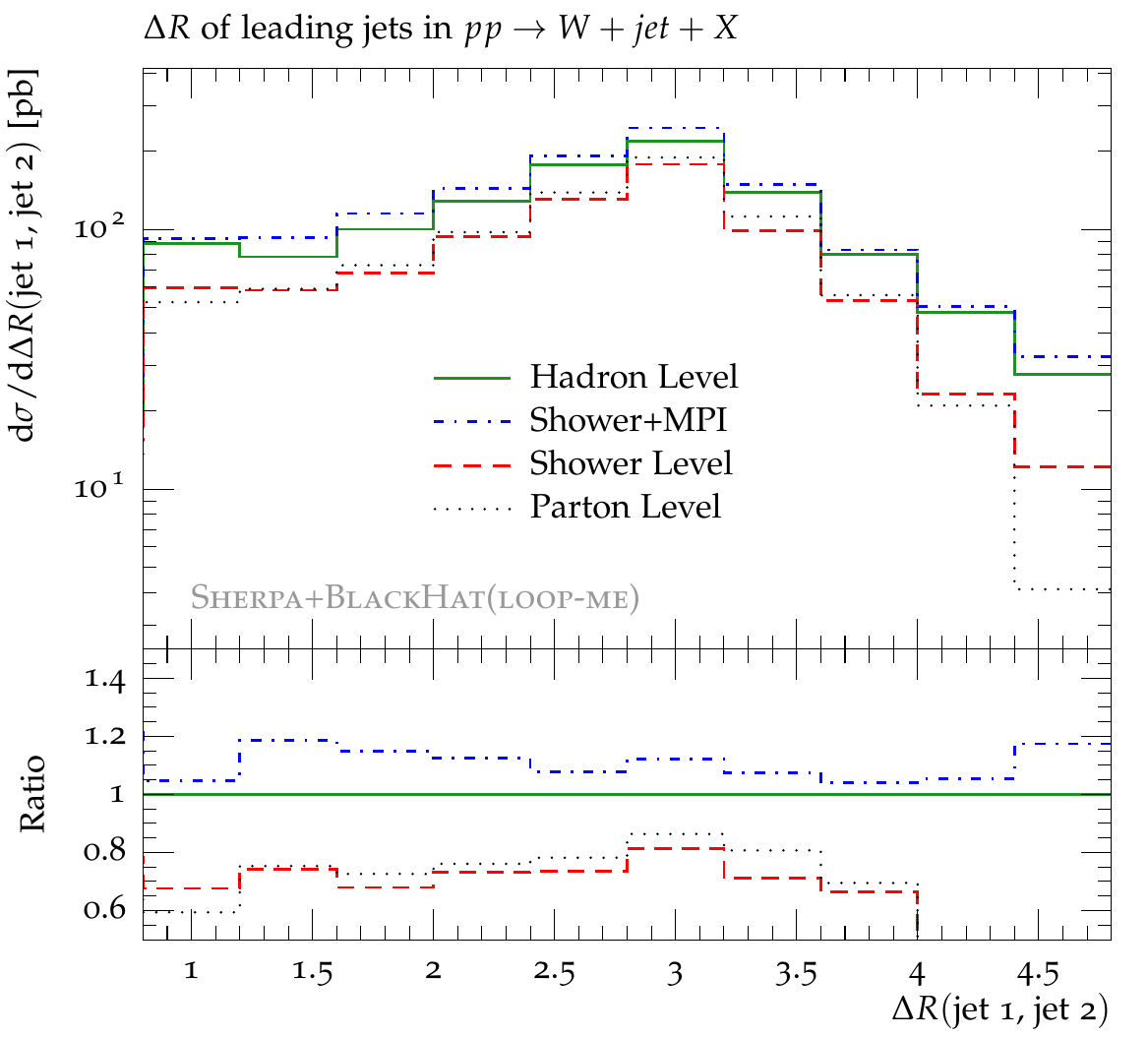}
  \caption{Correlations between $W$-boson and leading jets
    in $W[\to e\nu]+j$ production at the LHC.}
  \label{fig:nonpert:wjet12}
\end{figure}

\begin{figure}[p]
  \centering
  \includegraphics[width=0.45\textwidth]{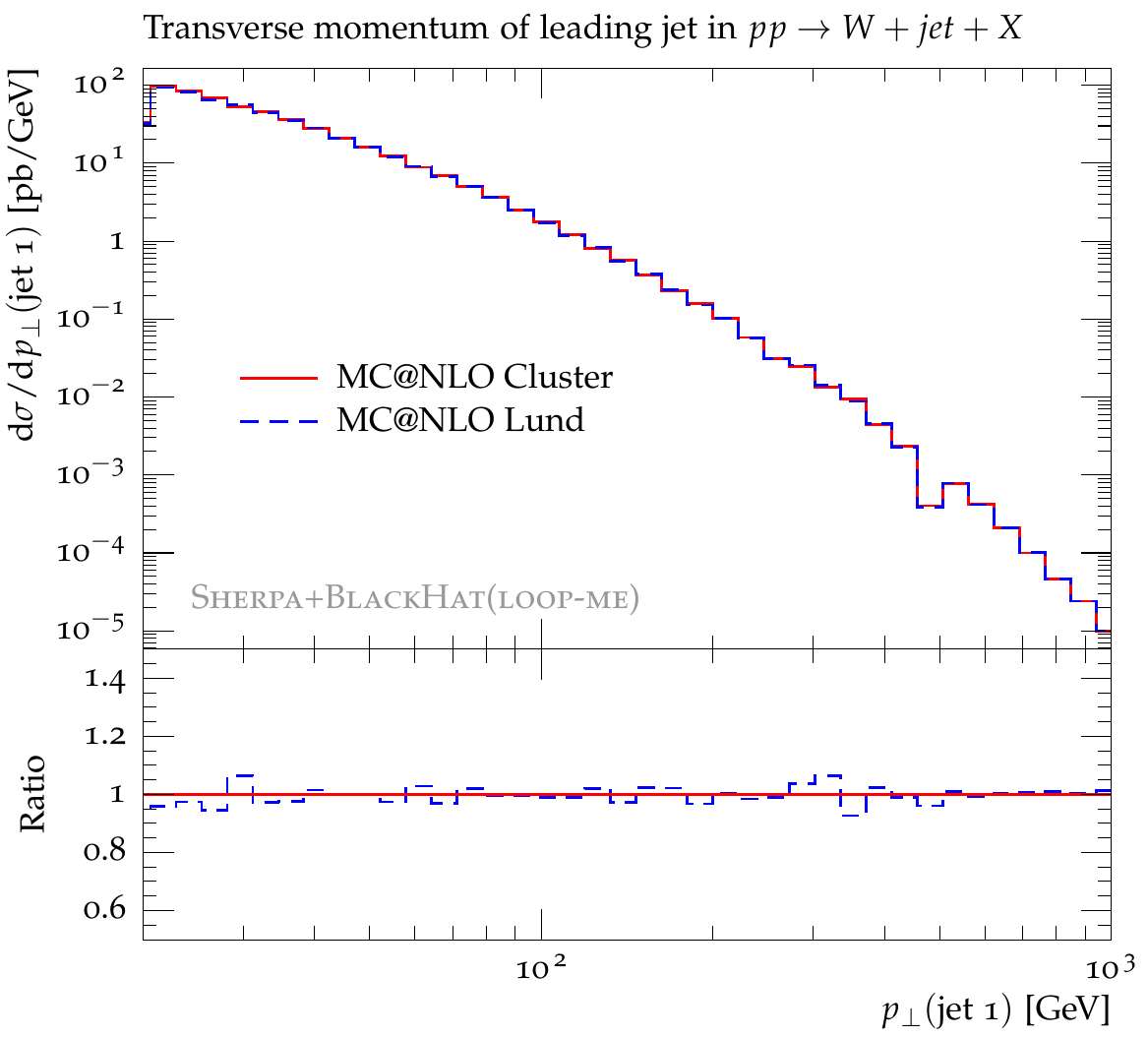}\nolinebreak
  \includegraphics[width=0.45\textwidth]{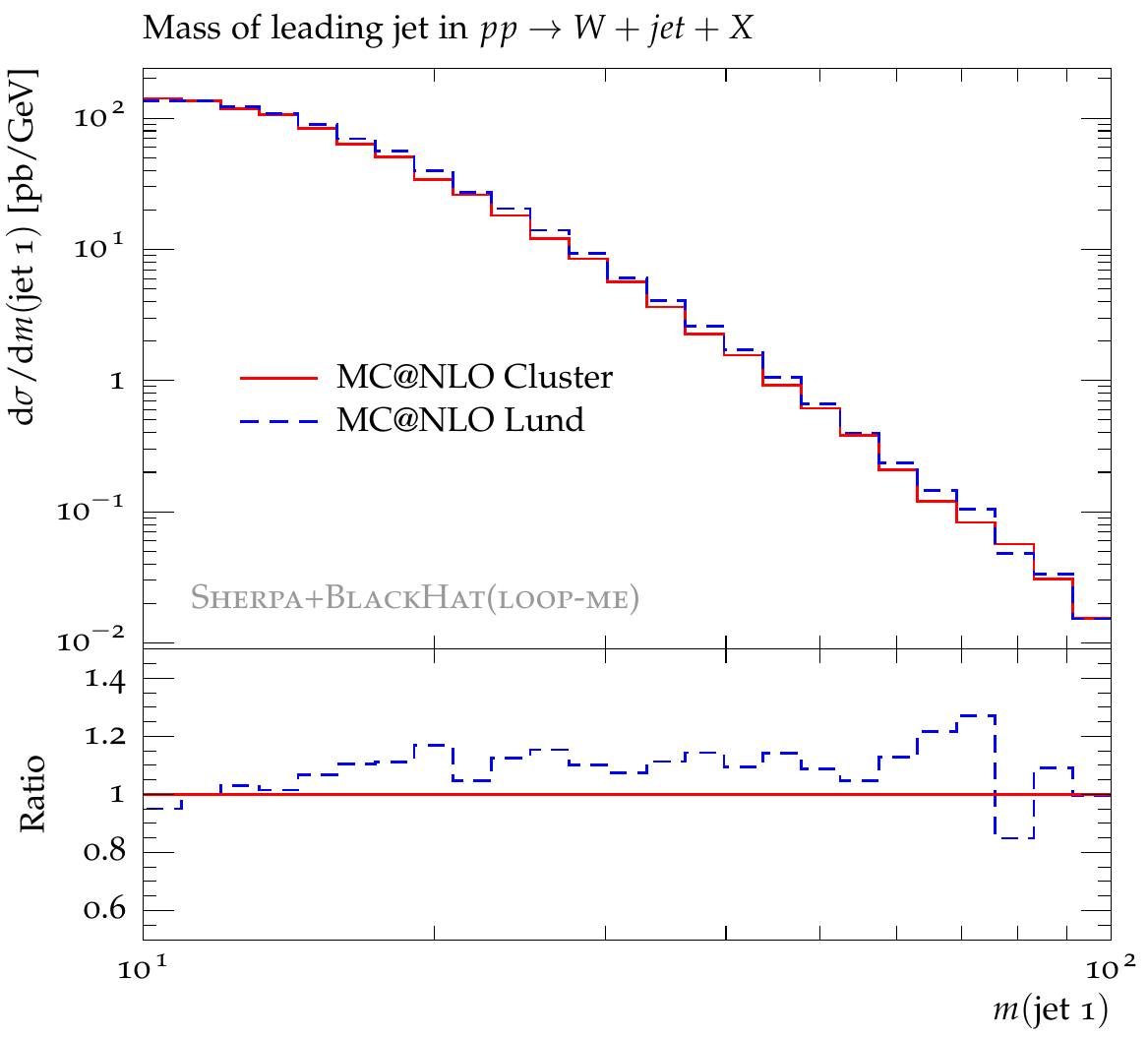}\\
  \includegraphics[width=0.45\textwidth]{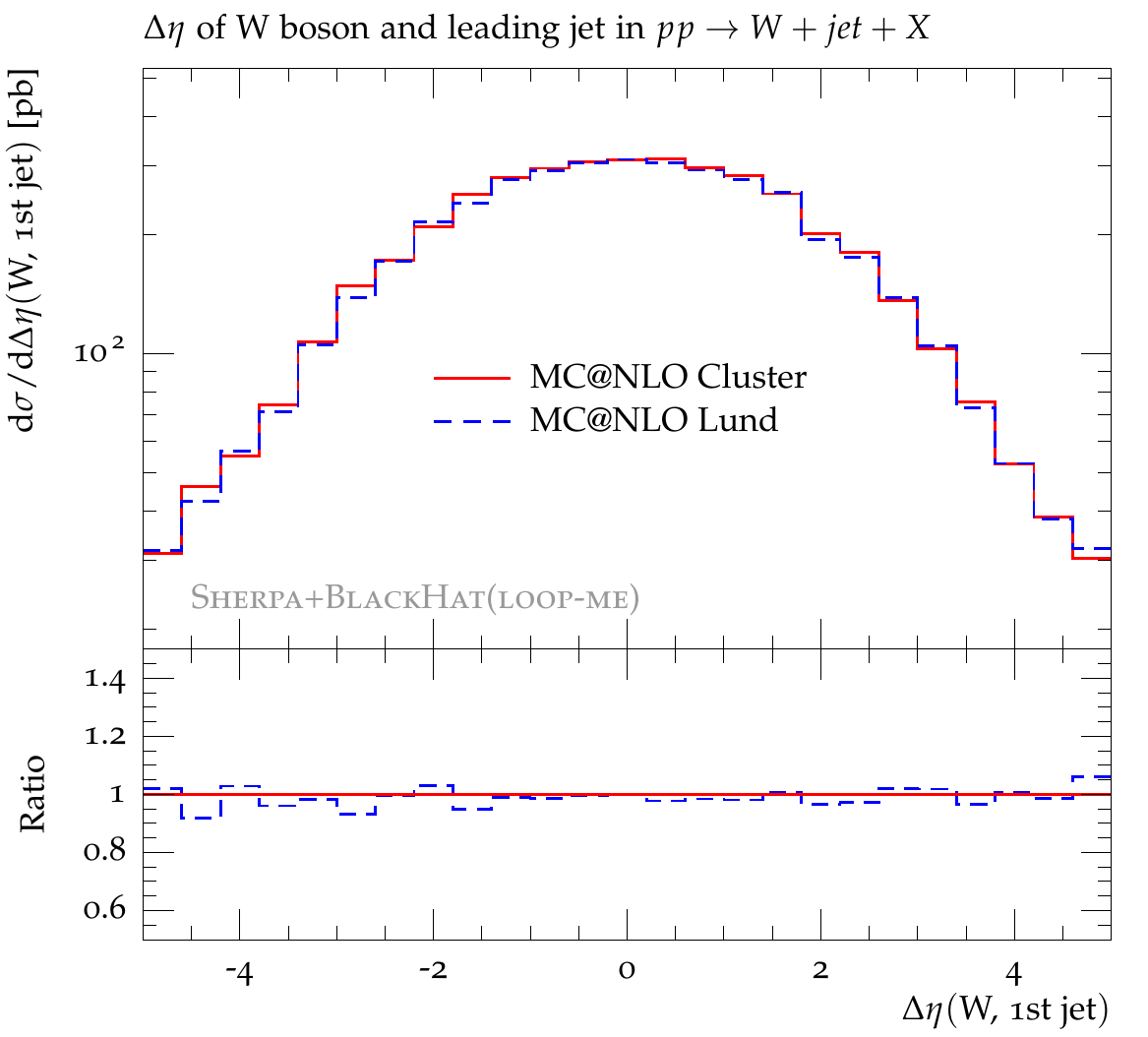}\nolinebreak
  \includegraphics[width=0.45\textwidth]{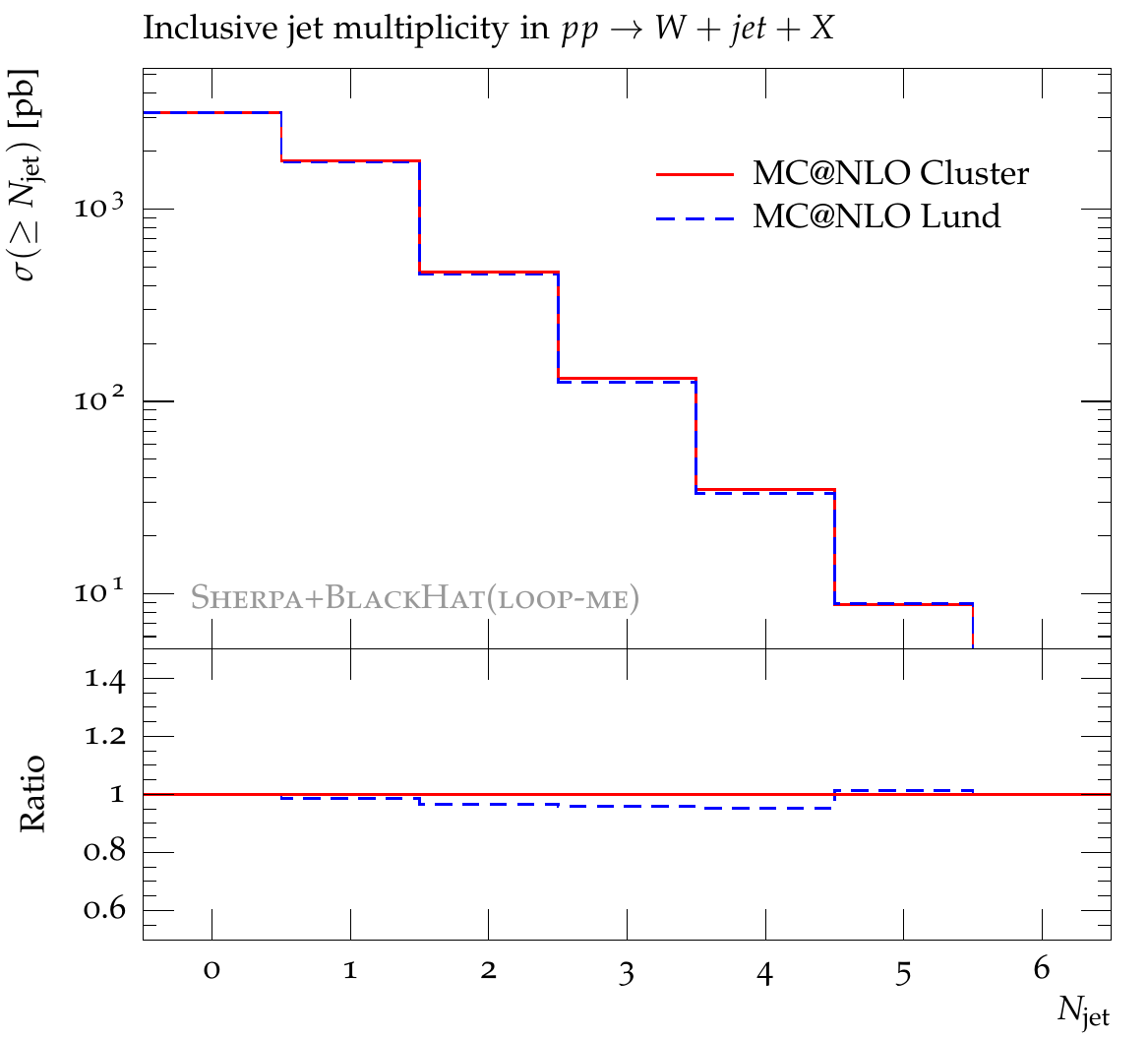}\\
  \caption{Hadronisation uncertainties for different observables studied in
    $W[\to e\nu]+j$ production at the LHC.}
  \label{fig:nonpert:lund}
\end{figure}

To this end, \MCatNLO simulations for $W$+jet production are compared with
a varying level of non-perturbative effects included:
\begin{description}
\item[``Parton Level'']\hfill \\
  Only the first emission off $\mb{S}$-events in \MCatNLO is generated in
  addition to the seed event. This is the same method that was used in the 
  comparison to fixed-order results in Sec.~\ref{sec:pheno:nlo}.
\item[``Shower Level'']\hfill \\
  All QCD emissions in the parton shower and QED emissions in the YFS approach are
  taken into account.
\item[``Shower+MPI'']\hfill \\
  Multiple parton interactions and intrinsic transverse momentum of the beam hadron
  are additionally allowed for.
\item[``Hadron Level'']\hfill \\
  Hadronisation and hadron decays are included to generate events at the particle level.
\end{description}
All other event generation parameters have been chosen identical to
Sec.~\ref{sec:pheno:nlo}.

Non-jet observables, like the rapidity and transverse momentum of the $W$-boson
and the rapidity and transverse momentum of the charged lepton are virtually 
unaffected by non-perturbative effects, as expected. This is shown
in Figs.~\ref{fig:nonpert:w} and~\ref{fig:nonpert:wlept}.

Jet observables are far more sensitive, as is exemplified by the transverse momentum
and rapidity spectra of the leading and next-to-leading jet in Fig.~\ref{fig:nonpert:wjetpt}.
They show effects of ``out-of-cone'' radiation at shower level, which are partially 
compensated by the simulation of multiple scattering effects. Hadronisation again leads 
to softer jet spectra, such that the various effects can compensate each other.
Although the tendency of the corrections stays the same, their precise magnitude
depends on the jet algorithm and its parameters and will have to be 
investigated separately for each analysis.
Figure~\ref{fig:nonpert:wjet12} exemplifies how non-perturbative effects can distort 
correlations between the two hardest jets.

The uncertainties inherent in the hadronisation model were probed by switching
from the \Sherpa default cluster fragmentation model~\cite{Winter:2003tt,*Krauss:2010xy}
to the Lund string model~\cite{Andersson:1983ia,*Andersson:1998tv} in the
implementation of \Pythia{}~\cite{Sjostrand:2006za}. Both models have been tuned 
to data from LEP and excellent agreement has been achieved.
For this comparison to be meaningful the same perturbative events (i.e.~identical
random seeds) were subjected to the two hadronisation models. Effects from
statistical fluctuations are thus cancelled.

The differences were found to be negligible for the
observables studied here, as displayed in Fig.~\ref{fig:nonpert:lund}, except
for the specifically hadronisation-sensitive jet mass, where variations up to 20\% occur.

\subsection{\texorpdfstring{$Z$}{Z}+jet production compared to Tevatron data}

Let us now turn to the comparison of \MCatNLO predictions to data and a simple
assessment of the systematic uncertainties associated with a significant part
of the non-perturbative effects. Central predictions are made at the hadron
level with \Sherpa{}'s default tune. An uncertainty band is generated
by allowing for a variation of the MPI parameters within boundaries given by
existing measurements of the underlying event. All event generation parameters
have been chosen analogous to the $W$+jet case above.

\begin{figure}[tbp]
  \includegraphics[width=0.33\textwidth]{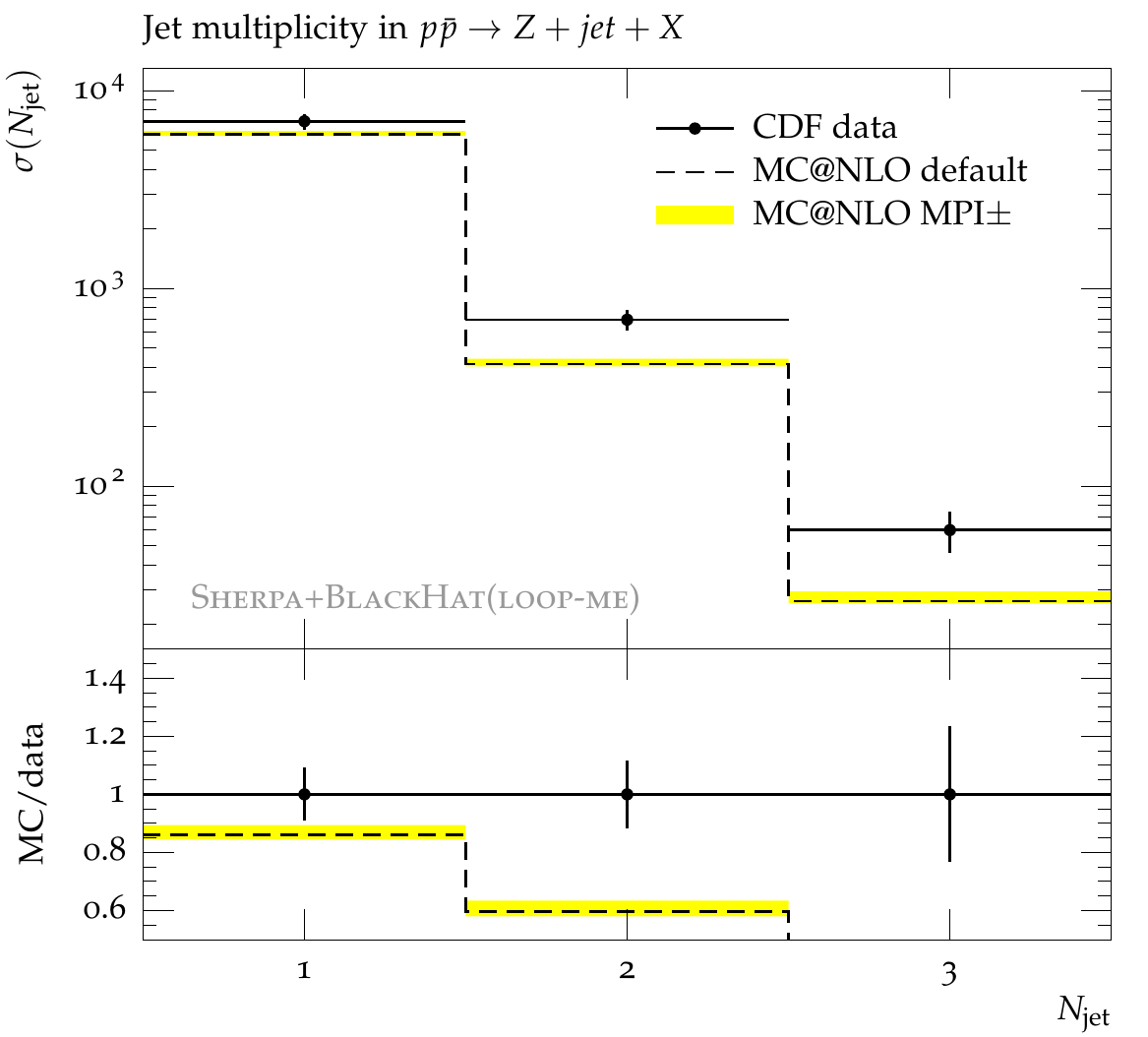}
  \includegraphics[width=0.33\textwidth]{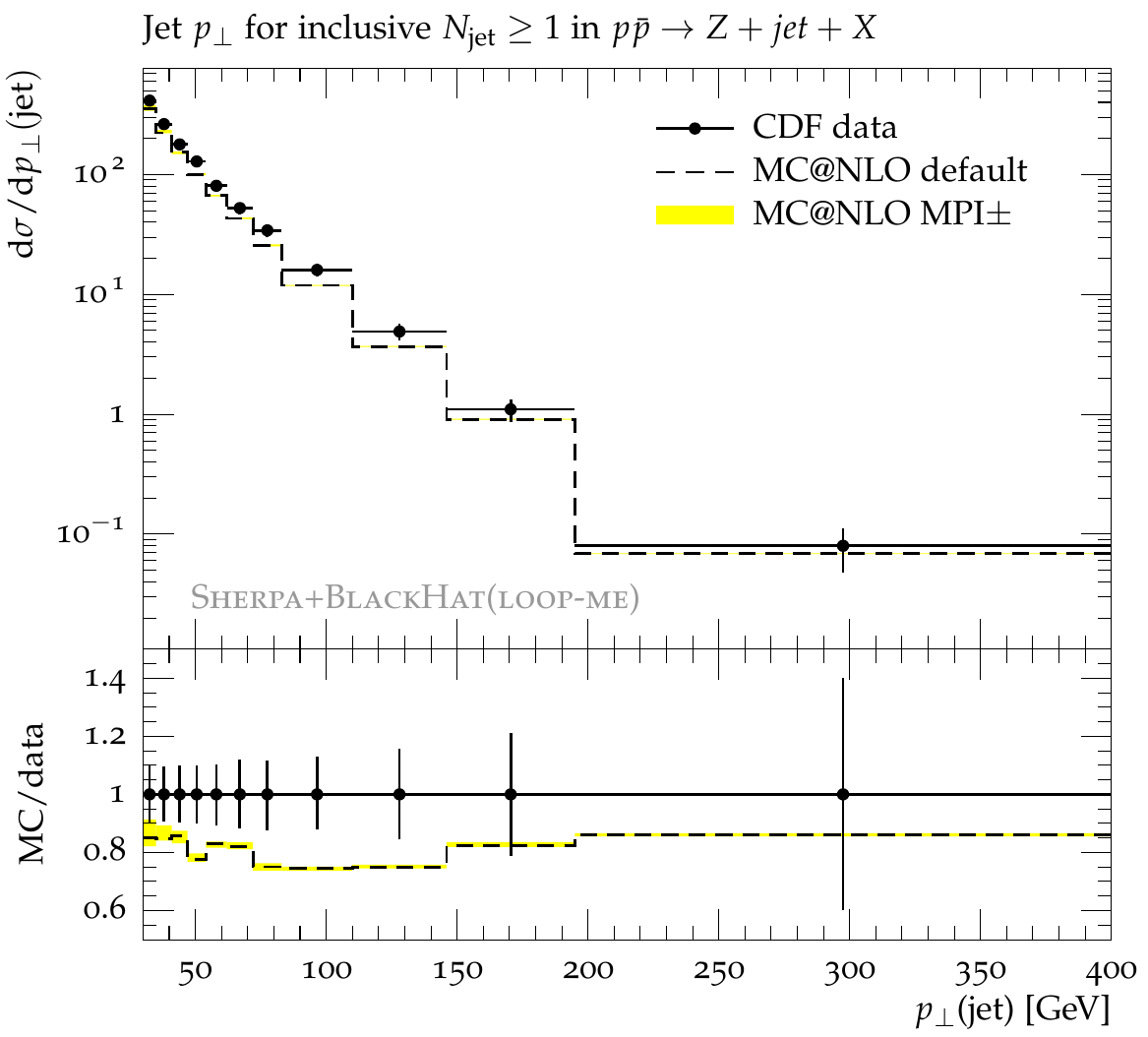}
  \includegraphics[width=0.33\textwidth]{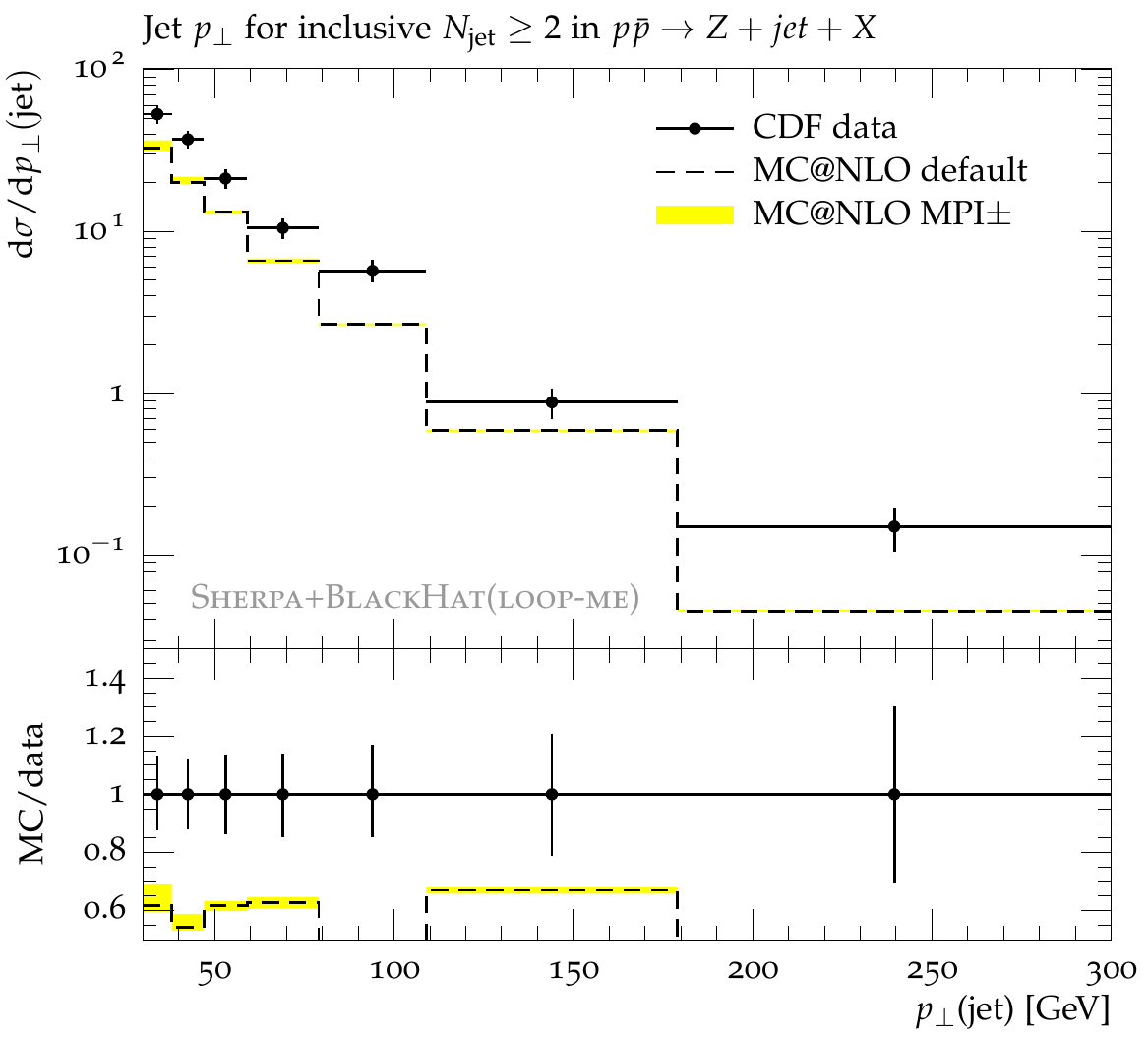}
  \caption{Inclusive jet cross sections and inclusive transverse momentum 
           distributions of all and all-but-the-hardest jets compared to CDF
           data \cite{:2007cp}.\label{Fig:Zj:CDF}}
\end{figure}

Figure~\ref{Fig:Zj:CDF} compares $Z$+jet production in the electron channel to a 
measurement from the CDF collaboration~\cite{:2007cp}. The reconstructed electrons are required to have 
transverse momenta $p_\perp>25$~GeV and an invariant mass of $66 < m_{ee} < 116$ GeV.
Jets are defined using the midpoint cone algorithm~\cite{Abulencia:2005yg} with $R=0.7$ 
and a split/merge fraction of 0.75. At least one jet with $p_\perp>30$ GeV and $|y|<2.1$
needs to be present and separated from both electrons by $\Delta R_{ej}>0.7$. 
The cross section in the one-jet bin is predicted slightly too low. 
The cross section of the two-jet bin is significantly underestimated, as it is determined
to leading order accuracy and subject to large uncertainties from the NLO+PS matching
procedure as discussed in Sec.~\ref{sec:perturbative}. The shape of the transverse momentum
distributions agrees well with data.

\begin{figure}[t]
  \centering
  \includegraphics[width=0.45\textwidth]{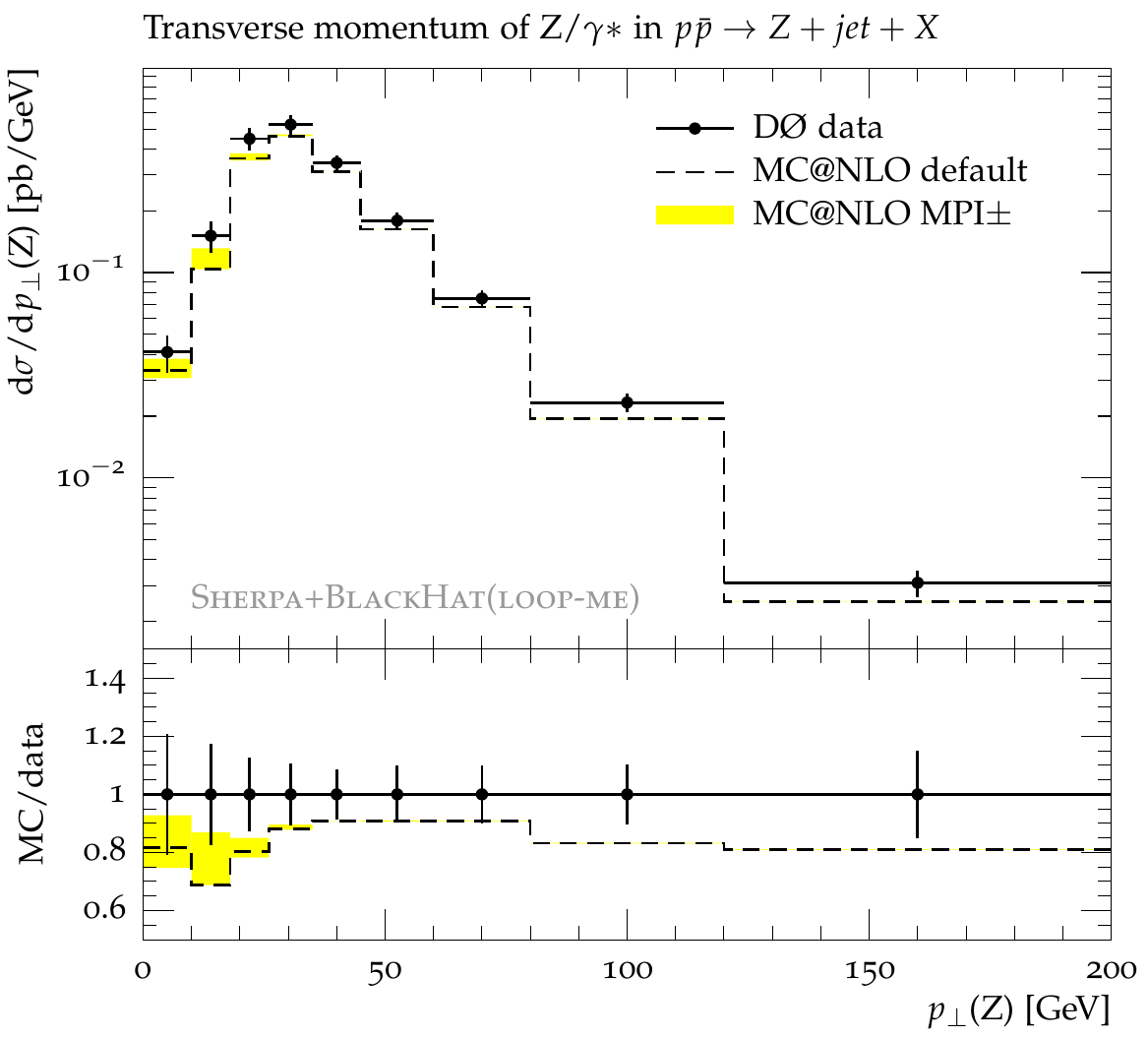}
  \includegraphics[width=0.45\textwidth]{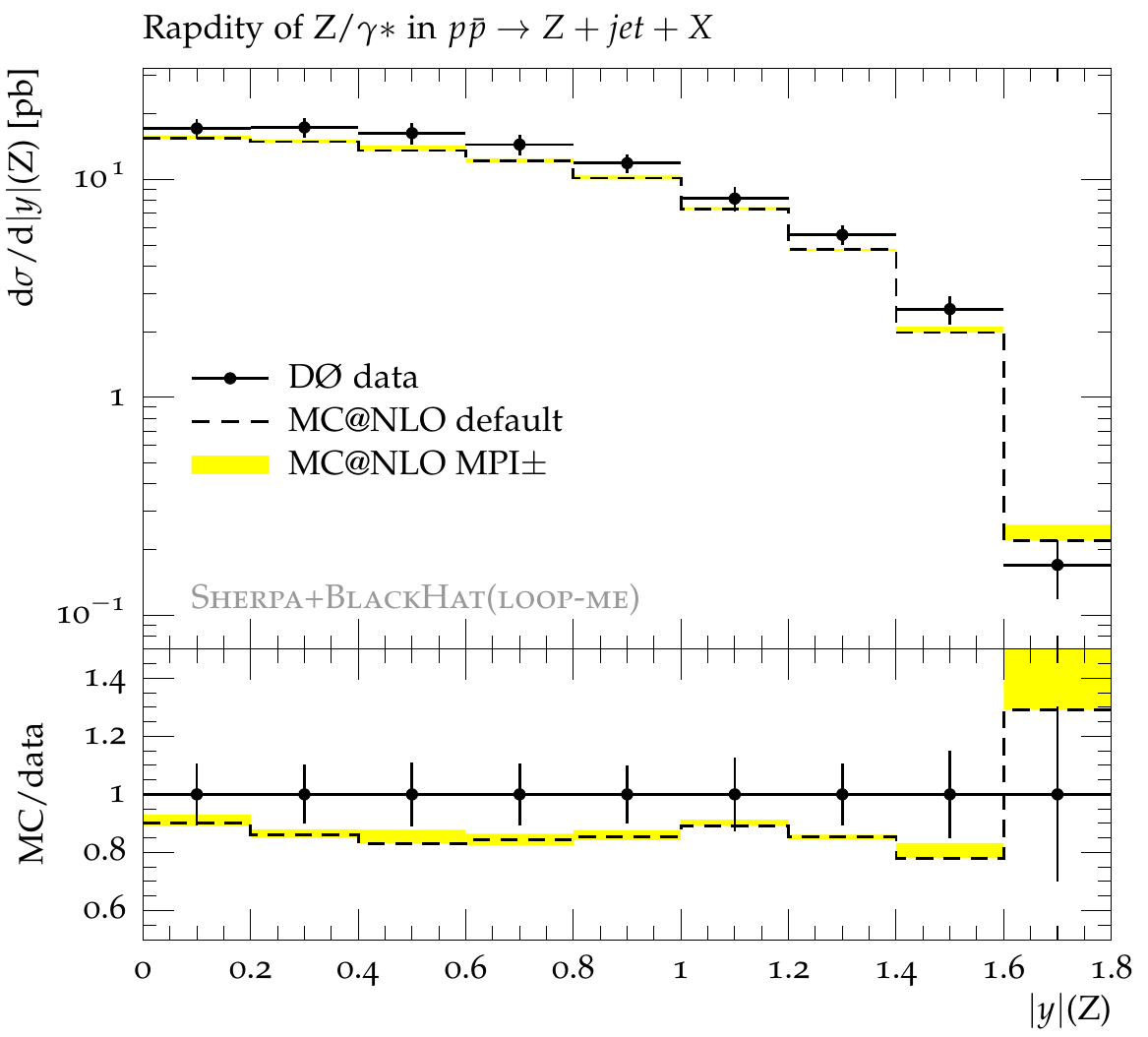}\\
  \includegraphics[width=0.45\textwidth]{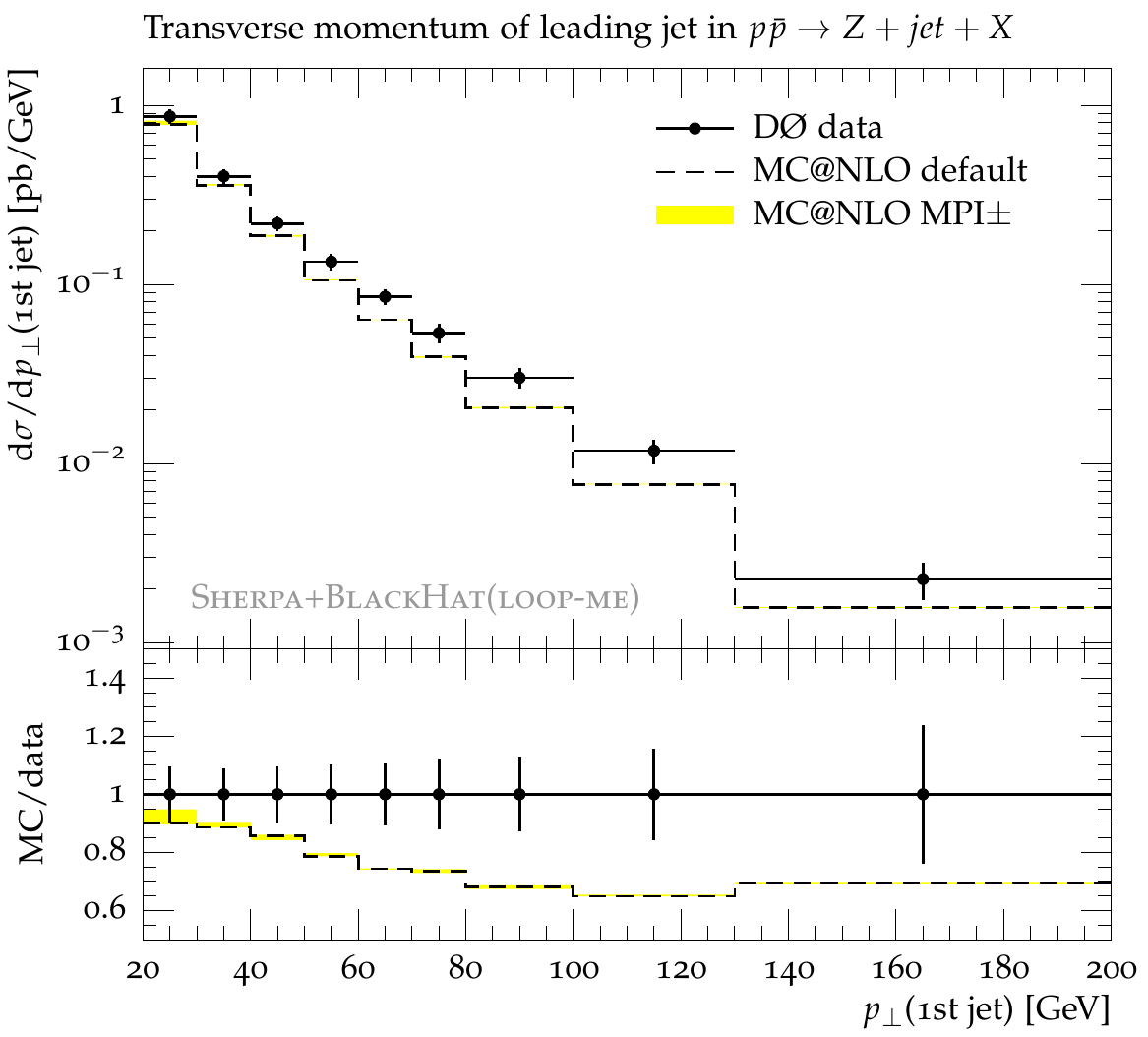}
  \includegraphics[width=0.45\textwidth]{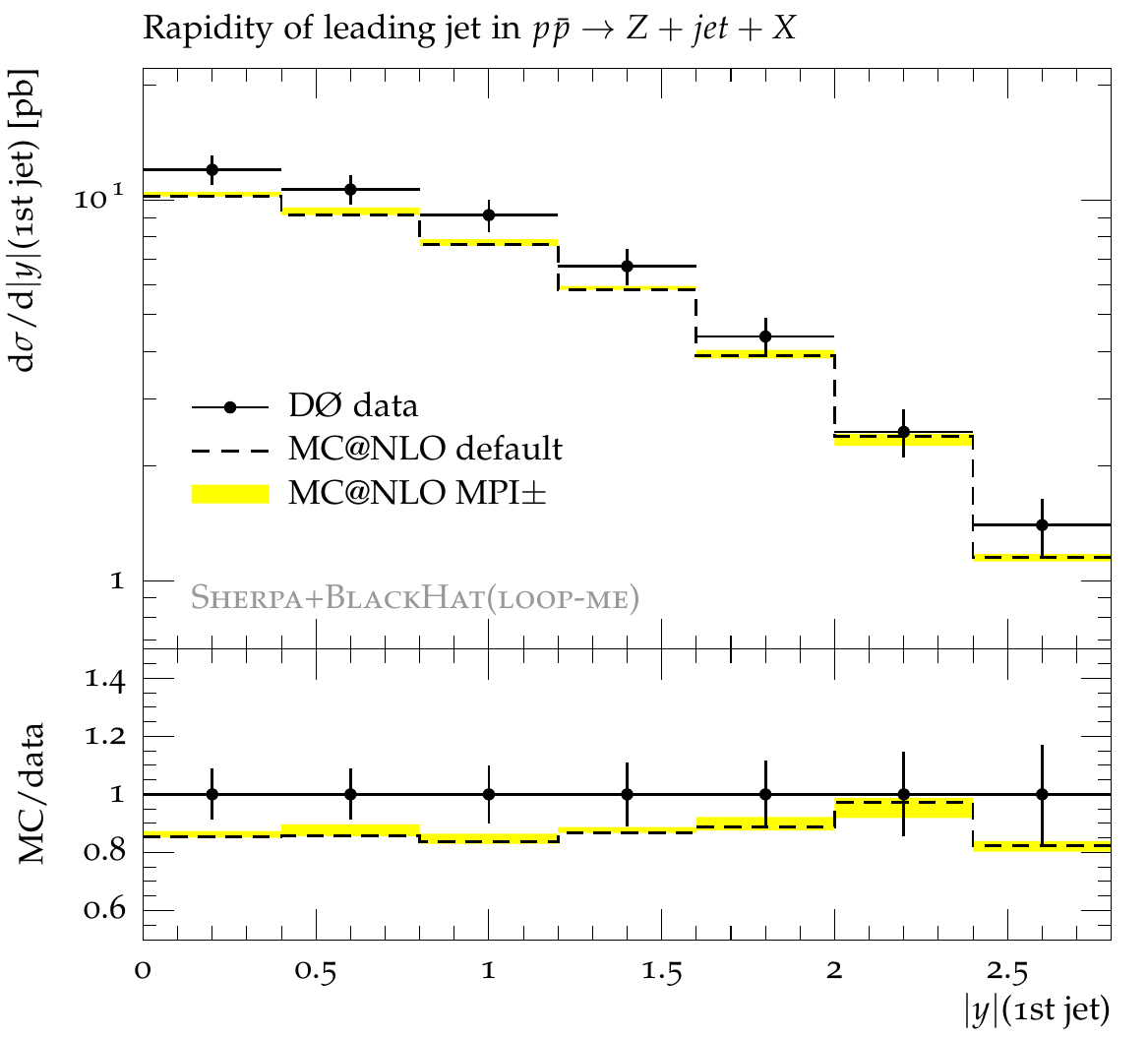}
  \caption{Transverse momentum and rapidity distributions of the reconstructed 
           $Z$ boson (top row), and the total cross section, the transverse 
           momentum and rapidity distributions of the hardest jet (bottom row) 
           in Drell-Yan production in association with at least one jet 
           compared to \DO data \cite{Abazov:2008ez}.\label{Fig:Zj:D0_Zptrap}}
\end{figure}

More characteristics of $Z$-boson plus jet production were investigated in a
recent \DO analysis \cite{Abazov:2008ez}. Events with two muons of invariant mass
$65 < m_{\mu\mu} < 115$ GeV and with at least one jet of $p_\perp > 20$ GeV and 
$|y|<2.8$ were collected at a center-of-mass energy of 1.96~TeV. Jets were defined 
using the \DO midpoint cone algorithm~\cite{Blazey:2000qt} with $R=0.5$ and a split/merge 
fraction of 0.5. Each jet had to be separated from both leptons by $\Delta R_{\mu j}>0.5$. 
A comparison of Monte-Carlo predictions with this measurement is shown 
in Fig.~\ref{Fig:Zj:D0_Zptrap}. The agreement is fair, except for the
$p_\perp$-spectrum of the first jet, where a deficiency of the 
Monte-Carlo result at intermediate and high $p_\perp$ is observed.

\begin{figure}[t]
  \includegraphics[width=0.33\textwidth]{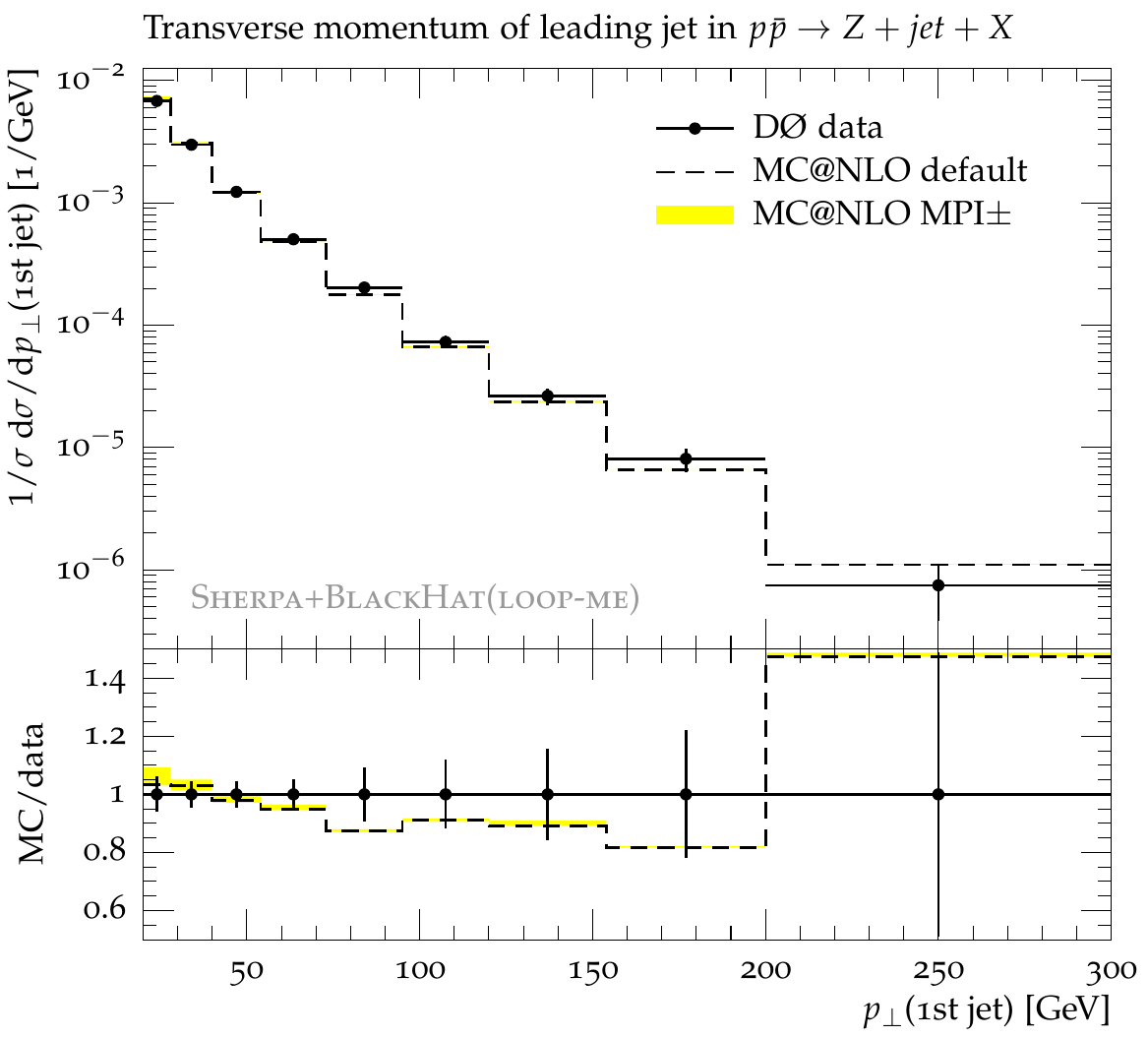}
  \includegraphics[width=0.33\textwidth]{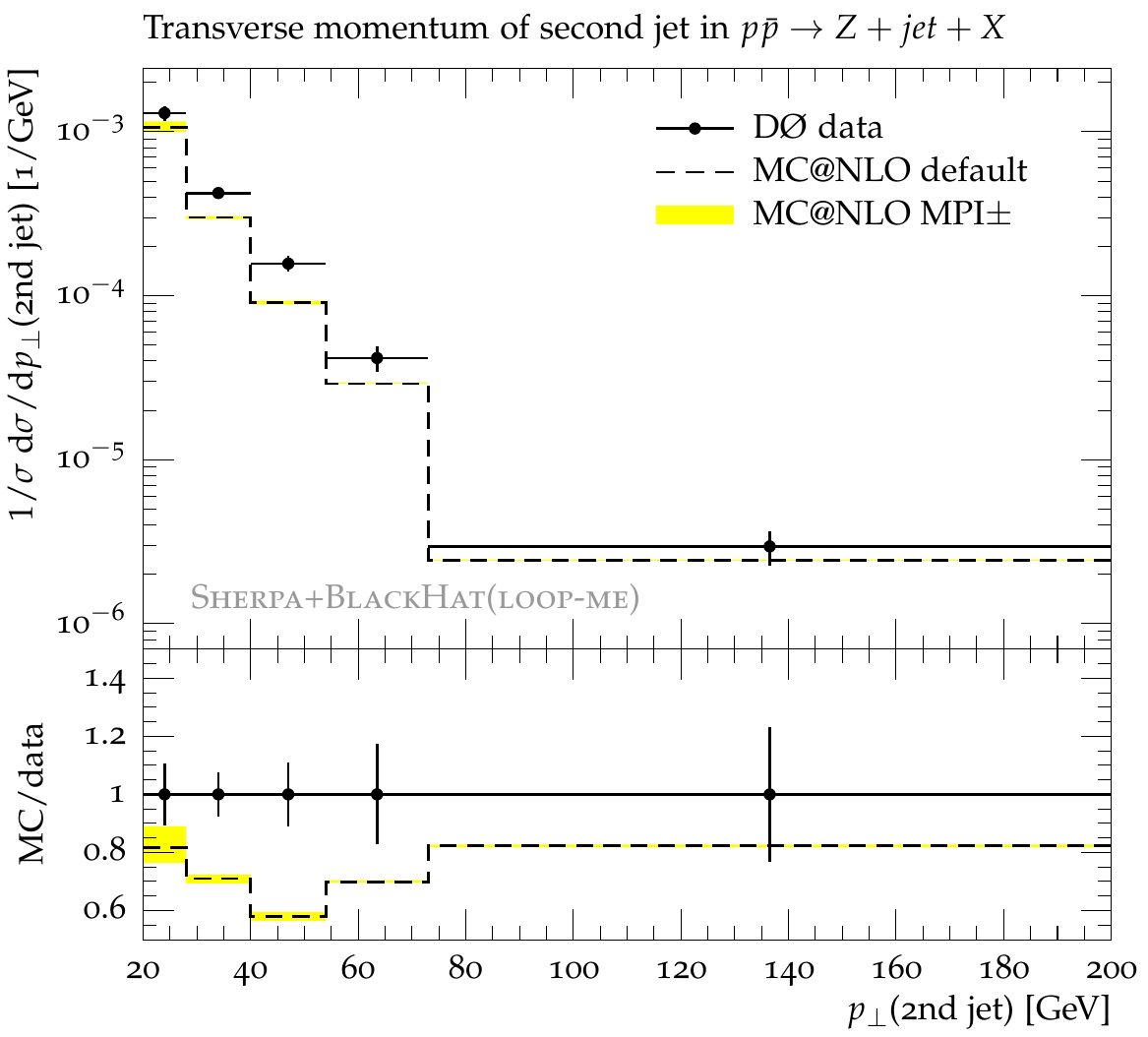}
  \includegraphics[width=0.33\textwidth]{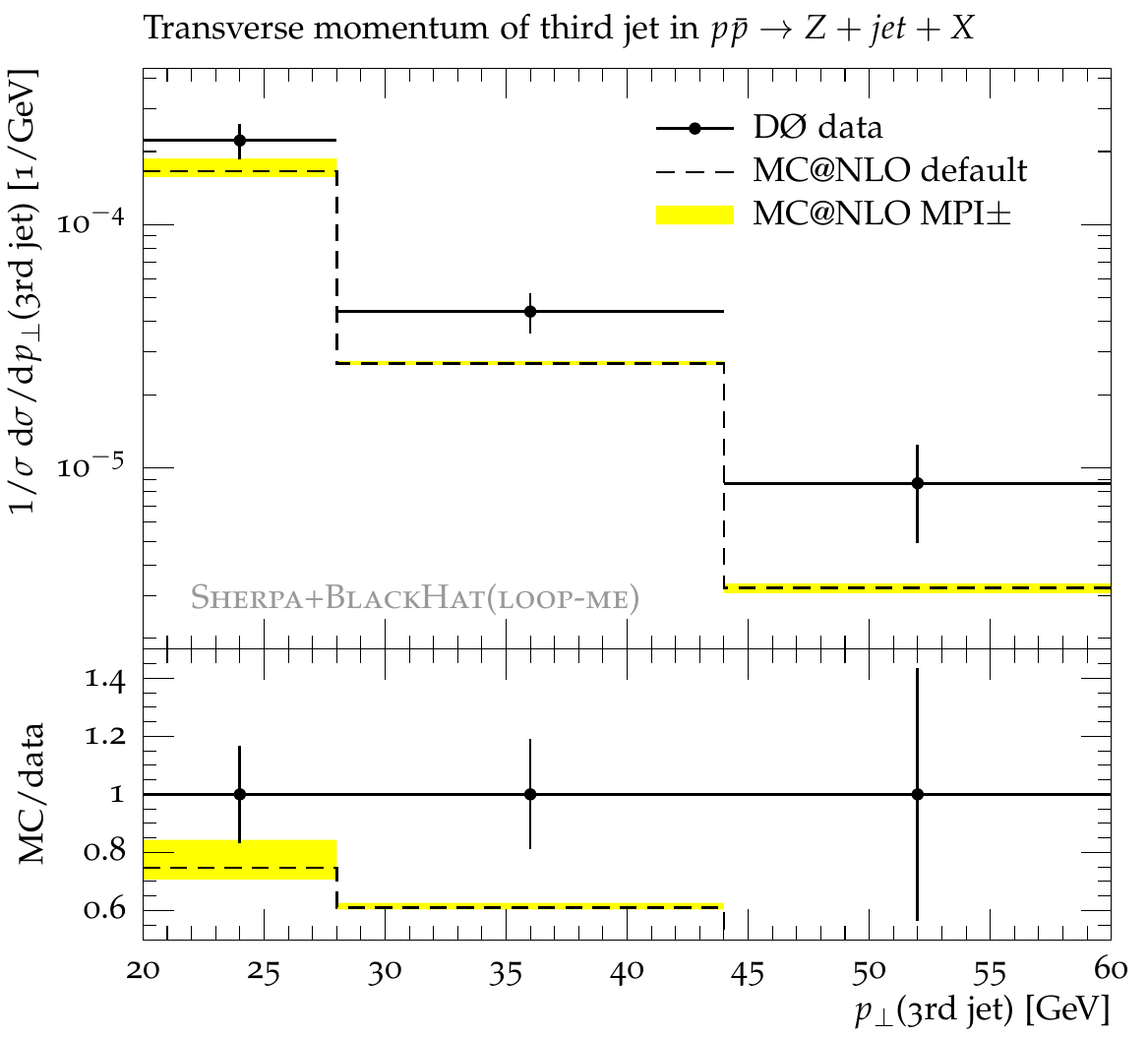}
  \caption{Transverse momentum distributions of the three hardest jets in 
           Drell-Yan production in association with at least one jet compared 
           to \DO data \cite{Abazov:2009av}. All \protect\Sherpa predictions 
           are scaled with a common factor to account for the unknown
           normalisation to the inclusive process. \label{Fig:Zj:D0_jetpt}}
\end{figure}

A further measurement of $Z$+jet production in the electron channel was presented
by the \DO collaboration in~\cite{Abazov:2009av}. Each electron is required to have
$p_\perp>25$ GeV and the mass window $65 < m_{ee} < 115$ GeV is enforced. Jets are defined 
using the midpoint cone algorithm~\cite{Blazey:2000qt} with $R=0.5$ and a split/merge 
fraction of 0.5. At least one jet with $p_\perp>20$ GeV and $|\eta|<2.5$ must be present 
in the event.
Experimental data were normalised to the inclusive Drell-Yan cross section.
This quantity is not predicted by the \MCatNLO simulation of $Z$+jet, and therefore
the \MCatNLO results are scaled by a global factor of 0.35 such that the normalisation
of the one-jet-rate agrees with data.
Figure~\ref{Fig:Zj:D0_jetpt} displays the comparison of these scaled \MCatNLO predictions
with the results from \DO. The transverse momentum shape of the leading jet is well
described by the simulation. The two-jet rate, described here at leading-order accuracy,
seems to be underestimated compared to data, but the shape of the sub-leading jet's
$p_\perp$ spectrum is relatively flat. For the third jet, which is generated in the
parton-shower approximation, both the rate and the shape of the spectrum are not
simulated correctly.

\begin{figure}[t!]
  \includegraphics[width=0.33\textwidth]{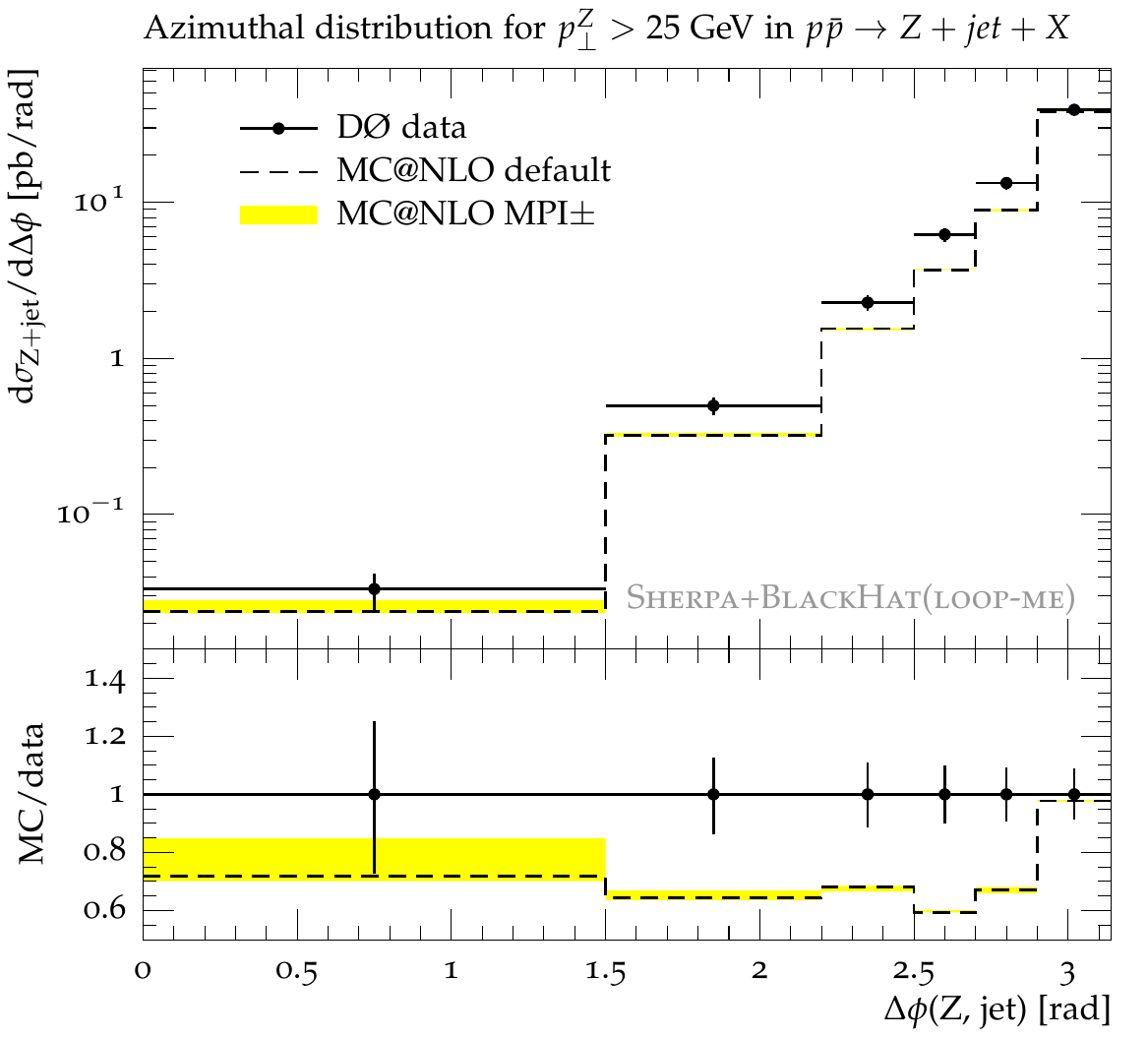}
  \includegraphics[width=0.33\textwidth]{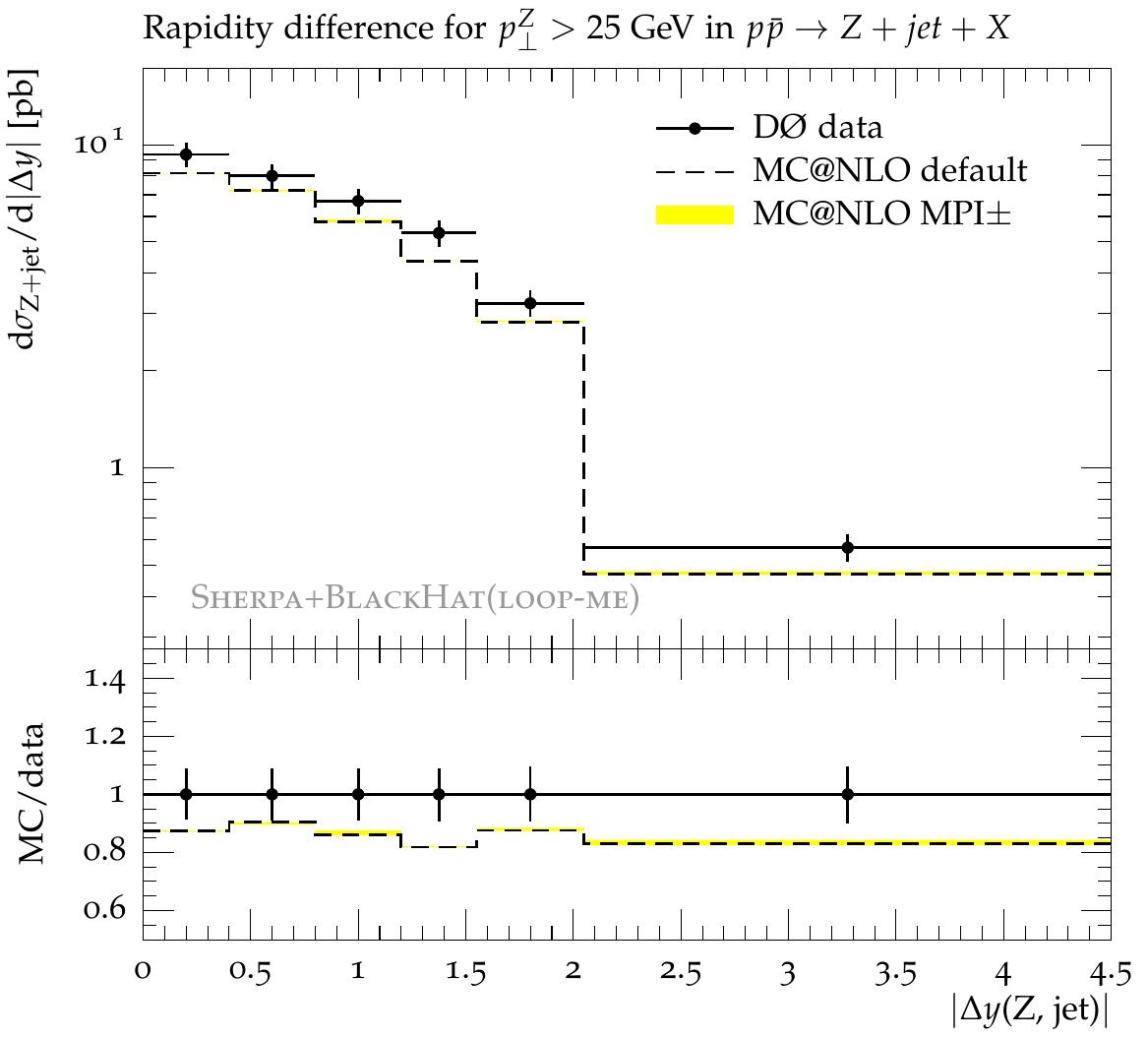}
  \includegraphics[width=0.33\textwidth]{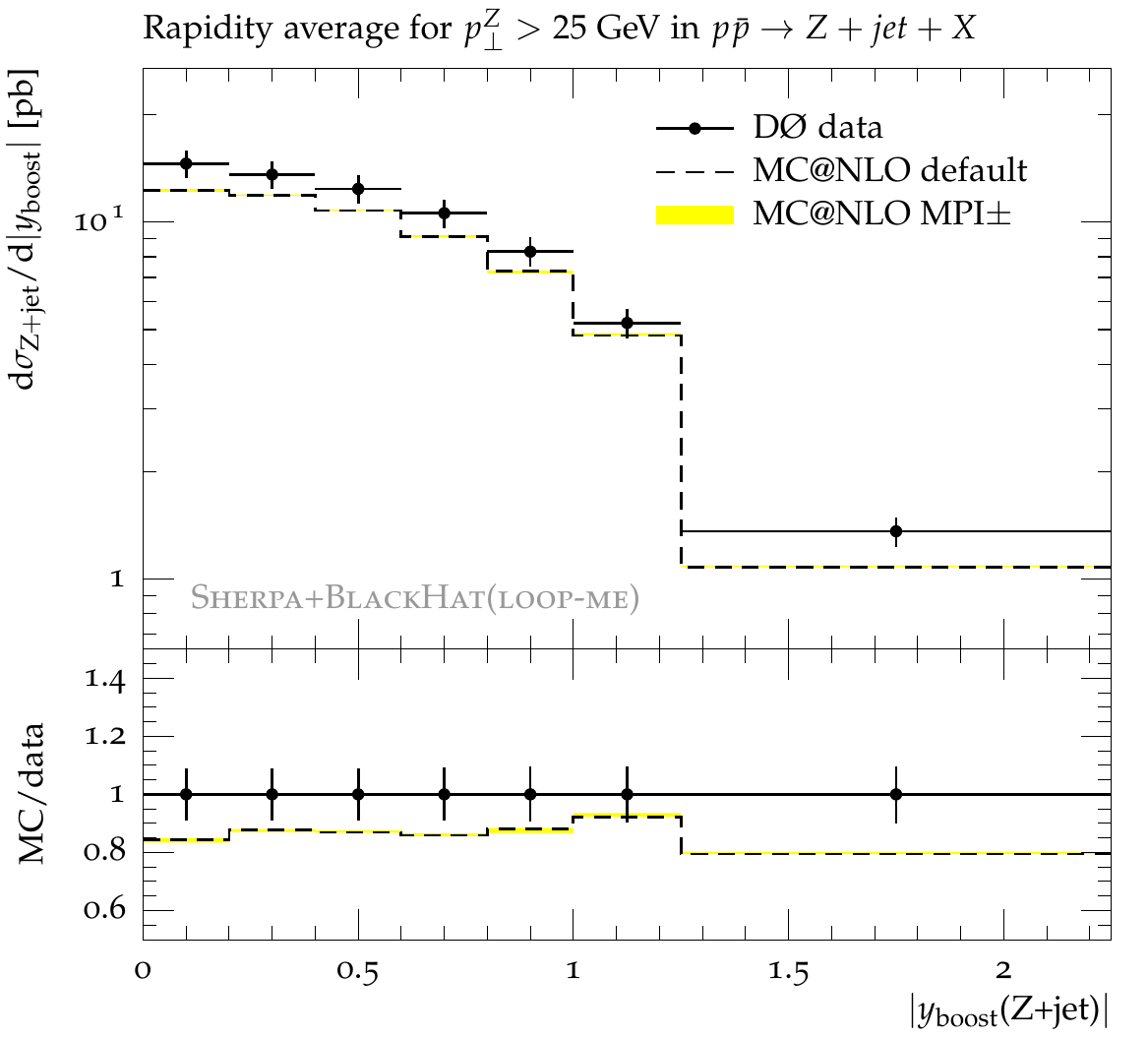}\\
  \includegraphics[width=0.33\textwidth]{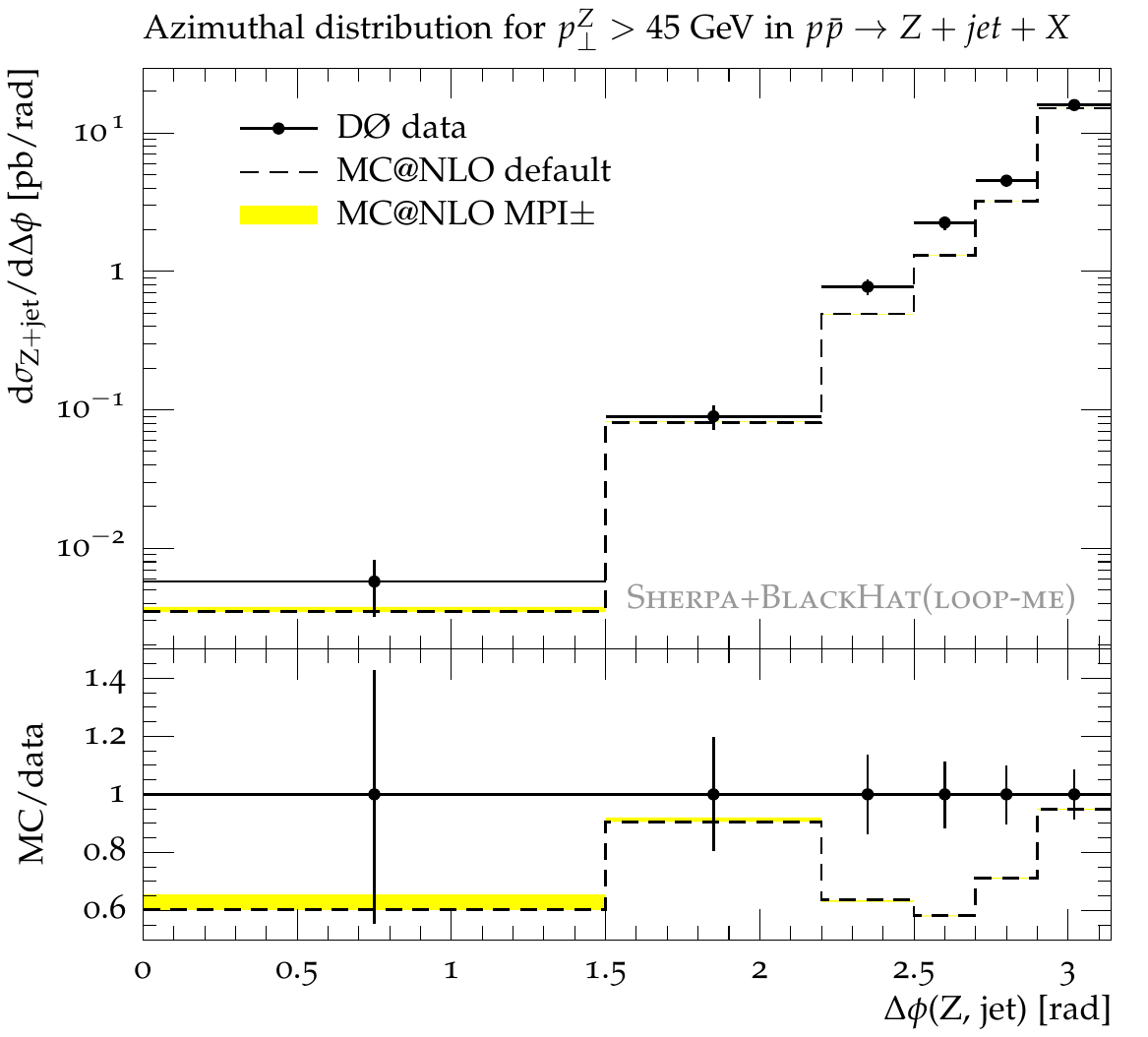}
  \includegraphics[width=0.33\textwidth]{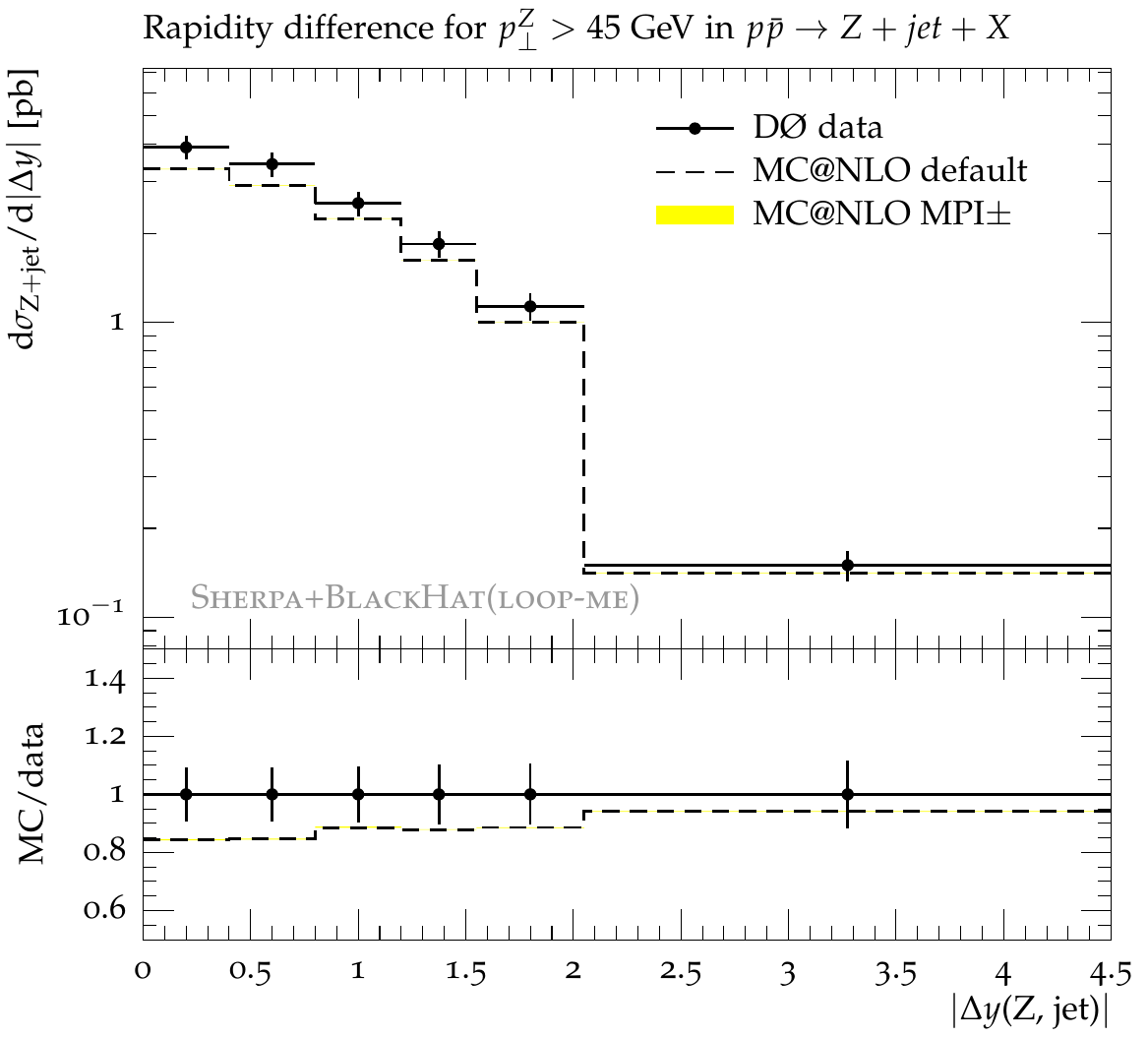}
  \includegraphics[width=0.33\textwidth]{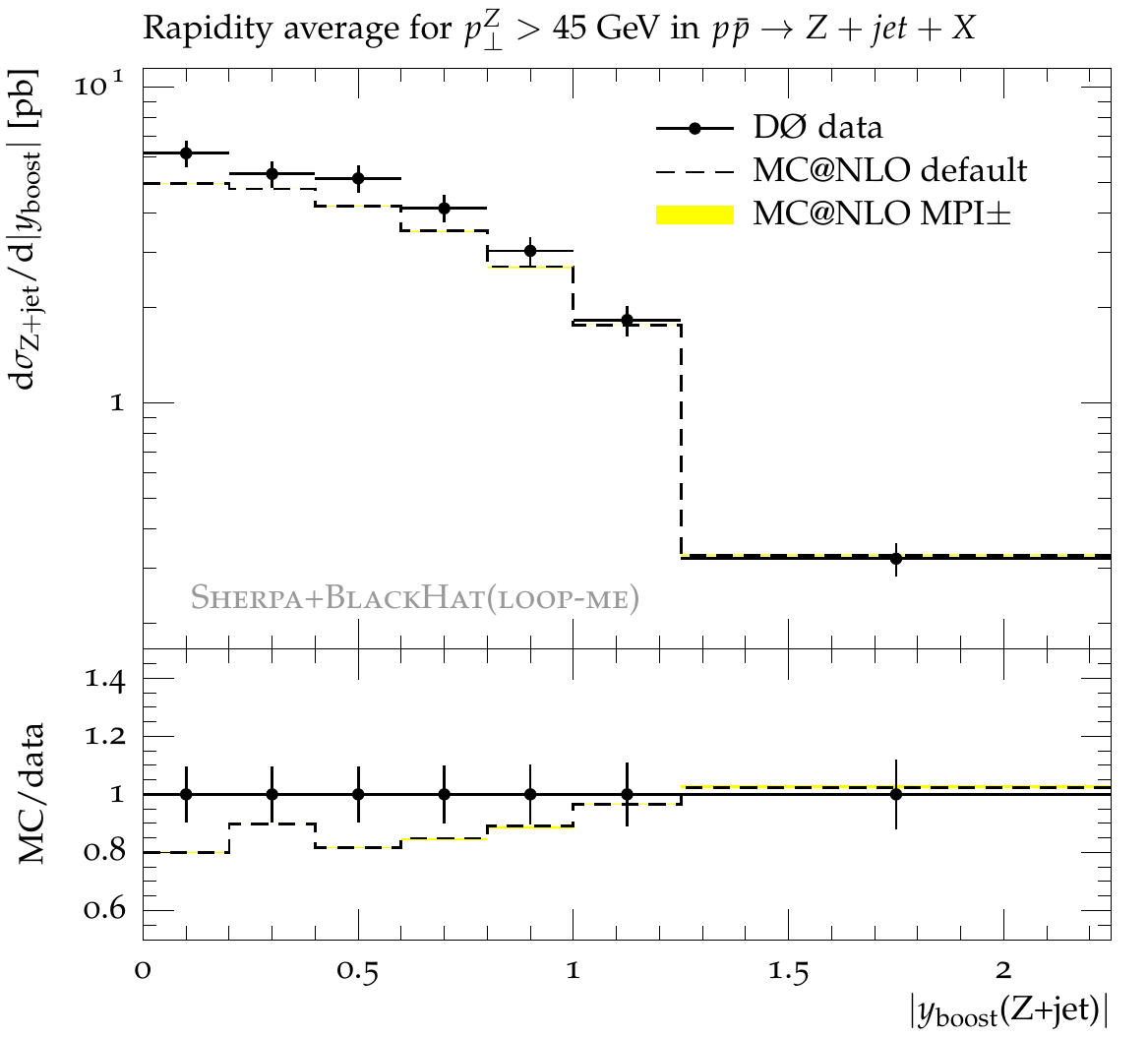}
  \caption{Azimuthal and rapidity difference distributions of the reconstructed 
           $Z$ boson and the hardest jet and their rapidity average in 
           Drell-Yan production in association with at least one jet compared 
           to \DO data \cite{Abazov:2009pp}.\label{Fig:Zj:D0_Zjcorr}}
\end{figure}

To quantify the success of the next-to-leading order calculation it is important not only
to investigate single-particle spectra, but also correlations between the $Z$-boson and the 
hardest jet. They might give some insight into the genuine one-loop effects in $Z$+jet 
production. Therefore, the analysis strategy of a measurement presented 
by the \DO collaboration in~\cite{Abazov:2009pp} is pursued. Opposite-sign muons with $p_\perp>15$ GeV 
and an invariant mass of $65 < m_{\mu\mu} < 115$ GeV are required in association with 
at least one jet of $p_\perp>20$ GeV and $|y|<2.8$. Jets are defined using the midpoint 
cone algorithm~\cite{Blazey:2000qt} with $R=0.5$ and a split/merge fraction of 0.5.
Two event samples are defined, one requiring the transverse momentum of the reconstructed
$Z$~boson to be above 25~GeV, the other requiring it to be above 45~GeV.
Figure \ref{Fig:Zj:D0_Zjcorr} compares \MCatNLO predictions to the measurement.
The shape of the rapidity distributions is matched fairly well, again with the
total rate prediction at the lower end of the uncertainty band. For the azimuthal
correlation distribution the shape shows significant deviations from data: Only
the back-to-back configuration is described well, but for
$\Delta\phi<\pi$ the \MCatNLO prediction underestimates the data. This signals the
uncertainty related to emissions beyond the hardest one, which are only
generated at leading-order or parton-shower accuracy and which are also subject to
large NLO+PS matching systematics as discussed in Sec.~\ref{sec:perturbative}.

\subsection{\texorpdfstring{$W$}{W}+jet production compared to LHC data}

\begin{figure}[t]
  \includegraphics[width=0.33\textwidth]{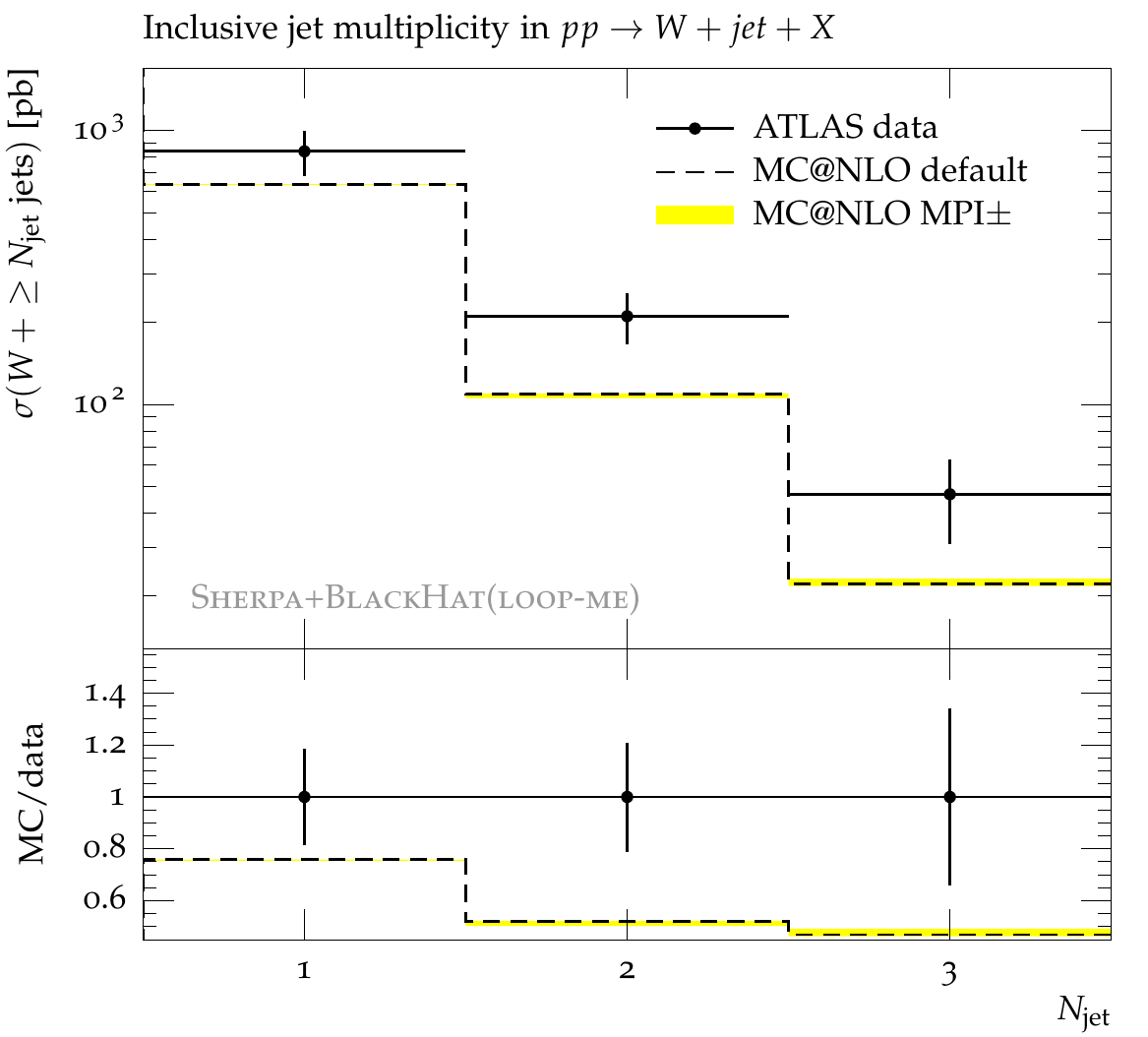}
  \includegraphics[width=0.33\textwidth]{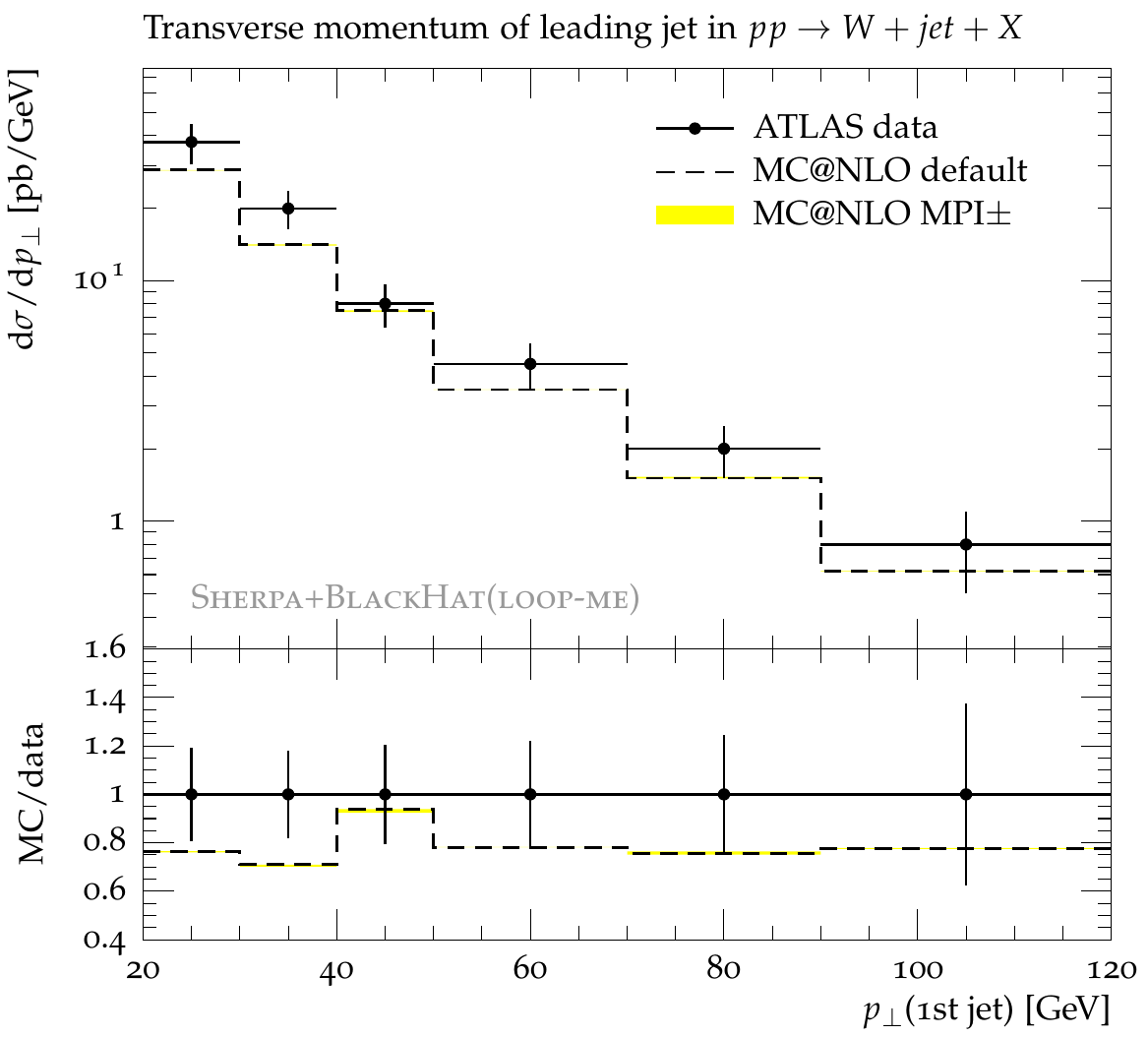}
  \includegraphics[width=0.33\textwidth]{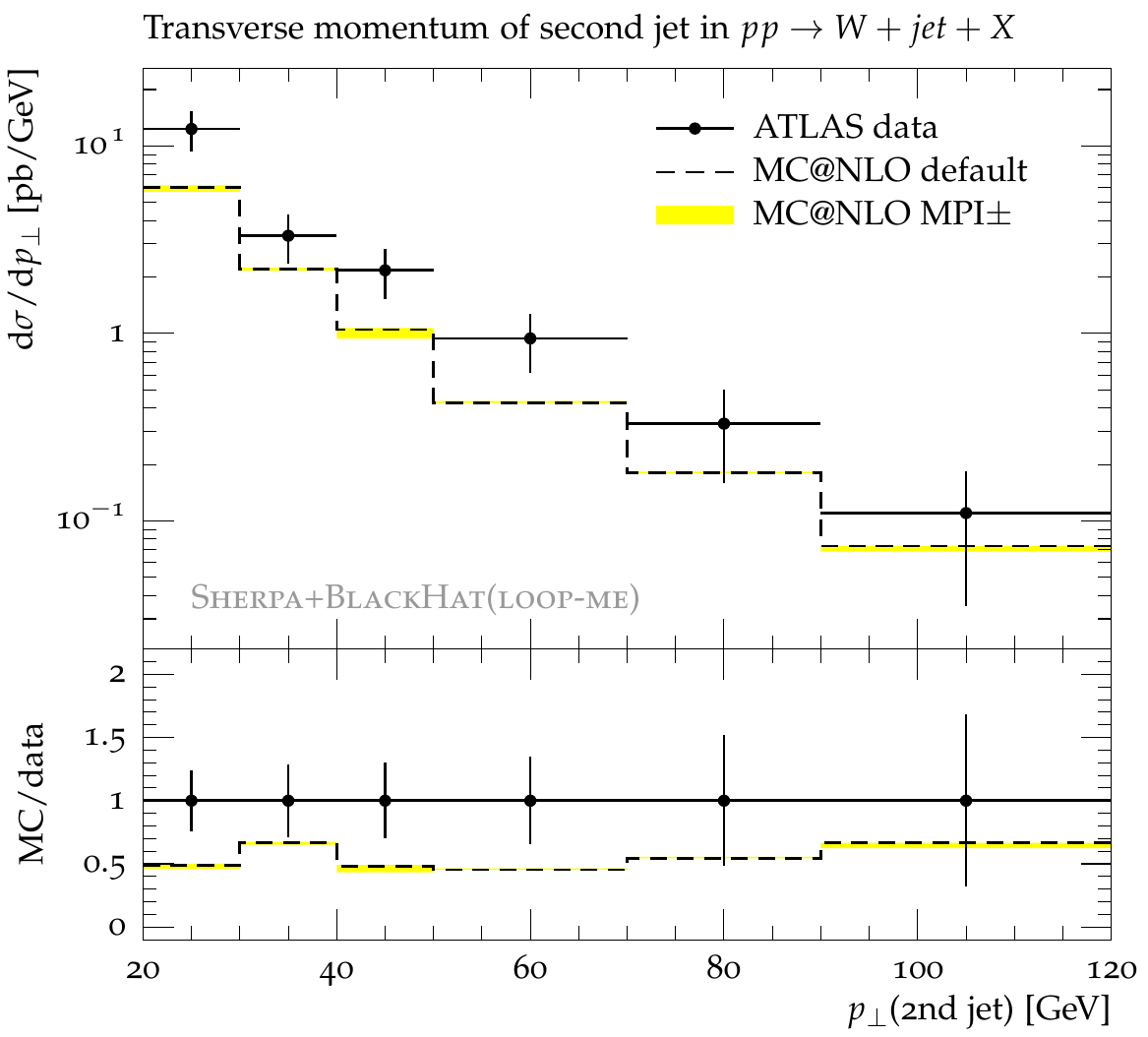}
  \caption{Inclusive jet cross sections and transverse momentum spectra of the
    hardest and second hardest jet in $W[\to e\nu]+j$ production at the LHC
    compared to \ATLAS data~\cite{Aad:2010pg}.
  }
  \label{fig:results:wjets}
\end{figure}

The production of a $W$ boson in association with at least one hard jet in 
proton-proton collisions at $\sqrt{s}=7$ TeV has been studied by the 
\ATLAS collaboration at the CERN LHC~\cite{Aad:2010pg}.

In the electron channel events are selected by requiring an electron with
$p_\perp>20$~GeV defined at the particle level to include all photon radiation
within a $\Delta R=0.1$ cone. Only electrons in the fiducial volume
$|\eta|<1.37$ or $1.52<|\eta|<2.47$ are taken into account.
$E_\perp^\text{miss}$ at the particle level has been defined through the
leading neutrino in the event which is required to have $p_{\perp,\nu}>25$~GeV.
The transverse mass cut is placed at
$m_T=\sqrt{2p_\perp^{\ell}p_\perp^{\nu}\left(1-\cos(\phi_\ell-\phi_\nu)\right)}>40$~GeV.

Jets are reconstructed using the anti-$k_t$ algorithm with $R=0.4$ and have
been taken into account if $p_\perp^{\text{jet}}>20$~GeV, $|\eta_{\text{jet}}|<2.8$
and $\Delta R(\ell, \text{jet})>0.5$. Muons, neutrinos and the leading electron
were excluded from the input of jet reconstruction.

The comparison in Figure~\ref{fig:results:wjets} shows good agreement of
the \Sherpa hadron level prediction with \ATLAS data in the shapes of
differential distributions on the one hand, and discrepancies in the prediction
of the total rate on the other hand. While the one-jet rate is still predicted
fairly well, the two and three-jet rates are significantly too low.
The latter are only predicted at leading-order and parton-shower accuracy and
also subject to the inherent uncertainties from NLO+PS matching.

%% file: text/conclusions.tex
\section{Conclusions}
\label{sec:conclusions}

In this publication, the \MCatNLO and \POWHEG methods to match NLO QCD 
matrix elements with parton showers have been compared,
with a special emphasis on some issues that generate large differences
between their predictions.  In particular, these issues are related to
\begin{itemize}
\item the treatment of sub-leading colour configurations in \MCatNLO
\item the choice of scales in \POWHEG and \MCatNLO and
\item the exponentiation of non-leading terms in \POWHEG.
\end{itemize}

Before discussing the findings related to these issues, it is worth
summarising some other results presented in this publication.  They
refer to the impact of scale variations at the parton-level of the processes
studied here, and to the effect of hadronisation and the underlying event 
on a number of observables:
\begin{itemize}
\item Scale uncertainties\\ 
  In $W/Z$(+jets) production, uncertainties due to scale variations are typically
  about 5-10\%, but they can increase up to about 25\% in the high-$p_\perp$ region
  of jet production, where higher jet multiplicities become relevant.
  In Higgs production processes scale uncertainties tend to be significantly larger,
  by at least a factor of 2, which is indicative of large higher-order corrections.
\item Underlying event corrections and uncertainties\\
  Multiple parton interactions tend to increase jet production rates, and they have
  a particularly large influence on observables which are sensitive to high 
  jet multiplicities, like $H_T$. They typically do not affect inclusive observables,
  with changes of about 5\% or below. Their associated uncertainty is small.
\item Hadronisation corrections and uncertainties\\
  Hadronisation corrections are similar in magnitude to parton-shower
  and multiple-scattering effects. The associated uncertainties are about 5\%,
  except for observables such as jet masses, where they can increase to about 20\%.
\end{itemize}

Let us now turn to the uncertainties induced by the NLO+PS matching.

Regarding the problem of infrared divergences in sub-leading colour 
configurations, it should be noted that up to now, only one critical process 
-- heavy quark pair production -- has been implemented in a publicly available 
and fully documented event generator.  In this case the problem was solved
by introducing a factor that modifies \changed{the integral in Eq.~\eqref{eq:split_bbar}
such that contributions from the region of soft-gluon emission are removed}.
A similar strategy seems to be employed in the recently presented
\aMCatNLO code, but it has not been published or made publicly available yet.
It is therefore not yet clear whether this technique is sufficiently
process-\-independent to allow for a general implementation of \MCatNLO.
In this publication a different approach has been followed, which aims instead 
at constructing an exact solution, irrespective of the process considered.
The basic idea is to correct the Casimir operators used in the parton shower
such that the full colour structure in the soft-gluon limit is obtained.
This introduces both positive and negative splitting functions, which is
unproblematic for NLO calculation but implies ``anti-probabilistic''
evolution in a parton shower. To resolve this issue, a weighting procedure 
has been introduced. It is anticipated that this modification will eventually allow the 
systematic inclusion of sub-leading colour terms in the parton shower.

The exponentiation of non-leading terms in \POWHEG is an issue that has been 
known for some time. Actual differences in how \POWHEG and \MCatNLO
predict the $p_\perp$ spectrum of the Higgs boson are documented
in the literature~\cite{Alioli:2008tz}.  They have typically been
attributed to a different treatment of higher order, in particular NNLO, 
corrections (see Sec.~4.3 of~\cite{Alioli:2008tz}).  We find that the difference
is related to exponentiation uncertainties, naturally affecting the next-to-next-to
leading order, but continuing to all orders.
\changed{The upper scale in the logarithms resummed in \POWHEG} is related to the total 
hadronic centre-of-mass energy, which is substantially different from the scale typically
used in \changed{the resummation programme \cite{Banfi:2004nk,Bozzi:2005wk},} where 
a value of the order of the Higgs boson mass is assumed. It is thus clear that a suitable 
constraint on the emission phase space in \POWHEG must be defined.  
\changed{This problem has already been solved by the \MCatNLO method.}
On a related note, the occurrence of the well-known dip around 
zero in the \MCatNLO simulation of the $y_j-y_H$ distribution, the rapidity difference 
of Higgs boson and accompanying jet, has been studied \changed{here}.  
Cutting on the phase space of 
the Catani-Seymour subtraction kernels and reflecting these cuts in the parton
shower, as enforced by \MCatNLO, yields dead zones of parton emission in
the latter.  While these dead zones are clearly different from the dead zones
in the \Herwig parton shower, the resulting effect is surprisingly similar.  
In fact, both including the full phase space and comparably tight cuts on the 
phase space leads to a vanishing dip; this finding suggests that the dip,
which is not present in the NLO calculation, must be attributed to 
\changed{true exponentiation uncertainties.}

\changed{To complete our implementation of the \MCatNLO method in the sense that only 
logarithms related to emissions on scales below the factorisation scale are exponentiated,}
physically more meaningful cuts on the phase space
of Catani-Seymour subtraction kernels are currently investigated.
Devising a similar technology for \POWHEG would result in two formally completely 
equivalent algorithms, but imply a larger computational effort for event
generation in \POWHEG due to the additional numerical integration in $\bar{B}$
and the matrix-element correction of the weighted parton shower, see~\cite{Hoeche:2010pf}.
For this reason, in favour of the \MCatNLO method, the further development of 
\POWHEG methods in the \Sherpa framework will be abandoned. 

%% file: text/acknowledgements.tex
\section*{Acknowledgements}
SH's work was supported by the US Department of Energy under contract 
DE--AC02--76SF00515, in part by the US National Science Foundation, grant 
NSF--PHY--0705682, (The LHC Theory Initiative), and in part by the US National 
Science Foundation under grant NSF--PHY--0551164.  MS's work was supported by 
the Research Executive Agency (REA) of the European Union under the Grant 
Agreement number PITN-GA-2010-264564 (LHCPhenoNet). FS's work was supported
by the German Research Foundation (DFG) via grant DI 784/2-1.

The authors would like to thank Massimiliano Grazzini and Stefano Catani for
fruitful discussion.
FK would also like to thank the Galileo Galilei Institute in Florence and 
the SLAC theory group for their hospitality while this work was finished.
SH thanks the CP$^\text{3}$-Origins Odense for hospitality.

%% file: text/appendix.tex
\section{Comment on the NLL accuracy of the \texorpdfstring{\POWHEG}{POWHEG} formula}
\label{sec:powershower}
\changed{
In \cite{Nason:2006hfa} it was proven that the \POWHEG formula yields
NLL accurate predictions, if the strong coupling is modified according 
to~\cite{Catani:1990rr}.
The proof rests on the crucial assumption that when logarithms of the form
$L=\log(p_\perp/Q)$ are resummed, the dependence on the resummation scale $Q$
introduces corrections beyond next-to-leading logarithmic order. 
This argument holds as long as $Q$ differs from the hard scale in the process,
say $m_h$ in Higgs-boson production, by a factor of order 
one~\cite{Dasgupta:2001eq,*Dasgupta:2002dc,Banfi:2004nk,Bozzi:2005wk,Bozzi:2010xn}.
As pointed out in~\cite{Banfi:2004nk} large spurious subleading logarithms
may otherwise be introduced, that are associated with the overall normalization
of observables, rather than radiative corrections. In other words, in the large-$k_T$
region the resummed results lose their predictivity and should be replaced by
conventional fixed-order calculations~\cite{Bozzi:2010xn}. Following these arguments,
the reasoning in \cite{Nason:2006hfa} is only correct as long as the resummation
scale and the hard scale are sufficiently similar. This is not the case in the 
\POWHEG method when it is applied to hadronic collisions where $Q\to E_\text{cms}$, 
the total hadronic centre-of-mass energy, while the hard scale is of order $m_h$,
for example.
}